%% file: draft.tex
\documentclass[sigplan]{acmart}
\renewcommand\footnotetextcopyrightpermission[1]{}
\setcopyright{acmcopyright}
\copyrightyear{2018}
\acmYear{2018}
\acmDOI{10.1145/1122445.1122456}

\acmConference{}{}{}
\acmBooktitle{}
\acmPrice{}
\acmISBN{}

\usepackage{enumitem}
\usepackage{verbatim}
\usepackage{booktabs}
\usepackage{soul}
\usepackage{xspace}
\usepackage{color}
\usepackage{xcolor}
\usepackage{upquote}
\usepackage{listings}
\usepackage{amsmath}
\usepackage{algorithm2e}
\usepackage{syntax}
\usepackage{adjustbox}
\usepackage{array}
\usepackage{rotating}
\usepackage{picins}
\usepackage{comment}
\usepackage{float}

\usepackage{subcaption}

\usepackage{longtable}
\usepackage{siunitx}

\setlist{noitemsep,leftmargin=10pt,topsep=2pt,parsep=2pt,partopsep=2pt}

\input{defs}

\newcommand{\ie}{{\em i.e.}, }

\newcommand{\heading}[1]{\vspace{4pt}\noindent\textbf{#1}:\enspace}
\newcommand{\ttt}[1]{\texttt{#1}}

\newcommand{\sys}{{\scshape KumQuat}\xspace}
\newcommand{\unix}{{\scshape Unix}\xspace}

\lstdefinelanguage{sh}{
  morekeywords={for, in, do, done, \|},
  keywordstyle=\color{purple}\ttfamily,
  ndkeywordstyle=\color{black}\ttfamily\bfseries,
  identifierstyle=\color{black},
  sensitive=false,
  comment=[l]{\#},
  commentstyle=\color{lightgray},
  stringstyle=\color{darkgray}\ttfamily,
  morestring=[b]',
  morestring=[b]",
  abovecaptionskip=0pt,
  aboveskip=0pt,
  belowcaptionskip=0pt,
  belowskip=0pt,
  frame=none
}

\lstdefinelanguage{js}{
  morekeywords={typeof, let, new, true, false, catch, function, return, null, catch, switch, var, if, in, while, do, else, case, break},
  keywordstyle=\color{purple}\ttfamily,
  keywordstyle=[3]\color{orange},
  keywords=[3]{SBX, PROC},
  ndkeywords={class, export, require, boolean, throw, implements, import, this},
  ndkeywordstyle=\color{cyan}\ttfamily,
  identifierstyle=\color{black},
  sensitive=false,
  comment=[l]{//},
  commentstyle=\color{lightgray},
  stringstyle=\color{blue}\ttfamily,
  basicstyle=\footnotesize\ttfamily,
  numberstyle=\tiny\color{gray},
  morestring=[b]',
  morestring=[b]",
}

\lstdefinelanguage{es}{
  morekeywords={string, number, bool, p, v, e, x},
  keywordstyle=\color{darkgray},
  keywordstyle=[3]\color{orange},
  keywords=[3]{SBX, PROC},
  ndkeywords={undefined, null, delete},
  ndkeywordstyle=\color{black}\ttfamily\bfseries,
  identifierstyle=\color{black},
  sensitive=false,
  comment=[l]{//},
  commentstyle=\color{lightgray},
  stringstyle=\color{darkgray}\ttfamily,
  morestring=[b]',
  morestring=[b]",
  abovecaptionskip=0em,
  aboveskip=0em,
  belowcaptionskip=0em,
  belowskip=0em,
  frame=none,
}

\newlist{mycases}{enumerate}{3} 
\setlist[mycases,1]{
  label=\underline{Case \arabic*:},
  ref={Case \arabic*},
  leftmargin=*
} 
\setlist[mycases,2]{
  label=\underline{Case \arabic{mycasesi}.\arabic*:},
  ref={Case \arabic{mycasesi}.\arabic*},
  leftmargin=*
} 
\setlist[mycases,3]{
  label=\underline{Case \arabic{mycasesi}.\arabic{mycasesii}.\arabic*:},
  ref={Case \arabic{mycasesi}.\arabic{mycasesii}.\arabic*},
  leftmargin=*
}

\usepackage{amsthm} 
\usepackage{thmtools, thm-restate} 

\declaretheoremstyle[headfont=\normalfont\bfseries,bodyfont=\normalfont]{mytheoremstyle}
\declaretheorem[style=mytheoremstyle]{theorem}
\declaretheorem[style=mytheoremstyle,numberwithin=section]{lemma}
\declaretheorem[style=mytheoremstyle,numberwithin=section]{proposition}

\declaretheoremstyle[bodyfont=\normalfont]{mydefinitionstyle}
\declaretheorem[style=mydefinitionstyle,numberwithin=section]{definition}

\declaretheoremstyle[headfont=\normalfont\itshape,bodyfont=\normalfont]{myremarkstyle}
\declaretheorem[numbered=no,style=myremarkstyle]{remark}
\declaretheorem[style=myremarkstyle]{example}

\usepackage{makecell}

\input{macro}

\begin{document}

\date{}

\title{Automatic Synthesis of Parallel Unix Commands and Pipelines with \sys}

\author{Jiasi Shen}
\affiliation{
  \institution{MIT EECS \& CSAIL}
  \country{USA}
}
\email{jiasi@csail.mit.edu}

\author{Martin Rinard}
\affiliation{
  \institution{MIT EECS \& CSAIL}
  \country{USA}
}
\email{rinard@csail.mit.edu}

\author{Nikos Vasilakis}
\affiliation{
  \institution{MIT EECS \& CSAIL}
  \country{USA}
}
\email{nikos@vasilak.is}

\input{abstract}

\maketitle

\input{introduction}

\input{example}

\input{design}

\input{results}

\input{related}

\input{acknowledgment}

\bibliographystyle{ACM-Reference-Format}
\bibliography{./bib}

\appendix

\input{appendix}

\end{document}

%% file: defs.tex
\usepackage{esvect}

\usepackage{stmaryrd}
\usepackage{amsmath}
\usepackage{pgf}
\usepackage{tikz}
\usetikzlibrary{arrows,automata,positioning, shapes, shapes.geometric, fit, calc}

\usepackage{amssymb}
\usepackage{proof}
\usepackage{enumitem}

\newcommand{\binopdef}     \oplus
\newcommand{\unopdef}      \ominus

\makeatletter
\DeclareRobustCommand*\cal{\@fontswitch\relax\mathcal}
\makeatother

    \newcommand{\infral} [3] {\infer[\textsc{#3}]{\begin{array}{c} #2 \end{array} }{ \begin{array}{c} #1  \end{array} } }

\makeatletter
\renewcommand\section{\@startsection{section}{1}{\z@}%
                       {-8\p@ \@plus -4\p@ \@minus -4\p@}%
                       {6\p@ \@plus 4\p@ \@minus 4\p@}%
                       {\normalfont\large\bfseries\boldmath
                        \rightskip=\z@ \@plus 8em\pretolerance=10000 }}
\renewcommand\subsection{\@startsection{subsection}{2}{\z@}%
                       {-8\p@ \@plus -4\p@ \@minus -4\p@}%
                       {6\p@ \@plus 4\p@ \@minus 4\p@}%
                       {\normalfont\normalsize\bfseries\boldmath
                        \rightskip=\z@ \@plus 8em\pretolerance=10000 }}
\renewcommand\subsubsection{\@startsection{subsubsection}{3}{\z@}%
                       {-4\p@ \@plus -4\p@ \@minus -4\p@}%
                       {-1.5em \@plus -0.22em \@minus -0.1em}%
                       {\normalfont\normalsize\bfseries\boldmath}}
\makeatother

%% file: macro.tex
\newcommand{\DSLCombiner}[1]{\mathsf{Combiner}_{#1}}
\newcommand{\DSLBasic}{\mathsf{RecOp}}
\newcommand{\DSLStruct}{\mathsf{StructOp}}
\newcommand{\DSLRun}[1]{\mathsf{RunOp}_{#1}}

\newcommand{\GBasic}{G_\text{rec}}
\newcommand{\GStruct}{G_\text{struct}}

\newcommand{\Eval}[4]{#1~#2~#3 \Longrightarrow_e\allowbreak #4}
\newcommand{\SetLegal}[1]{L(#1)}
\newcommand{\BoolTable}[1]{T(#1)}
\newcommand{\BoolEquivCap}[2]{#1 \equiv_{\cap} #2}
\newcommand{\BoolSat}[2]{P(#1, #2)}
\newcommand{\BoolEnough}[2]{E(#1, #2)}
\newcommand{\BoolEnoughBasic}[1]{E_\text{rec}(#1)}
\newcommand{\BoolEnoughStruct}[1]{E_\text{struct}(#1)}
\newcommand{\Count}[2]{C(#1, #2)}
\newcommand{\Size}[1]{\left| #1 \right|}
\newcommand{\SetSat}[2]{P_{#1}(#2)}

\newcommand{\ttx}[1]{\ttt{x}_{#1}}
\newcommand{\tty}[1]{\ttt{y}_{#1}}

\newcommand{\CombN}[3]{G^{#1,#2}_{#3}}

\newcommand{\Sat}[0]{\sim}
\newcommand{\Correct}[3]{\BoolSat{#1}{#2(#3)}}

%% file: abstract.tex
\begin{abstract}
We present \sys, a system for automatically generating
data parallel implementations of \unix shell commands and pipelines. The generated parallel versions split input streams, execute multiple instantiations of the original pipeline commands to process the splits in parallel, then combine the resulting parallel outputs to produce the final output stream. \sys automatically synthesizes the combine operators, with a domain-specific combiner language acting as a strong regularizer that promotes efficient inference of correct combiners.

We evaluate \sys on 70 benchmark scripts that together have a total of 427 stages.
\sys synthesizes a correct combiner for 113 of the 121 unique commands that appear in these benchmark scripts. The synthesis times vary between 39 seconds and 331 seconds with a median of 60 seconds. We present experimental results that show that these combiners enable the effective parallelization of our benchmark scripts.

\end{abstract}

%% file: introduction.tex
\section{Introduction}
\label{intro}

The \unix shell, working in tandem with the wide range of commands it supports, provides a convenient programming environment for many stream processing computations. Shell commands---which can be written in multiple languages---typically execute sequentially on a single processor. This sequential execution often leaves available data parallelism, in which a command operates on different parts of an input stream in parallel, unexploited. This observation has motivated the development of systems that exploit data parallelism in shell pipelines~\cite{posh,pash}. A key prerequisite is obtaining the combiners required to merge the resulting multiple parallel output streams correctly into a single output stream. Previous systems rely on developers to manually implement such combiners and associate them with their corresponding shell commands~\cite{posh,pash}.

We present a new system, \sys, for automatically exploiting data parallelism available in \unix pipelines. Working with the commands in the pipeline as black boxes, \sys automatically generates inputs that explore the behavior of the command to infer and automatically generate a combiner for the command. This capability enables \sys to automatically generate data parallel versions of \unix pipelines, including pipelines that contain new commands or command options for which combiners were previously unavailable.

\sys targets commands that can be expressed as data parallel divide-and-conquer computations with two phases:\footnote{There is no requirement that the actual internal implementation must be structured as a divide-and-conquer computation --- because \sys interacts with the command as a black box, the requirement is instead only that the computation that it implements can be expressed in this way.}
  the first phase executes the original, unmodified command in parallel on disjoint parts of the input;
  the second phase combines the partial results from the first phase to obtain the final output. \sys generates candidate combiners, then repeatedly feeds selected inputs to parallelized versions of the command that use the candidate combiners. A comparison of the resulting outputs with corresponding outputs from the original serial version of the command enables \sys to identify a correct combiner for the command. A domain-specific combiner language acts as a strong regularizer that promotes efficient learning of correct combiners. The resulting (automatically generated) parallel computation executes directly in the same environment and with the same program and data locations as the original sequential command.

This paper makes the following contributions:
  \begin{itemize}
      \item {\bf Algorithm:} We present a new algorithm that automatically synthesizes  combiners for parallel and distributed versions of \unix commands. The resulting synthesized combiners enable the automatic generation of parallel and distributed versions of \unix commands and pipelines. 
     \item {\bf Domain-Specific Language:} We present a domain-spe\-cif\-ic language 
      for combiner operators. This language supports both the class of combiner operators
      relevant to this domain and an efficient algorithm for 
      automatically synthesizing these combiner operators.
      \item {\bf Correct Combiners:} We present theorems (\autoref{thm:synthesizebasic} and \autoref{thm:synthesizestruct}) that characterize when the combiner synthesis algorithm will identify a correct combiner for a given set of parallel output streams. 
      
      We also present an analysis of the interaction between our input generation algorithm, our benchmark commands, and our combiner synthesis algorithm. This analysis identifies why \sys generates correct combiners for the 113 of 121 benchmark commands for which correct combiners exist, identifies command patterns that ensure correct combiner synthesis and correct data parallel execution, and provides insight into the rationale behind the \sys design and the reasons why the \sys design can effectively exploit data parallelism available in its target class of commands. 
 
      \item {\bf Experimental Results:} We present experimental results that characterize the effectiveness of \sys on a set of 70 benchmark scripts that together have a total of 477 commands, among which 121 are unique and process an input stream.
The results show that \sys can effectively synthesize combiners for the majority of our benchmark commands and that these synthesized combiners enable effective parallelizations of our benchmark \unix pipelines.
  \end{itemize}

\noindent

%% file: example.tex
\section{Example}
\label{bg}

\autoref{fig:wf} presents an example pipeline that we use to illustrate  \sys. The pipeline implements a computation that counts the frequency of words in an input document. The six commands in the pipeline (1) read the input document, (2) break the document into lines of words, 
(3) translate the words into lower case, (4) sort the words, (5) remove duplicates and prepend each unique word with a count, and (6) sort the words on their counts in reverse order. 

The pipeline conforms to a standard \unix model that
structures computations as pipelines of building-block commands that process character streams. The example commands process the streams as lines of words, with the lines and words separated by delimiters, in the example newline and space. As the commands process lines, words, or characters, they apply a function to each unit and either output the result of the function (commands ``\ttt{tr -cs A-Za-z '\textbackslash n'}'' and ``\ttt{tr A-Z a-z}''), sort the units according to a certain order (commands ``\ttt{sort}'' and ``\ttt{sort -rn}''), or accumulate a result that is output when the command finishes reading the input stream and terminates (command ``\ttt{uniq -c}'').

To exploit the data parallelism available in this computation, \sys splits the input data stream into substreams, then instantiates the commands to process the input substreams in parallel. The result is a set of parallel output substreams that must then be combined to obtain the single final output stream of the command.

Different commands often require different combine operators.
The combine operator for
command ``\ttt{tr A-Z a-z}'' simply concatenates the output substreams. 
The combine operator for command ``\ttt{tr -cs A-Za-z '\textbackslash n'}'' concatenates the output substreams, then reruns the command on this concatenated stream.\footnote{Simple concatenation is incorrect due to potential empty lines at the split boundary.}
Note that \ttt{tr} commands with different flags may have different combine operators.
The combine operators for sort commands apply an appropriate merge function, which may depend on the sort flag that specifies the comparison function.
The ``\ttt{uniq -c}'' command produces a stream of (count, word) pairs. Given two streams $\tty{1}$ and $\tty{2}$, the combine operator 
compares the word in the last line of $\tty{1}$ with the word in the first line of $\tty{2}$. If they are the same, it concatenates $\tty{1}$ and $\tty{2}$ but combines the last and first lines to include the sum of the two word counts. Otherwise it simply concatenates $\tty{1}$ and $\tty{2}$. As these examples highlight, selecting an appropriate combine operator for each command is a critical step for obtaining a correct parallel execution.

The default \sys parallel computation applies the combine operator after the parallel execution of each command to obtain a single output stream for that command. In many cases, however, it is possible to enhance parallel performance by eliminating intermediate combine operators so that the output substreams for one command feed directly into the input substreams for the direct parallel execution of the next command. \sys therefore applies an optimization that automatically eliminates intermediate combine operators when possible (\autoref{sec:opt}).

\begin{figure}
\begin{lstlisting}[language=sh, numbers=none]
  cat $IN | tr -cs A-Za-z '\n' | tr A-Z a-z |
      sort | uniq -c | sort -rn
\end{lstlisting}
\caption{Example pipeline that computes word frequencies~\cite{pash,bentley1986literate}.}
\label{fig:wf}
\end{figure}

\heading{Model of Computation}
\sys targets commands $f$ that have a combine operator $g$ that satisfies
$$f(\ttx{1}~{+}{+}~\ttx{2}) = g(f(\ttx{1}), f(\ttx{2}))$$ for all input streams $\ttx{1},\ttx{2}$, where the streams are (potentially recursively) structured as units separated by delimiters.
\sys currently targets character streams structured as lines with the newline delimiter, so that $\ttx{1}$ and $\ttx{2}$ terminate with newlines. ${+}{+}$ denotes string concatenation. A key step in the parallelization of $f$ is the synthesis of a correct combiner $g$ for $f$. To focus the synthesis algorithm on a productive space of candidate combiners, \sys works with combiners expressible in a domain-specific combiner language (\autoref{fig:dsl}).

\heading{Combiner Synthesis} 
To infer a combiner $g$ for a command $f$, \sys works with a set of candidate combiner functions. In the current \sys implementation, this set consists of all combiner functions with seven or fewer nodes in the DSL abstract syntax tree. \sys repeatedly generates input streams $\ttx{1}$ and $\ttx{2}$, feeds the input streams to the serial and parallel versions of $f$ instantiated with candidate combiners $g$, and compares the resulting serial and parallel outputs to discard candidate combiners $g$ that do not satisfy the equation $f(\ttx{1}~{+}{+}~\ttx{2}) = g(f(\ttx{1}), f(\ttx{2}))$.
The input stream generation algorithm is designed to produce inputs that quickly find and discard incorrect combiners (see below). With these generated inputs, we have found that the combiner synthesis algorithm typically converges quickly to a few semantically equivalent correct combiners (\autoref{sec:results}).

\heading{Input Generation}
\sys uses a set of {\em input shapes} to specify the format of each generated input stream.
An input shape specifies the number of lines in the input stream, the number of words per line, and the number of characters per word. The input shape also specifies how diverse these input units are, in terms of the percentage of distinct lines, words, and characters.
These input shapes are designed to generate meaningful inputs for commands that conform to our model of computation. A goal is to efficiently generate  counterexample inputs that cause the command to produce counterexample outputs that enable \sys to identify and discard incorrect candidate combiners. 

The design of input shapes is inspired by the observation that certain input shapes cause commands to produce outputs that enable \sys to identify and discard incorrect combiners. For example, when  $f = (\ttt{tr -cs A-Za-z '\textbackslash n'})$ and $g = \mathbf{concat}$, a counterexample input has $\ttx{1}$ ending with a newline and $\ttx{2}$ starting with a newline.
In this case, $f(\ttx{1})$ also ends with a newline and $f(\ttx{2})$ also starts with a newline, so $g(f(\ttx{1}), f(\ttx{2}))$ has two consecutive newlines at the concatenation point.
But because ``\ttt{tr -cs A-Za-z '\textbackslash n'}'' eliminates consecutive newlines, these two consecutive newlines do not appear in  $f(\ttx{1}~{+}{+}~\ttx{2})$.
Therefore $f(\ttx{1}~{+}{+}~\ttx{2}) \ne g(f(\ttx{1}),f(\ttx{2}))$ and \sys eliminates $\mathbf{concat}$ as a potential combiner. Such counterexample inputs can be generated by input shapes whose number of words per line and number of characters per word are small.

As another example, when $f = (\ttt{uniq -c})$ and $g = \mathbf{concat}$, a counterexample input has $\ttx{1}$ ending with a nonempty line $l$ and $\ttx{2}$ starting with the same line $l$.
In this case, $f(\ttx{1})$ ends with a line with a padded integer $n_1$ on the left and the content $l$ on the right.
Meanwhile, $f(\ttx{2})$ starts with a line with a padded integer $n_2$ on the left and the content $l$ on the right.
Hence $g(f(\ttx{1}), f(\ttx{2}))$ has two consecutive lines at the concatenation point, whose contents on the right are both $l$.
But because ``\ttt{uniq -c}'' merges consecutive duplicate lines, these two consecutive lines whose right sides equal do not appear in   $f(\ttx{1}~{+}{+}~\ttx{2})$.
Therefore  $f(\ttx{1}~{+}{+}~\ttx{2})\ne g(f(\ttx{1}),f(\ttx{2}))$ and \sys eliminates $\mathbf{concat}$ as a potential combiner.
Such counterexample inputs can be generated by input shapes whose percentage of distinct lines is small.

The synthesis algorithm starts with a predefined seed input shape, around which the algorithm generates a space of mutated input shapes.
For each such input shape, \sys generates a set of input streams and feeds them to the original command.
Some of these input streams may cause \sys to discard candidate combiners that violate the divide-and-conquer property.
The sizes of the sets of discarded candidates for different input shapes induce a gradient over the input shapes. \sys follows this gradient to find input shapes that maximize its ability to quickly find and discard incorrect combiners. \sys continues this process until it executes several gradient steps that do not discard any remaining candidate combiners. In our example this process quickly produces the correct combiners described above for the commands in our example pipeline (and typically also for the commands in our benchmark scripts, see \autoref{sec:results}).

\begin{figure}[t]
\includegraphics[width=.7\linewidth]{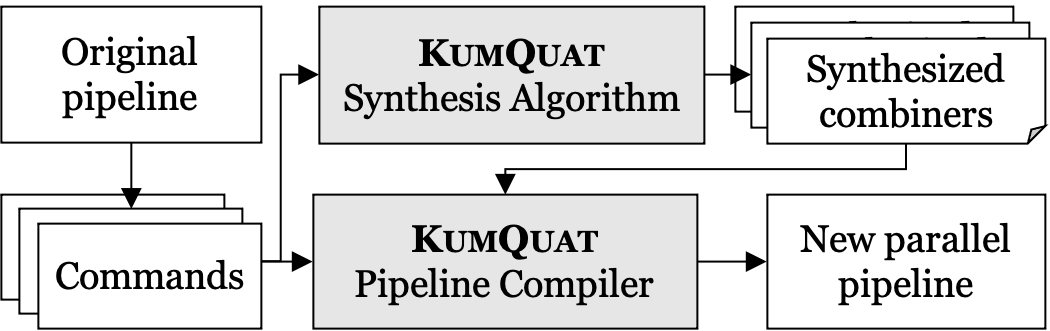}
\caption{\sys system workflow includes splitting an original serial pipeline into commands, synthesizing combiners for each of these commands, and reassembling the commands and synthesized combiners into a new parallel pipeline.}
\label{fig:system}
\end{figure}

\heading{New Data-Parallel Pipeline}
\sys parses the original pipeline, splits it into individual commands, synthesizes combiners for these commands, and compiles them into a new data parallel pipeline (\autoref{fig:system}). In our example \sys synthesizes combiners for all commands in the pipeline. The combiner for the command ``\ttt{tr -cs A-Za-z '\textbackslash n'}'' is $\mathbf{rerun}$. Because this command does not significantly reduce the size of the output stream (in comparison with the input stream), parallelizing the command with the $\mathbf{rerun}$ combiner reduces the overall performance. \sys therefore executes this command sequentially, with the input file piped directly into the command. All other commands execute in parallel.
\sys applies the intermediate combiner elimination optimization to eliminate the combiner for the ``\ttt{tr A-Z a-z}'' command. The resulting optimized pipeline has one sequential stage and three parallel stages (one of which executes the ``\ttt{tr A-Z a-z}'' and ``\ttt{sort}'' commands with no intermediate combiner).

On a 3GB benchmark input, the serial execution time is 2089 seconds.
The unoptimized parallel execution time with 16 way parallelism is 196 seconds, with a parallel speedup of $10.7\times$.
The optimized parallel execution time is 146 seconds with a speedup of $14.4\times$ (see \autoref{sec:results}).

%% file: design.tex
\section{Design}
\label{core}

\sys is designed to work with commands that implement deterministic computations over
input streams that are structured as units separated by delimiters.
\sys targets commands $f$ that have a combine operator $g$ in a domain-specific combiner language
that satisfies
$$f(\ttx{1}~{+}{+}~\ttx{2}) = g(f(\ttx{1}), f(\ttx{2}))$$ for all input streams $\ttx{1},\ttx{2}$, where the streams terminate with newlines.
\sys generates input streams to collect  outputs from $f$, which are used to eliminate incorrect candidate combiners.
Note that the combiners do not directly operate on the input streams generated by \sys, but instead operate on the command outputs.
We present details of the DSL semantics, definitions, and theorems in the appendix.

\input{dsl}

\input{synthesizer}

\input{soundness}

\input{rationale}

\input{optimization}

%% file: dsl.tex
\subsection{The \sys Combiner DSL}
\label{sec:dsl}

\begin{figure}[t]
\centering
    \input{lang}
    \caption{Combiners synthesizable by \sys.
}
\label{fig:dsl}
\end{figure}

To capture the space of possible combiners, \sys defines and uses a domain-specific language (DSL) presented in \autoref{fig:dsl}.
A combiner in \sys's  DSL is an expression, which is a binary operation that accepts two streams $\tty{1}$ and $\tty{2}$ as arguments.

\begin{definition}
A stream is a string that ends with a newline character $`\textbackslash n`$,
$\mathsf{Stream} = \{ x ~{+}{+}~ `\textbackslash n` \mid x \in \mathsf{String} \}$.
\end{definition}
\begin{definition}
A command $f: \mathsf{Stream} \rightarrow \mathsf{Stream}$ is a function that takes a stream as input and produces a stream as output.
\end{definition}

The DSL has three classes of operators: $\DSLBasic$, $\DSLStruct$, and $\DSLRun{f}$.
$\DSLBasic$ defines recursive operators and includes numeric addition ($\mathbf{add}$), string concatenation ($\mathbf{concat}$), selection ($\mathbf{first}$ and $\mathbf{second}$), and delimiter-based composite operators ($\mathbf{front}$, $\mathbf{back}$, $\mathbf{fuse}$).
The $\mathbf{front}$ (or $\mathbf{back}$) operator removes a delimiter at the front (or back) of $\tty{1}$ and $\tty{2}$, applies a child operator, and attaches the original delimiter to the front (or back) of the combined result.
The $\mathbf{fuse}$ operator applies a child operator piecewise on elements in $\tty{1}$ and $\tty{2}$ that are separated by a delimiter, after which piecewise results are connected back with the original delimiter.

$\DSLStruct$ defines operators that apply $\DSLBasic$ operators on structured streams ($\mathbf{stitch}$, $\mathbf{stitch2}$, and $\mathbf{offset}$).
These operators depend on the values at certain locations in $\tty{1}$ and $\tty{2}$.
The $\mathbf{stitch}$ (or $\mathbf{stitch2}$) operator compares $\tty{1}$'s last line with $\tty{2}$'s first line, then behaves differently conditioned on whether these two lines (or whether the second field from these two lines) equal.
The $\mathbf{offset}$ operator uses the first field in the last line of $\tty{1}$ to adjust the first field in every line of $\tty{2}$.

$\DSLRun{f}$ defines operators that require command executions ($\mathbf{rerun}_f$ and $\mathbf{merge}$).
The $\mathbf{rerun}_f$ command reexecutes command $f$ on the concatenataion of $\tty{1}$ and $\tty{2}$.
The $\mathbf{merge}$ command invokes a standard \unix merge command that takes two pre-sorted streams and interleaves them into a sorted merged stream.
The ``<flags>'' parameter represents a set of known flags specific to command $f$.

\input{language-less}

%% file: lang.tex
\[
\begin{array}{rcl}
g \in \DSLCombiner{f} &:=& b ~\vert~ s ~\vert~ r\\
b,b_1,b_2 \in \DSLBasic &:=& \mathbf{add} ~\vert~ \mathbf{concat}
                ~\vert~ \mathbf{first} ~\vert~ \mathbf{second} \\
                &\vert& \mathbf{front}~d~b ~\vert~ \mathbf{back}~d~b
                ~\vert~ \mathbf{fuse}~d~b\\
s \in \DSLStruct &:=& \mathbf{stitch}~b ~\vert~ \mathbf{stitch2}~d~b_1~b_2
                 ~\vert~ \mathbf{offset}~d~b\\
r \in \DSLRun{f} &:=& \mathbf{rerun}_f ~\vert~ \mathbf{merge}~\text{<flags>}\\
d \in \mathsf{Delim} &:=& `\textbackslash n`~\vert~`\textbackslash
t`~\vert~`~`~\vert~`,`\\
\end{array}
\]

%% file: language-less.tex
\begin{figure*}[t]
\centering
    \input{semantics-less}
\caption{
  Semantics of the combiner DSL (selected rules).
}
\label{fig:semantics}
\end{figure*}

\autoref{app:fig:semantics} presents several rules of the big-step execution semantics for the DSL.
The transition function $\Rightarrow$ maps a DSL expression to its output value.

%% file: semantics-less.tex
\footnotesize
\[
\begin{array}{c}

\begin{array}{cc}
    \begin{array}{c}
        \infral{
            i_1 = \mathsf{strToInt}~\tty{1}
            \qquad
            i_2 = \mathsf{strToInt}~\tty{2}
        }
        {\Eval{\mathbf{add}}{\tty{1}}{\tty{2}}
              {\mathsf{intToStr}~(i_1+i_2)}}
        {}
    \\ \\
            \infral{}
        {\Eval{\mathbf{concat}}{\tty{1}}{\tty{2}}
              {\tty{1}~{+}{+}~\tty{2}}}
        {}
        \\ \\
        \infral{}
        {\Eval{\mathbf{first}}{\tty{1}}{\tty{2}}
              {\tty{1}}}{}
        \\ \\
        \infral{
        \Eval{b}{(\mathsf{delFront}~d~\tty{1})}{(\mathsf{delFront}~d~\tty{2})}
                {v}
        }
        {\Eval{(\mathbf{front}~d~b)}{\tty{1}}{\tty{2}}
              {d ~{+}{+}~v}}
        {}
    \end{array}
    &
    \begin{array}{c}
        \infral{
            h_1, t_1 = \mathsf{splitFirst}~d~\tty{1}
            \qquad
            h_2, t_2 = \mathsf{splitFirst}~d~\tty{2}
            \qquad
            t_1 \ne \mathsf{nil}
            \qquad
            t_2 \ne \mathsf{nil}
            \\
            d \in t_1
            \qquad
            d \in t_2
            \qquad
            \Eval{b}{h_1}{h_2}{v}
            \qquad 
            \Eval{(\mathbf{fuse}~d~b)}{t_1}{t_2}{v'}
        }
        {\Eval{(\mathbf{fuse}~d~b)}{\tty{1}}{\tty{2}}
              {v~{+}{+}~d~{+}{+}~v'}}
        {}
        \\
        \\
        \infral{
            y_1', l_1 = \mathsf{splitLastLine}~\tty{1}
            \qquad
            l_2, y_2' = \mathsf{splitFirstLine}~\tty{2}
            \qquad
            l_1 = l_2
            \qquad
            \Eval{b}{l_1}{l_2}{v}
        }
        {\Eval{(\mathbf{stitch}~b)}{\tty{1}}{\tty{2}}
              {y_1'~{+}{+}~`\textbackslash n`~{+}{+}~v~{+}{+}~`\textbackslash n`~{+}{+}~y_2'}}
        {}
        \\ \\
        \begin{array}{cc}
            \infral{
                v = (\mathsf{unixMerge}~\text{<flags>})~\tty{1}~\tty{2}
            }
            {\Eval{(\mathbf{merge}~\text{<flags>})}{\tty{1}}{\tty{2}}{v}}
            {}
            &
            \infral{
            }
            {\Eval{\mathbf{rerun}_f}{\tty{1}}{\tty{2}}
                  {f~(\tty{1}~{+}{+}~\tty{2})}}
            {}
        \end{array}
    \end{array}
\end{array}

\\
\vspace{5pt}
\\
b, b_1, b_2 \in \DSLBasic
\qquad
d \in \mathsf{Delim}
\qquad
\tty{1}, \tty{2}, y_1', y_2', v, v', v_1, v_2, h, h_1, h_2, t, t_1, t_2, l_1, l_2 \in \mathsf{String}
\qquad
i_1, i_2 \in \mathsf{Int}
\end{array}
\]
\vspace{-5pt}
\normalsize

%% file: synthesizer.tex
\subsection{Combiner Synthesis}
\label{combinersynthesis}

The synthesizer starts with an initial search space of candidate combiners in $\DSLCombiner{f}$.
The algorithm generates a set of input streams, uses them to execute $f$, observes the outputs, and uses the observations to do two things: (1) remove implausible candidates and (2) choose an input shape for the next round of input generation.
We formalize these definitions below.

\begin{definition}
An input pair $\langle \ttx{1}, \ttx{2} \rangle$ consists of two strings $\ttx{1},\ttx{2} \in \mathsf{String}$.
An output tuple $\langle \tty{1},\tty{2},\tty{12} \rangle$ consists of three strings $\tty{1},\tty{2},\tty{12} \in \mathsf{String}$.
\end{definition}

\begin{definition}
An input stream pair $\langle \ttx{1},\ttx{2} \rangle$ consists of two streams $\ttx{1}, \ttx{2} \in \mathsf{Stream}$.
An observation $\langle \tty{1},\tty{2},\tty{12} \rangle$ consists of three streams $\tty{1},\tty{2},\tty{12} \in \mathsf{Stream}$.
\end{definition}

\begin{definition}
Executing command $f$ with an input stream pair $\langle \ttx{1}, \ttx{2} \rangle$ produces the observation $\langle f(\ttx{1}), f(\ttx{2}), f(\ttx{1}~{+}{+}~\ttx{2}) \rangle$.
For a set of input stream pairs $X$, $f(X)$ denotes the set of observations obtained from executing $f$ with $X$,
$f(X) = \{ \langle f(\ttx{1}), f(\ttx{2}), \allowbreak f(\ttx{1}~{+}{+}~\ttx{2}) \rangle \mid \langle \ttx{1},\ttx{2} \rangle \in X\}$.
\end{definition}

Our current \sys implementation allows the user to specify the initial search space with the maximum AST size in the combiner DSL.

\begin{definition}
The size of a combiner $g \in \DSLCombiner{f}$ is denoted as $\Size{g}$ and defined as two (each combiner operates on two arguments) plus the number of times that the AST of $g$ applies a production to expand a ``$\DSLBasic$'', ``$\DSLStruct$'', or ``$\DSLRun{f}$'' symbol.
\end{definition}

\begin{definition}
$G_n = \{ g \in \DSLCombiner{f} \mid \Size{g} \le n \}$  denotes the set of combiners that are under size $n$.
\end{definition}

We next define legal inputs and plausible combiners.

\begin{definition}\label{def:dsllegal}
For $g \in \DSLCombiner{f}$, $\SetLegal{g}$ denotes the set of legal strings for which $g$ is defined.
For example:
{\footnotesize \input{proof/legalexample}}%
For any $\tty{1},\tty{2} \in \SetLegal{g}$, the evaluation $\Eval{g}{\tty{1}}{\tty{2}}{v}$ succeeds for some $v \in \mathsf{String}$.
\end{definition}

\begin{definition}\label{def:plausible}\label{thm:satlegal}
A combiner $g \in \DSLCombiner{f}$ is plausible for output tuples $Y$,
denoted as $\BoolSat{g}{Y}$, if
$\tty{1},\tty{2} \in \SetLegal{g}$
and
$\Eval{g}{\tty{1}}{\tty{2}}{\tty{12}}$
for all $\langle \tty{1},\tty{2},\tty{12} \rangle \in Y$.
\end{definition}

\begin{definition}\label{def:correctwrtpairs}
A combiner $g \in \DSLCombiner{f}$ is correct with respect to input pairs $X$ if $\BoolSat{g}{f(X)}$.
$\CombN{f}{X}{n} = \{ g \in G_n \mid \Correct{g}{f}{X} \}$ denotes the combiners for command $f$ that are under size $n$ and are correct with respect to input pairs $X$.
\end{definition}

\input{candidates}

\input{algsynth}

\input{alginput}

\input{mutation}

\heading{Preprocessing}\label{sec:literals}
The current \sys implementation preprocesses command scripts to obtain a set of literals for generating inputs and input shapes.
For example, ``\ttt{grep 'light.\*light'}'' does not produce any outputs unless the input stream contains lines that match the regular expression \ttt{'light.\*light'}.
\sys extracts this regular expression and generates a dictionary of strings that match. It then uses this dictionary as elements for generating input streams based on input shapes.
The command ``\ttt{sed 100q}'' copies input to output when the input stream contains at most 100 lines. When the input stream contains more lines, the command removes the trailing lines.
\sys obtains the number 100 as a literal and uses it to generate initial input shapes where one dimension is close to this number.
\sys then mutates this initial input shape to obtain a range of different input streams that exercise different behavior in the original command.

\sys also checks whether the original command can process three test input streams without errors: a list of unsorted English words separated by newlines, the same list of words but sorted, and a list of legal file names separated by newlines.
Most of our benchmark commands can process all three test input streams without errors.
Benchmark commands that use ``\ttt{comm}'' print an error with the first test input stream but succeed with the second.
Based on this outcome, \sys generates only sorted input streams for these commands during combiner synthesis.
Benchmark commands that use ``\ttt{xargs}'' print an error with the first two test input streams but succeed with the third.
Based on this outcome, \sys configures a dictionary of legal file names and uses the dictionary to generate input streams for these commands.

\heading{Multiple Plausible Combiners}
Recall that \autoref{alg:synth} returns the set of plausible combiners (\autoref{def:plausible} and \autoref{def:correctwrtpairs}).
Let $G$ be the set of returned plausible combiners.
If $G = \emptyset$, the synthesizer reports an error.
If $G$ contains exactly one combiner, the synthesizer returns the combiner directly.

If $G$ contains more than one combiner, the synthesizer builds a composite combiner using the following subset of $G$.
If $G \cap \DSLBasic \ne \emptyset$, the synthesizer uses the set $G \cap \DSLBasic$.
Otherwise, if $G \cap \DSLStruct \ne \emptyset$, the synthesizer uses the set $G \cap \DSLStruct$.
Otherwise, $G \cap \DSLRun{f} \ne \emptyset$ and the synthesizer uses the set $G \cap \DSLRun{f}$.
Let this nonempty subset be $\{ g_1, g_2, \ldots, g_m\}$ ($m \ge 1$).
\sys uses it to construct a composite combiner as follows.
For any $\tty{1},\tty{2}$, if they belong to the domain of $g_1$, then return $g_1(\tty{1},\tty{2})$.
Else if they belong to the domain of $g_2$, return $g_2(\tty{1},\tty{2})$.
\ldots
Otherwise, return $g_m(\tty{1},\tty{2})$.
We show in \autoref{sec:soundness} that, if the correct combiner for $f$ is among a certain set, then the order in which these combiners are composed together does not matter---the resulting composite combiner is semantically equivalent regardless of the order.
Alternatively, if the domain of one of these plausible combiners is the superset of any other plausible combiner's domain, then it suffices to return only the combiner with the largest domain.

%% file: proof/legalexample.tex
\begin{align*}
\SetLegal{\mathbf{add}} &= [`0`-`9`]^+ \\
\SetLegal{\mathbf{front}~d~b} &= \{ d~{+}{+}~y \mid y \in \SetLegal{b} \} \\
\SetLegal{\mathbf{fuse}~d~b} &= \{ y_1~{+}{+}~d~{+}{+}~y_2~{+}{+}~d~{+}{+}~\ldots~{+}{+}~d~{+}{+}~y_k \\
                             &\phantom{= \{}
                              \mid y_1 \ne \mathsf{nil}, y_k \ne \mathsf{nil}, \text{ and } y_i \in \SetLegal{b} \text{ and } d \not\in y_i \\
                             &\phantom{= \{ \mid~}
                              \text{ for all }i=1,\ldots,k,
                              \text{ where } k \ge 2\}
\end{align*}

%% file: candidates.tex
\heading{Combiner Synthesis}
\autoref{alg:synth} presents \sys's combiner synthesis algorithm, which takes a black-box command $f$ and an integer $n$.
It starts by preparing a set $C_0$ of the initial search space.
The algorithm then performs multiple rounds of filtering on the candidates.

Variable $C_r$ holds the set of combiners that are correct with respect to all seen input pairs, \ie the set $\CombN{f}{I}{n}$ where $I = \bigcup_{r'=1}^{r} I_{r'}$.
\autoref{alg:synth} terminates if either
  (1) no candidate combiners remain, in which case it returns $\mathsf{nil}$ and reports an error, or 
  (2) no progress is made in a number of rounds, in which case it returns the set of plausible combiners.

%% file: algsynth.tex
\begin{algorithm}[tb]
  \small
  \KwData{Command $f$, max combiner size $n$}
  \KwResult{Synthesized plausible combiners}

  $C_0 \leftarrow \mathit{AllCandidates}(n)$

  \For{$r = 1, 2, \ldots$}{

      $I_r \leftarrow \mathit{GetEffectiveInputs}(f, C_{r-1}, \mathit{RandomShape}())$

      $C_r \leftarrow \mathit{FilterCandidates}(f, C_{r-1}, I_r)$

      \If{$C_r = \emptyset$}{
          \textbf{return} $\mathsf{nil}$
      }

      \If{\textbf{not} $\mathit{MakingProgress}([C_0,\ldots,C_r])$}{
          \textbf{return} $C_r$
      }
  }

  \caption{Procedure $\mathit{Synthesize}$, which implements \sys's core synthesis algorithm.
The procedure takes a command and synthesizes a combiner for the command that is correct with respect to a range of generated input pairs.}
  \label{alg:synth}
\end{algorithm}

%% file: alginput.tex
\begin{algorithm}[tb]
  \small
  \KwData{Command $f$, candidate combiners $C$, input shape $s_0$}
  \KwResult{Input stream pairs, generated from mutating $s_0$, for eliminating incorrect candidates in $C$}

  $I \leftarrow $ Empty set

  \For{$m = 1,\ldots,M$}{

     \For{$j = 1,\ldots,12$}{

         $s_{m-1}^j \leftarrow \mathit{MutateShape}(s_{m-1}, j)$

         $I_{m-1}^j \leftarrow \mathit{GetInputStreamPairs}(s_{m-1}^j)$

         Add $I_{m-1}^j$ to $I$

     }

     $j' \leftarrow \mathit{IndexBestMutation}(C, I_{m-1}^1,\ldots,I_{m-1}^{12})$

     $s_m \leftarrow s_{m-1}^{j'}$
  }

  \textbf{return} $I$
    
  \caption{Procedure $\mathit{GetEffectiveInputs}$, which mutates input shapes to generate input stream pairs.}
  \label{alg:input}
\end{algorithm}

%% file: mutation.tex
\heading{Input Generation}
We next present how the inputs are generated for \autoref{alg:synth} to filter candidates.
A key goal of input generation is to generate a variety of input streams that exercise a wide range of the functionality of the command $f$.
The \sys input generation algorithm
is driven by mutations to an input shape, from which \sys generates random inputs.
The mutations are chosen by how effectively their resulting inputs eliminate incorrect candidate combiners.

\begin{definition}
An \emph{input shape} $s = \langle s_L, s_W, s_C\rangle \in \mathit{Shape}$ specifies the configurations for three \emph{dimensions} of an input: the lines in each input as separated by newline characters ($s_L \in \mathit{Config}$), the words in each line as separated by spaces ($s_W \in \mathit{Config}$), and the characters in each word ($s_C \in \mathit{Config}$).
The configuration for each dimension is of the form $\langle l, u, d \rangle$ and specifies three bounds: the minimum element count ($l \in \mathit{Int}$), the maximum element count ($u \in \mathit{Int}$), and the percentage of distinct elements ($d \in \mathit{Percent}$) on that dimension.
\end{definition}

\begin{definition}
A stream $x$ \emph{satisfies} an input shape $s \in \mathit{Shape}$, denoted as $x \Sat s$, if $x$ conforms to the bounds specified in $s$.
An input stream pair $\langle \ttx{1},\ttx{2} \rangle$ satisfies an input shape $s \in \mathit{Shape}$, denoted as $\langle \ttx{1},\ttx{2} \rangle \Sat s$, if $(\ttx{1}~{+}{+}~\ttx{2}) \Sat s$.
\end{definition}

\autoref{alg:input} presents \sys's input generation algorithm.
The procedure takes a black-box command $f$, a set of candidate combiners $C$, and an initial input shape $s_0$;
  it mutates the input shape iteratively, generating input streams along the way.

The iterative mutation process is inspired by gradient descent.
The $m$-th iteration ($m=1,\ldots,M)$ mutates the input shape $s_{m-1}$ using one of twelve potential mutations.
These potential mutations are along three dimensions (lines, words, and characters) and four directions (more/fewer elements, more/less varied).
Procedure $\mathit{MutateShape}$ takes an initial input shape and a mutation index, then returns a new input shape mutated as specified.
For the $j$-th potential mutation ($j=1,\ldots,12$), \autoref{alg:input} uses the mutated input shape $s_{m-1}^j$ to generate a set of input stream pairs $I_{m-1}^j$.
In other words, the variable $I_{m-1}^j$ satisfies $\langle i_1,i_2\rangle \Sat s_{m-1}^j$ for all $\langle i_1,i_2\rangle \in I_{m-1}^j$.

\autoref{alg:input} then evaluates the effectiveness of all of the input shape mutations.
It returns the index, $j'$, of the most effective set.
The $j'$-th mutation then produces the input shape for the next iteration, $s_m = s_{m-1}^{j'}$.
The procedure repeats these operations for $M$ iterations.
Finally the procedure returns the set of all observed input pairs, $I = \bigcup_{j,m} {I_{m-1}^j}$.

%% file: soundness.tex
\subsection{Conditions for Synthesizing Correct Combiners}
\label{sec:soundness}

We present theorems that characterize when the combiner synthesis algorithm will identify a correct combiner for a given set of parallel output streams.

Broadly speaking, when the combiner involves numerical addition ($\mathbf{add}$), the corresponding stream fragments on which the numerical addition applies are required to be nonzero in some observations.
When the combiner involves string concatenation ($\mathbf{concat}$), the corresponding fragments on which the string concatenation applies are required to be nonempty in some observations.
When the combiner involves selection ($\mathbf{first}$ and $\mathbf{second}$), the corresponding stream fragments on which the selection applies are required to contain non-delimiter and non-zero characters in some observations.

Combiners that process formatted streams may nest these three classes of basic operators inside more complex operators ($\mathbf{front}$, $\mathbf{back}$, $\mathbf{fuse}$, $\mathbf{stitch}$, $\mathbf{stitch2}$, $\mathbf{offset}$).
For these combiners, their requirements for sufficient observations include a specification of the formatting as well as a specification of the deformatted fragments.

\begin{definition}\label{def:equivcap}
For $g_1, g_2 \in \DSLCombiner{f}$, $g_1$ and $g_2$ are equivalent by intersection, denoted as $\BoolEquivCap{g_1}{g_2}$, if
for all $\tty{1},\tty{2} \in \SetLegal{g_1} \cap \SetLegal{g_2}$,
$\Eval{g_1}{\tty{1}}{\tty{2}}{v}$ and $\Eval{g_2}{\tty{1}}{\tty{2}}{v}$ for some $v$.
\end{definition}

\begin{definition} \label{def:correctsat}
A combiner $g \in \DSLCombiner{f}$ is correct for command $f$ if 
$\BoolSat{g}{f(X)}$ holds for all input stream pairs $X$.
\end{definition}

\begin{definition}
We define two sets of representative combiners for command $f$,
$\GBasic = \{ g_\text{a}, g_\text{c}, g_\text{f}, g_\text{s}, g_\text{ba}, g_\text{fa}, g_\text{bfa}, g_\text{fbfa}, \allowbreak g_\text{fc} \} \subset \DSLBasic$
and
$\GStruct = \{ g_\text{sf}, g_\text{saf}, g_\text{oa} \} \subset \DSLStruct$,
whose elements include:
$g_\text{c} = \mathbf{concat}$,
$g_\text{ba} = (\mathbf{back}~d~\mathbf{add})$,
$g_\text{sf} = (\mathbf{stitch}~\mathbf{first})$,
and
$g_\text{saf} = (\mathbf{stitch2}~d~\mathbf{add}~\mathbf{first})$.
\end{definition}

\begin{definition}\label{def:boolenough}
For combiner $g \in \GBasic \cup \GStruct$ and any set of output tuples $Y$,
$\BoolEnough{g}{Y}$ denotes a conservative predicate that is true only if $Y$ is sufficient for eliminating incorrect candidates when the correct combiner is $g$.
\end{definition}

\begin{definition}
For any set of output tuples $Y$, $\BoolEnoughBasic{Y}$ denotes a conservative predicate that is true only if $Y$ is sufficient for eliminating incorrect candidates when the correct combiner $g \in \GBasic$.
\end{definition}

\begin{definition}
For any set of output tuples $Y$, $\BoolTable{Y}$ denotes a predicate that is true only if $Y$ is interpretable as a table.
\end{definition}

\begin{definition}
For any set of output tuples $Y$, $\BoolEnoughStruct{Y}$ denotes a conservative predicate that is true only if $Y$ is sufficient for eliminating incorrect candidates when the correct combiner $g \in \GStruct$.
\end{definition}

\begin{definition} \label{def:setsat}
For a command $f$, integer $k$, and set of output tuples $Y$,
the set of plausible combiners
$\SetSat{k}{Y} = \{ g \in \DSLCombiner{f} \mid \Size{g} \le k \text{ and } \BoolSat{g}{Y} \}$.
\end{definition}

\begin{theorem} \label{thm:equivcapbasic}
For any combiner $g \in \GBasic$,
set of output tuples $Y$ such that $\BoolSat{g}{Y}$ and $\BoolEnough{g}{Y}$,
and
$g' \in \DSLBasic$,
we have
$\BoolSat{g'}{Y}$
implies
$\BoolEquivCap{g'}{g}$.
\end{theorem}

\begin{theorem} \label{thm:synthesizebasic}
For any command $f$, set of input streams $X$, combiner $g \in \GBasic$, combiner $g' \in \DSLBasic$, and integer $k$,
if the following conditions hold:
\begin{itemize}
\item $\BoolEnoughBasic{f(X)}$,
\item $g$ is correct for $f$,
\item $\tty{1},\tty{2} \in \SetLegal{g'}$ for all $\langle \tty{1},\tty{2},\tty{12} \rangle \in f(X)$, and
\item $k \ge \Size{g'}$,
\end{itemize}
then
$g' \in \SetSat{k}{f(X)} \cap \DSLBasic$ if and only if $\BoolEquivCap{g'}{g}$.
\end{theorem}

\begin{theorem} \label{thm:equivcapstruct}
For any combiner $g \in \GStruct$,
set of output tuples $Y$ such that $\BoolSat{g}{Y}$ and $\BoolEnough{g}{Y}$,
and
$g' \in \DSLStruct$,
we have
$\BoolSat{g'}{Y}$
implies
$\BoolEquivCap{g'}{g}$.
\end{theorem}

\begin{theorem} \label{thm:synthesizestruct}
For any command $f$, set of input streams $X$, combiner $g \in \GStruct$, combiner $g' \in \DSLStruct$, and integer $k$,
if the following conditions hold:
\begin{itemize}
\item $\BoolEnoughStruct{f(X)}$,
\item $g$ is correct for $f$,
\item $\tty{1},\tty{2} \in \SetLegal{g'}$ for all $\langle \tty{1},\tty{2},\tty{12} \rangle \in f(X)$, and
\item $k \ge \Size{g'}$,
\end{itemize}
then
$g' \in \SetSat{k}{f(X)} \cap \DSLStruct$ if and only if $\BoolEquivCap{g'}{g}$.
\end{theorem}

%% file: rationale.tex
\subsection{Input Generation and Correct Combiners}
\label{sec:rationale}

Our target commands often consist of two components: unit-based computation and delimiter-based formatting.
Unit-based computation often determines how a combiner applies the $\mathbf{add}$, $\mathbf{concat}$, $\mathbf{first}$, and $\mathbf{second}$ operators to certain fragments of the output substreams.
Delimiter-based formatting often determines how a combiner uses the $\mathbf{front}$, $\mathbf{back}$, $\mathbf{fuse}$, $\mathbf{stitch}$, $\mathbf{stitch2}$, and $\mathbf{offset}$ operators.
We focus on three broad classes of unit-based computation that appear in our benchmark commands.

\heading{Counting Lines, Words, or Characters}
Many \unix commands output formatted counts of certain lines, words, or characters. Benchmark commands that implement this pattern include
``\ttt{wc -l},''
``\ttt{grep -c [regex]},''
and
``\ttt{uniq -c}.''

For each of these \ttt{wc} and \ttt{grep} commands, a correct combiner is $(\mathbf{back}~d~\mathbf{add})$.
By \autoref{thm:equivcapbasic} and \autoref{thm:synthesizebasic}, as long as the command outputs 
satisfy the requirements for $(\mathbf{back}~d~\mathbf{add})$, any synthesized plausible combiner in $\DSLBasic$ will be equivalent to $(\mathbf{back}~d~\mathbf{add})$ when processing streams that belong to the combiner's domain.
For the \ttt{uniq} command, a correct combiner is $(\mathbf{stitch2}~`~`~\mathbf{add}~\mathbf{first})$.
By \autoref{thm:equivcapstruct} and \autoref{thm:synthesizestruct}, as long as the command outputs collected by \sys satisfy the requirements for $(\mathbf{stitch2}~`~`\allowbreak~\mathbf{add}\allowbreak~\mathbf{first})$, any synthesized plausible combiner in $\DSLStruct$ will be equivalent to $(\mathbf{stitch2}~`~`~\mathbf{add}~\mathbf{first})$ when processing streams that belong to the combiner's domain.

Both of these correct combiners use $\mathbf{add}$ nested inside other operators. Because these other operators process formatting, the remaining deformatted fragments in the output substreams are therefore processed by $\mathbf{add}$.
Here we focus on identifying the correct $\mathbf{add}$ operator for processing these deformatted fragments.
The requirement is (conceptually) observing nonzero values in these fragments.

\sys generates input streams with various numbers of lines. Many of these lines cause the \ttt{wc} counter(s) to be nonzero. For ``\ttt{grep -c [regex]},'' the \sys preprocessing extracts literals that it uses to generate input streams that contain matching values that cause the \ttt{grep} counter to be nonzero. 
The resulting output streams therefore contain nonzero characters even after removing the formatting, which satisfy the requirements for identifying $\mathbf{add}$ correctly as a building block of the final synthesized combiners.

For ``\ttt{uniq -c}'' the combiner contains a conditional in the $\mathbf{stitch2}$ operator that applies the $\mathbf{add}$ operator only when the right-hand content in the last line of $\tty{1}$ equals the right-hand content in the first line of $\tty{2}$.
These contents correspond to the last line of $\ttx{1}$ and the first line of $\ttx{2}$.
\sys generates input streams $\ttx{1}, \ttx{2}$ with varying percentages of distinct lines, some of which enable the command to produce output streams $\tty{1}, \tty{2}$ that contain deformatted fragments that are processed by the $\mathbf{add}$ operator.
Because these deormatted fragments are always nonzero for this \ttt{uniq} command, they satisfy the requirements for identifying $\mathbf{add}$ correctly as a building block of the final synthesized combiner.

\heading{Mapping Input Lines to Disjoint Output Lines}
Many \unix commands
apply a function to map each input line to a sequence of output lines. 
Benchmark commands that implement this pattern include:
``\ttt{tr '[a-z]' 'P'},''
``\ttt{tr -c '[A-Z]' '\textbackslash n'},''
``\ttt{sed s/\textbackslash \$/'0s'/},''
``\ttt{cut -c 1-4},''
``\ttt{cut -d ',' -f 3,1},''
``\ttt{awk "length >= 16"},''
``\ttt{grep 'light.\textbackslash *light'},''
``\ttt{grep -v '\textasciicircum 0\$'},''
``\ttt{xargs cat},''
and
``\ttt{xargs file}.''

For each of these commands, a correct combiner is $\mathbf{concat}$.
By \autoref{thm:equivcapbasic} and \autoref{thm:synthesizebasic}, as long as the command outputs collected by \sys satisfy the requirements for $\mathbf{concat}$, any synthesized plausible combiner in $\DSLBasic$ will be equivalent to $\mathbf{concat}$ when processing streams that belong to the combiner's domain.
In this case, the requirement is (conceptually) observing nonempty output streams.

For the \ttt{tr}, \ttt{sed}, and \ttt{cut} commands, \sys generates input streams with various numbers of lines, many of which cause these commands to produce nonempty outputs.
For the \ttt{awk}, \ttt{grep}, and \ttt{xargs} commands, \sys uses preprocessing to determine that it will generate input streams based on certain literals or file names (\autoref{sec:literals}).
The resulting input streams therefore enable these commands to produce nonempty outputs, which satisfy the requirements for synthesizing $\mathbf{concat}$ correctly.

\heading{Selecting Elements}
Some \unix commands select
certain elements from a list, output selected elements, and discard others. Benchmark commands include:
``\ttt{uniq}''
and
``\ttt{uniq -c}.''
$(\mathbf{stitch}~\mathbf{first})$ is a correct combiner for ``\ttt{uniq}''; $(\mathbf{stitch2}~`~`~\mathbf{add}~\mathbf{first})$ is a correct combiner 
for ``\ttt{uniq -c}''.

These combiners use $\mathbf{first}$ or $\mathbf{second}$ nested inside  deformatting operators so that correct fragments are processed by $\mathbf{first}$ or $\mathbf{second}$.
Here we focus on how to identify the correct $\mathbf{first}$ or $\mathbf{second}$ operator.
The requirement is (conceptually) observing non-delimiter and non-zero characters in these fragments and observing such fragments for the two operands to differ.

These \ttt{uniq} commands select one line out of any two adjacent lines that equal.
The combiners contain conditional statements in the $\mathbf{stitch}$ and $\mathbf{stitch2}$ operators that apply $\mathbf{first}$ or $\mathbf{second}$ only when certain contents are equal.
The combiner for ``\ttt{uniq}'' applies the $\mathbf{first}$ operator when the last line in $\tty{1}$ equals the first line in $\tty{2}$.
The combiner for ``\ttt{uniq -c}'' applies the $\mathbf{first}$ operator when the right-hand content in the last line of $\tty{1}$ equals the right-hand content in the first line of $\tty{2}$.
\sys generates input streams $\ttx{1}, \ttx{2}$ with varying percentages of distinct lines, some of which enable these commands to produce output streams $\tty{1}, \tty{2}$ that contain deformatted fragments that are processed by the $\mathbf{first}$ operator.
Also, because \sys generates input streams with varying numbers of words per line, some of the generated lines contain nonzero and nondelimiter characters which enable the command to produce such characters in the deformatted fragments as well.
These deformatted fragments therefore satisfy the requirements for identifying $\mathbf{first}$ and $\mathbf{second}$
as a building block of the final synthesized combiner.

%% file: optimization.tex
\subsection{Pipeline Optimization}
\label{sec:opt}

\begin{figure}[t]
\begin{subfigure}{\linewidth}
\includegraphics[width=\linewidth]{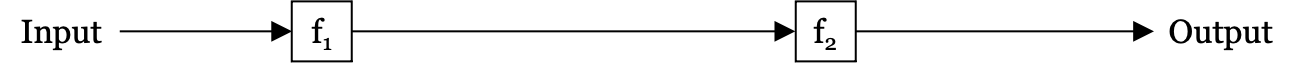}
\vspace{-10pt}
\caption{Serial pipeline that consists of two commands $f_1, f_2$}
\label{fig:reassemble:serial}
\vspace{5pt}
\end{subfigure}
\begin{subfigure}{\linewidth}
\includegraphics[width=\linewidth]{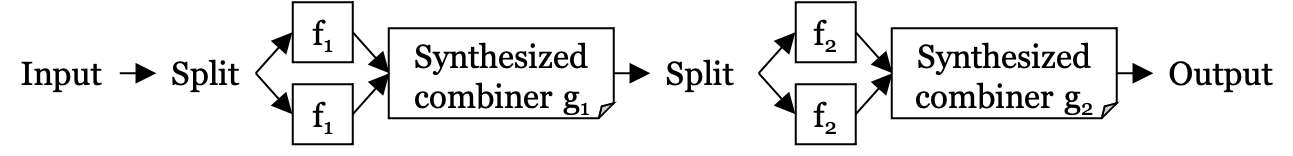}
\vspace{-10pt}
\caption{Unoptimized parallel pipeline that executes a combiner after each parallel command}
\label{fig:reassemble:naive}
\vspace{5pt}
\end{subfigure}
\begin{subfigure}{\linewidth}
\includegraphics[width=\linewidth]{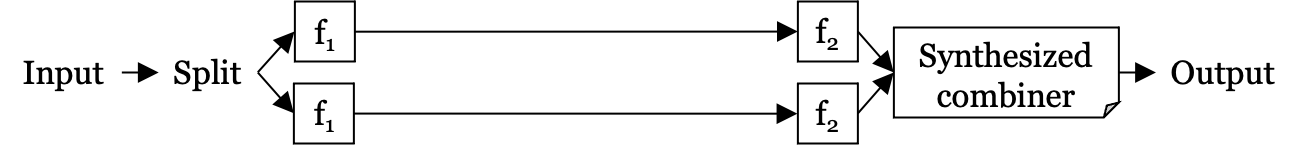}
\vspace{-10pt}
\caption{Optimized parallel pipeline that executes multiple commands in parallel after eliminating intermediate combiners}
\label{fig:reassemble:defer}
\end{subfigure}
\caption{\sys reassembles a new parallel pipeline by splitting the input stream, running parallel copies of the original commands on the input substreams, and combining the output substreams.}
\label{fig:reassemble}
\end{figure}

\heading{Eliminating Intermediate Combiners}
\sys eliminates unnecessary intermediate combiners as follows.

\begin{theorem}\label{thm:defer}
For any commands $f_1,f_2$, input streams $\ttx{1},\ttx{2} \in \mathsf{Stream}$, correct combiner $g_1$ for $f_1$, correct combiner $g_2$ for $f_2$
if $g_1=\mathbf{concat}$ and $f_1(\ttx{1}), f_1(\ttx{2}) \in \mathsf{Stream}$
then
$$g_2(f_2(\tty{1}), f_2(\tty{2})) = g_2(f_2(f_1(\ttx{1})), f_2(f_1(\ttx{2})))$$
for any $\tty{1},\tty{2} \in \mathsf{Stream}$ such that $f_1(\ttx{1}~{+}{+}~\ttx{2}) = \tty{1}~{+}{+}~\tty{2}$.
\end{theorem}

\autoref{fig:reassemble} shows the effect of this optimization---rather than combining after every pipeline stage, the optimized parallel pipeline combines only once.
In general, if the combiner for a stage (command $f_1$) concatenates the parallel output substreams, then we can eliminate the combiner $g_1$ for this stage and feed the output substreams directly to the next parallel stage $f_2$ as input substreams.
The final combined output stream applies only the combiner $g_2$ for the second stage (\autoref{fig:reassemble:defer}).
By \autoref{thm:defer}, this output is identical to the unoptimized output if $g_1$ were in place (\autoref{fig:reassemble:naive}).

A prerequisite for this optimization is that the command $f_1$ must produce output streams that terminate with newlines.
One of our benchmark commands violate this precondition: the command ``\ttt{tr -d '\textbackslash n'}'' removes all newline characters.
Hence the optimization in \autoref{thm:defer} does not apply to this command (\sys still parallelizes this command with the $\mathbf{concat}$ combiner).

\heading{Combining Multiple Substreams}
Although \sys synthesizes combiners that process two streams, the commands may be executed with $k$ way parallelism that produces $k$ output substreams ($k>2$).
\sys generalizes the following combiners to apply to all $k$ substreams at the same time.
The $\mathbf{merge}~\text{<flags>}$ combiner is implemented in \sys as an invocation of a \unix script ``\ttt{sort -m <flags> \$*}'' which merges multiple sorted streams at the same time.
The $\mathbf{concat}$ combiner is implemented as the script ``\ttt{cat \$*}'' which concatenates multiple streams.
The $\mathbf{rerun}$ combiner can also be implemented by concatenating all substreams at the same time and rerunning the original command only once.
For other combiners, the current \sys implementation applies the combiner on two substreams repeatedly until only one substream remains.

%% file: results.tex
\section{Experimental Results}
\label{sec:results}

\input{tables/performance-less}

We evaluate \sys on the following benchmarks:

\begin{itemize}
\item\textbf{Mass-transit analytics during COVID-19:}
This benchmark set contains 4 scripts
that were used to analyze real telemetry data from bus schedules during the COVID-19 response in a large European city~\cite{oasa-article}.
The pipelines compute several average statistics on the transit system per day---such as daily serving hours and daily number of vehicles.
Each script has 1 pipeline.
Each pipeline has between 7 and 8 stages.\footnote{We count pipelines as groups of two or more commands connected by \unix pipes. We count pipeline stages as commands in the pipeline excluding initial ``\ttt{cat}'' commands that read input files.\label{fn:stages}}
These scripts
operate on a fixed 3.4GB dataset that contains mass-transport data collected over a single year.

\item\textbf{Natural language processing:}
This benchmark set contains 22 scripts
from Kenneth's Unix-for-Poets~\cite{poets}, updated in 2016 by a Stanford linguistic class~\cite{poets2}.
These scripts
calculate natural-language processing metrics such as n-grams, morphs, counts, and frequencies. 
Each script has between 1 and 3 pipelines.
Each pipeline has between 2 and 8 stages.
These scripts are applied to 1823 books that total 927MB from Project Gutenberg~\cite{hart1971project}.

\item\textbf{Classic \unix One-liners:}
This benchmark set contains 10
pipelines written by \unix experts:
  a few pipelines are from \unix legends~\cite{bentley1986literate, bentley1985spelling, mcilroy1978unix}, one from a book on \unix scripting~\cite{taylor2004wicked}, and a few are from top Stackoverflow answers~\cite{bigrams}.
Each script has between 1 and 2 pipelines, except for a script that has only one command.
Each pipeline has between 2 and 8 stages.
Inputs are script-specific and average 1.6GB per benchmark.

\item\textbf{Unix50 from Bell Labs:}
This benchmark set contains 34 pipelines solving the \unix50 game~\cite{unix50}, designed to highlight \unix's modular philosophy~\cite{mcilroy1978unix}, found on GitHub~\cite{unix50sol}
Each script has 1 pipeline, except for a script that has only one command.
Each pipeline has between 2 and 10 stages.
Inputs are script-specific and average 1.1GB per benchmark.

\end{itemize}

\heading{Experimental Setup}
To evaluate the pipeline performance, we implemented an infrastructure that can execute each stage in a pipeline to completion before starting to execute the next stage.
The infrastructure configures any stage that invokes the \unix \ttt{sort} utility to be serial (using the option ``\ttt{--parallel=1}'').
Each stage's output is redirected to a file, which is read by the next stage as input.
This infrastructure provides a parameter for specifying the amount of parallelism for each parallelizable stage.

We performed experiments on a server with 0.5TB of memory and 80 $\times$ 2.27GHz Intel(R) Xeon(R) E7-8860, Debian GNU/Linux 9, GNU Coreutils 8.26-3, and Python 3.8.2.
We note that our benchmarks never come close to exhausting the server's available memory.

\heading{Performance Results}
The 70 benchmark scripts have a total of 477 commands and 427 pipeline stages.\footnote{See \autoref{fn:stages}.}
\autoref{tab:performance-cut-by-hand}
presents the performance results for our automatically parallelized pipelines
for the two longest-running scripts in each benchmark.
In general, shorter scripts have smaller parallel speedup.  We present full results in the appendix.

The first two columns present the benchmark and script names.
The next column (\textbf{Parallelized}) presents the number of stages automatically parallelized by \sys, $k$, and the number of stages in the original pipeline, $n$, as a pair ``$k/n$'' for each pipeline in the parentheses. Here we also report the single commands that are not in pipelines, as pairs ``$k/1$''. The pair before the parentheses presents the sum over all pipelines in the script.
The next column (\textbf{Eliminated}) presents the number of parallelized stages whose combiners are eliminated by \sys during optimization. Again, the numbers in the parentheses correspond to pipelines in the script. The number before the parentheses presents the sum over all pipelines.
Among all benchmark scripts, \sys parallelizes 325 of the the 427 stages (76.1\%) with synthesized combiners.
The optimization eliminates 144 of these combiners (44.3\%).
These results highlight the ability of \sys to effectively extract the parallelism implicitly present in the benchmark pipelines. 

The next column (\textbf{$T_\text{orig}$}) presents the execution time for the original benchmark script, which exploit the default \unix pipelined parallelism and deploy the default \unix sort, which exploits 8 way parallelism. The next two columns (\textbf{$u_1$} and \textbf{$u_{16}$}) present the execution time for the \emph{unoptimized} pipeline with 1 and 16 way parallelism for each data parallel command. Since these generated pipelines always wait for each stage to terminate before starting the next stage, $u_1$ is the serial execution time.
The last column (\textbf{$T_{16}$}) presents the execution time for the \emph{optimized} pipeline with 16 way command parallelism.
Among the benchmark scripts whose serial execution time is at least 3 minutes, the unoptimized parallel speedup ranges between 3.5$\times$ and 14.9$\times$, with a median speedup of 8.5$\times$.
Among these benchmark scripts, the optimized parallel speedup ranges between 3.8$\times$ and 26.9$\times$, with a median speedup of 11.3$\times$ (we attribute the superlinear speedup to pipelined parallelism exploited across consecutive parallelized commands with no intermediate combiner).

\heading{Synthesis Results}
We summarize the synthesis results below and present full results in the appendix.
The benchmarks contain 133 unique command/flag combinations (we refer to them as ``commands'' below).
Among these commands, 121 are data-processing commands that read an input stream.\footnote{The remaining 12 unique commands include 2 function calls, 3 commands that do not process data streams (\ttt{ls}, \ttt{mkfifo}, and \ttt{rm}), and 7 commands that process multiple input streams.}
\sys synthesizes a combiner for 113 of the 121 unique commands, with no combiner synthesized for the remaining 8 commands.
The 8 unsupported commands include 7 commands for which no correct combiner exists and 1 command that requires a specific field of the input file to equal ``2''.

The synthesis times vary between 39 seconds and 331 seconds with a median of 60 seconds.
The most common synthesized plausible combiners, including their equivalents, are:
$\mathbf{concat}$ (synthesized 81 times),
$\mathbf{rerun}$ (30 times),
$\mathbf{merge}(\texttt{*})$ (16 times),
and
$(\mathbf{back}~`\textbackslash n`~\mathbf{add})$ (12 times).
Other synthesized combiners involve operators $\mathbf{first}$, $\mathbf{second}$, $\mathbf{fuse}$, $\mathbf{stitch}$, and $\mathbf{stitch2}$.
For each benchmark command, the synthesized plausible combiners are all equivalent when operating on the command's outputs.
\sys uses the synthesized combiners to parallelize the benchmark scripts.
The generated parallel pipelines all produce correct outputs (same outputs as the original scripts).

%% file: tables/performance-less.tex
\begin{table*}[t]
\footnotesize
\caption{Performance results for the two longest-running scripts from each benchmark}
\label{tab:performance-cut-by-hand}
\begin{tabular}{llp{2.5cm}p{1.4cm}rrrr}
\toprule

\input{tables/parallel-cut-by-hand}

\bottomrule

\end{tabular}
\end{table*}

%% file: tables/parallel-cut-by-hand.tex
\textbf{Benchmark} & \textbf{Script Name} & \textbf{Parallelized} & \textbf{Eliminated} & \textbf{$T_\text{orig}$} & \textbf{$u_{1}$} & \textbf{$u_{16}$} & \textbf{$T_{16}$} \\
\midrule
analytics-mts & 2.sh (vehicle days on road) & $8/8$ $(8/8)$ & $3$ $(3)$ & $335$\,s $(1.1\times)$ & $379$\,s & $41$\,s $(9.3\times)$ & $28$\,s $(13.5\times)$ \\
analytics-mts & 3.sh (vehicle hours on road) & $8/8$ $(8/8)$ & $3$ $(3)$ & $408$\,s $(1.0\times)$ & $427$\,s & $51$\,s $(8.4\times)$ & $38$\,s $(11.3\times)$ \\
oneliners & set-diff.sh & $5/8$ $(0/1,3/3,2/2,0/1,0/1)$ & $3$ $(0,2,1,0,0)$ & $879$\,s $(1.5\times)$ & $1308$\,s & $144$\,s $(9.1\times)$ & $128$\,s $(10.2\times)$ \\
oneliners & wf.sh & $4/5$ $(4/5)$ & $1$ $(1)$ & $1155$\,s $(1.8\times)$ & $2089$\,s & $196$\,s $(10.7\times)$ & $145$\,s $(14.4\times)$ \\
poets & 4_3b.sh (count_trigrams) & $4/9$ $(2/4,0/1,0/1,2/3)$ & $1$ $(1,0,0,0)$ & $862$\,s $(1.2\times)$ & $1049$\,s & $275$\,s $(3.8\times)$ & $279$\,s $(3.8\times)$ \\
poets & 8.2_2.sh (bigrams_appear_twice) & $4/9$ $(2/4,0/1,2/3,0/1)$ & $1$ $(1,0,0,0)$ & $645$\,s $(1.4\times)$ & $921$\,s & $177$\,s $(5.2\times)$ & $91$\,s $(10.2\times)$ \\
unix50 & 21.sh (8.4: longest words w/o hyphens) & $3/3$ $(3/3)$ & $1$ $(1)$ & $428$\,s $(1.7\times)$ & $733$\,s & $64$\,s $(11.4\times)$ & $49$\,s $(14.9\times)$ \\
unix50 & 23.sh (9.1: extract word PORT) & $6/6$ $(6/6)$ & $4$ $(4)$ & $111$\,s $(1.8\times)$ & $202$\,s & $23$\,s $(8.8\times)$ & $10$\,s $(19.8\times)$ \\

%% file: related.tex
\section{Related Work}
\label{prior}

We discuss related work in parallel execution of shell commands and scripts, synthesis of divide-and-conquer computations, program synthesis driven by provided input/output examples,
and synthesis of \unix commands. 

\heading{POSH and PaSh}
The POSH and PaSh systems parallelize and distribute Unix shell scripts~\cite{posh, pash}. Both systems require combiners and both systems work with manually coded combiners.
\sys eliminates the need for manually coded combiners, enabling such systems to immediately work with new commands (or new combinations of command flags) that require new combiners without the need to manually develop new combiners.

\heading{Synthesis of MapReduce Programs from Examples} \cite{mrsynth:16} present a technique for automatically synthesizing complete MapReduce programs given a partial specification in the form of a set of input/output examples. \sys, in contrast, supports (but does not synthesize) commands with much more sophisticated semantics than the synthesized map computations in \cite{mrsynth:16}. By working with existing shell commands, \sys also eliminates the need for the user to provide input/output examples. 

Preserving the order in which components appear in output streams is required to correctly implement the streaming semantics of \unix pipelines. \sys preserves this required order by incorporating metadata into the combiners and the parallelization. \cite{mrsynth:16}, in contrast, does not support ordered streams --- it targets computations that do not have ordering constraints and can produce output components in any order.

\heading{Automatic Parallelization of Divide and Conquer Computations} Some research in this area uses program analysis, typically over loops that access dense arrays or matrices, to generate parallel divide and conquer computations~\cite{RuginaR99,GuptaMS99}. Other research works with a complete characterization of the semantics of the original sequential computation~\cite{dq:17, dq:19}. \sys, in contrast,  synthesizes combiners for black-box streaming computations. \sys can therefore successfully target much more complex computations implemented in arbitrary programming languages. A trade-off is that the correctness of the \sys combiner synthesis algorithm relies on assumptions about the computation that the black-box components implement.

\heading{\unix Synthesis}
Prior work on synthesis for \unix shell commands and pipelines~\cite{macho:13,bhansali1993synthesis} is guided by examples or natural-language specifications.
Instead of automatically generating parallel or distributed versions of an
existing command or pipeline, the goal is to synthesize the sequential 
command itself from examples or natural language specs.

\heading{Commands and Shells}
There is a series of systems that aid developers in running commands or script fragments in a parallel or distributed fashion.
These range from simple \unix utilities~\cite{Tange2011a, yoo2003slurm, rush} to parallel/distributed shells~\cite{plan9:90, dgsh:17} to data-parallel frameworks that incorporate \unix commands~\cite{hadoop-streaming, dryad}
These tools require developers to modify programs to make use of the tools' APIs. \sys,  in contrast, aims to provide an automated solution that works directly on sequential scripts.

\section{Conclusion}

\sys synthesizes combiners that enable the exploitation of data parallelism in \unix commands and pipelines. Our experimental results show that the \sys input generation and combiner synthesis algorithms effectively identify correct combiners for our benchmark scripts and that these combiners enable the effective parallelization of these scripts.

%% file: acknowledgment.tex
\begin{acks}
  We are thankful to Shivam Handa, Kai Jia, Charles Jin, Konstantinos Kallas, Konstantinos Mamouras, and Claudia Zhu for interesting discussions.
\end{acks}

%% file: appendix.tex
\clearpage
\onecolumn

We present the combiner DSL semantics in \autoref{sec:app:semantics},
a correctness result in \autoref{sec:app:soundness},
performance results in \autoref{sec:app:results},
and combiner synthesis results in \autoref{sec:app:synthesis}.

\section{Appendix: DSL Semantics}
\label{sec:app:semantics}
\input{appendix/language}

\section{Appendix: Conditions for Synthesizing Correct Combiners}
\label{sec:app:soundness}
\input{appendix/soundness}

\section{Appendix: Performance Results}
\label{sec:app:results}

\autoref{tab:pipeline-counts} presents the pipeline stages that are automatically parallelized by \sys.
The first two columns present the benchmark and script names.
The next column (\textbf{Parallelized}) presents the number of stages automatically parallelized by \sys, $k$, and the number of stages in the original pipeline, $n$, as a pair ``$k/n$'' for each pipeline in the parentheses. Here we also report the single commands that are not in pipelines, as pairs ``$k/1$''. The pair before the parentheses presents the sum over all pipelines in the script.
The next column (\textbf{Eliminated}) presents the number of parallelized stages whose combiners are eliminated by \sys during optimization. Again, the numbers in the parentheses correspond to pipelines in the script. The number before the parentheses presents the sum over all pipelines.

\autoref{tab:performance-all-speedup} compares the parallel execution times with the original script execution times.
The first two columns present the benchmark and script names.
The next column (\textbf{$T_\text{orig}$}) presents the execution time for the original unmodified benchmark script.
The next column (\textbf{$u_1$}) presents the serial execution time.
The next column (\textbf{$u_{16}$}) presents the \emph{unoptimized} parallel execution time with 16 way parallelism.
The next column (\textbf{$T_{16}$}) presents the \emph{optimized} parallel execution time with 16 way parallelism.

\autoref{tab:performance-all-naive} presents the parallel execution times for \emph{unoptimized} pipelines with 1, 2, 4, 8, and 16 way parallelism.

\autoref{tab:performance-all-defer} presents the parallel execution times for \emph{optimized} pipelines with 1, 2, 4, 8, and 16 way parallelism.

Among all benchmark scripts, the unoptimized parallel speedup ranges between 0.5$\times$ and 14.9$\times$, with a median speedup of 5.3$\times$.
The optimized parallel speedup ranges between 0.6$\times$ and 26.9$\times$, with a median speedup of 7.1$\times$ (we attribute the superlinear speedup to pipelined parallelism exploited across consecutive parallelized commands with no intermediate combiner).
All scripts that exhibit a slowdown have a serial execution time under 10 seconds.

\autoref{tab:performance-3min} presents the performance results for benchmark scripts whose serial execution time is at least 3 minutes. In general, shorter scripts have smaller parallel speedup.

\input{tables/performance-all}

\section{Appendix: Combiner Synthesis Results}
\label{sec:app:synthesis}

\autoref{tab:plausible-combiners} summarizes all plausible combiners identified by \sys during the combiner synthesis for the benchmarks.
The first column presents the number of times the combiner appears as plausible across all benchmark scripts.
The second column presents the combiner in the \sys combiner DSL, with variables $a,b$ denoting input arguments and $*$ denoting any flags.
For each command in the benchmarks, the synthesized plausible combiners are all equivalent when operating on the command's outputs.

\autoref{tab:unsupported} presents all of the benchmark commands for which \sys did not synthesize a combiner.

\autoref{tab:synthesis}
presents comprehensive synthesis results from our benchmark set of commands. The first column presents the name of the command $f$. The second presents the command itself. The third presents the size of the candidate combiners with a finite scope of eight. Inside parentheses is the breakdown of the candidates into the three combiner classes $\DSLBasic, \DSLStruct, \DSLRun{f}$. The fourth column presents the wall-clock synthesis time in seconds.  The fifth column presents the final set of synthesized plausible combiners expressed in the combiner DSL, where variables $a, b$ denote the first and the second input streams, respectively.  The last column presents the number of these plausible combiners.

\input{tables/synthesis-combiners}

\input{tables/synthesis-errors}

\input{tables/synthesis}

%% file: appendix/language.tex
\begin{figure*}[t]
\centering
    \input{appendix/semantics}
\caption{
  \textbf{DSL Semantics.}
  The semantics of synthesizable combiners $g \in \DSLCombiner{f}$ for command $f$.
A combiner accepts two strings $\tty{1}, \tty{2}$ that are the outputs from two executions of $f$.
A plausible combiner $g$ for $f$ must satisfy
$\Eval{g}{\tty{1}}{\tty{2}}{f(\ttx{1}~{+}{+}~\ttx{2})}$
for all of the observed input pairs $\langle\ttx{1},\ttx{2}\rangle$, where $\tty{1} = f(\ttx{1})$ and $\tty{2} = f(\ttx{2})$.
}
\label{app:fig:semantics}
\end{figure*}

\autoref{app:fig:semantics} presents the big-step execution semantics for the DSL.
The transition function $\Rightarrow$ maps a DSL expression to its output value.
$\mathsf{strToInt}$ converts a string into an integer.
$\mathsf{intToStr}$ converts an integer into a string.
${+}{+}$ concatenates two strings.
$\mathsf{unixMerge}$ takes a comparator flag and two strings, then uses the flag to execute the ``\ttt{sort -m}'' command to merge the two strings.
$\mathsf{delFront}$ and $\mathsf{delBack}$ each takes a delimiter and a string. $\mathsf{delFront}$ removes the specified delimiter from the beginning of the string, while $\mathsf{delBack}$ removes the delimiter at the end of the stream.
$\mathsf{nil}$ denotes an empty string.
$\mathsf{splitFirst}$, $\mathsf{splitLast}$, and $\mathsf{splitLastNonempty}$ each takes a delimiter and a string.
$\mathsf{splitFirst}$ splits the string into elements separated by the delimiter, then returns the first element as the first output. It connects the remaining elements using the delimiter as the second output.
$\mathsf{splitLast}$ likewise splits the string with the delimiter, then returns the last element as the second output and returns the remaining substring as the first output.
$\mathsf{splitLastNonempty}$ splits the string with the delimiter, then returns the last nonempty element.
$\mathsf{delPad}$ removes leading spaces from a string, then returns the number of removed spaces as the first output and returns the remaining substring as the second output.
$\mathsf{calcPad}$ takes an integer and two strings, where the integer denotes the number of spaces that pad the first string. It returns the padding needed for the second string.
$\mathsf{addPad}$ inserts padding before a string.

%% file: appendix/semantics.tex
\footnotesize
\[
\begin{array}{c}

\begin{array}{cc}
    \begin{array}{c}
        \infral{
            i_1 = \mathsf{strToInt}~\tty{1}
            \qquad
            i_2 = \mathsf{strToInt}~\tty{2}
        }
        {\Eval{\mathbf{add}}{\tty{1}}{\tty{2}}
              {\mathsf{intToStr}~(i_1+i_2)}}
        {}
    \\ \\
            \infral{}
        {\Eval{\mathbf{concat}}{\tty{1}}{\tty{2}}
              {\tty{1}~{+}{+}~\tty{2}}}
        {}
        \\ \\
        \infral{}
        {\Eval{\mathbf{first}}{\tty{1}}{\tty{2}}
              {\tty{1}}}{}
        \\ \\
        \infral{}
        {\Eval{\mathbf{second}}{\tty{1}}{\tty{2}}
              {\tty{2}}}{}
        \\ \\
        \infral{
        \Eval{b}{(\mathsf{delFront}~d~\tty{1})}{(\mathsf{delFront}~d~\tty{2})}
                {v}
        }
        {\Eval{(\mathbf{front}~d~b)}{\tty{1}}{\tty{2}}
              {d ~{+}{+}~v}}
        {}
        \\ \\
        \infral{
        \Eval{b}{(\mathsf{delBack}~d~\tty{1})}{(\mathsf{delBack}~d~\tty{2})}
                {v}
        }
        {\Eval{(\mathbf{back}~d~b)}{\tty{1}}{\tty{2}}
              {v~{+}{+}~d}}
        {}
        \\ \\
        \infral{
            h_1, t_1 = \mathsf{splitFirst}~d~\tty{1}
            \qquad
            h_2, t_2 = \mathsf{splitFirst}~d~\tty{2}
            \\
            t_1 \ne \mathsf{nil}
            \qquad
            t_2 \ne \mathsf{nil}
            \qquad
            d \not\in t_1
            \qquad
            d \not\in t_2
            \\
            \Eval{b}{h_1}{h_2}{v}
            \qquad 
            \Eval{b}{t_1}{t_2}{v'}
        }
        {\Eval{(\mathbf{fuse}~d~b)}{\tty{1}}{\tty{2}}
              {v~{+}{+}~d~{+}{+}~v'}}
        {}
        \\ \\
        \infral{
            h_1, t_1 = \mathsf{splitFirst}~d~\tty{1}
            \qquad
            h_2, t_2 = \mathsf{splitFirst}~d~\tty{2}
            \\
            t_1 \ne \mathsf{nil}
            \qquad
            t_2 \ne \mathsf{nil}
            \qquad
            d \in t_1
            \qquad
            d \in t_2
            \\
            \Eval{b}{h_1}{h_2}{v}
            \qquad 
            \Eval{(\mathbf{fuse}~d~b)}{t_1}{t_2}{v'}
        }
        {\Eval{(\mathbf{fuse}~d~b)}{\tty{1}}{\tty{2}}
              {v~{+}{+}~d~{+}{+}~v'}}
        {}
        \\ \\
        \infral{
        }
        {\Eval{\mathbf{rerun}_f}{\tty{1}}{\tty{2}}
              {f~(\tty{1}~{+}{+}~\tty{2})}}
        {}
        \\ \\
        \infral{
            v = (\mathsf{unixMerge}~\text{<flags>})~\tty{1}~\tty{2}
        }
        {\Eval{(\mathbf{merge}~\text{<flags>})}{\tty{1}}{\tty{2}}{v}}
        {}
    \end{array}
    &
    \begin{array}{c}
        \infral{
            \tty{1} = `\textbackslash n`
            \text{ or }
            \tty{2} = `\textbackslash n`
        }
        {\Eval{(\mathbf{stitch}~b)}{\tty{1}}{\tty{2}}
              {\tty{1}~{+}{+}~\tty{2}}}
        {}
        \\
        \\
        \infral{
            y_1', l_1 = \mathsf{splitLastLine}~\tty{1}
            \qquad
            l_2, y_2' = \mathsf{splitFirstLine}~\tty{2}
            \qquad
            l_1 \ne l_2
        }
        {\Eval{(\mathbf{stitch}~b)}{\tty{1}}{\tty{2}}
              {\tty{1}~{+}{+}~\tty{2}}}
        {}
        \\
        \\
        \infral{
            y_1', l_1 = \mathsf{splitLastLine}~\tty{1}
            \qquad
            l_2, y_2' = \mathsf{splitFirstLine}~\tty{2}
            \qquad
            l_1 = l_2
            \qquad
            \Eval{b}{l_1}{l_2}{v}
        }
        {\Eval{(\mathbf{stitch}~b)}{\tty{1}}{\tty{2}}
              {y_1'~{+}{+}~`\textbackslash n`~{+}{+}~v~{+}{+}~`\textbackslash n`~{+}{+}~y_2'}}
        {}
        \\
        \\
        \infral{
            y_1', l_1 = \mathsf{splitLastLine}~v_1
            \qquad
            h_1, t_1 = \mathsf{splitFirst}~d~(\mathsf{delPad}~l_1)
            \\
            l_2, y_2' = \mathsf{splitFirstLine}~v_2
            \qquad
            h_2, t_2 = \mathsf{splitFirst}~d~(\mathsf{delPad}~l_2)
            \qquad
            t_1 \ne t_2
        }
        {\Eval{(\mathbf{stitch2}~d~b_1~b_2)}{v_1}{v_2}
              {\tty{1}~{+}{+}~\tty{2}}}
        {}
        \\
        \\
        \infral{
            y_1', l_1 = \mathsf{splitLastLine}~v_1
            \qquad
            h_1, t_1 = \mathsf{splitFirst}~d~(\mathsf{delPad}~l_1)
            \\
            l_2, y_2' = \mathsf{splitFirstLine}~v_2
            \qquad
            h_2, t_2 = \mathsf{splitFirst}~d~(\mathsf{delPad}~l_2)
            \\
            t_1 = t_2
            \qquad
            \Eval{b_1}{h_1}{h_2}{h}
            \qquad
            \Eval{b_2}{t_1}{t_2}{t}
            \qquad
            v = \mathsf{addPad}~(h~{+}{+}~d~{+}{+}~t)
        }
        {\Eval{(\mathbf{stitch2}~d~b_1~b_2)}{v_1}{v_2}
              {y_1'~{+}{+}~`\textbackslash n`~{+}{+}~v~{+}{+}~`\textbackslash n`~{+}{+}~y_2'}}
        {}
        \\
        \\
    \infral{}
    {\Eval{(\mathbf{helper}~d~b)}{h_1}{\mathsf{nil}}
          {\mathsf{nil}}}
    {}
    \\
    \\
    \infral{
            l_2, y_2' = \mathsf{splitFirstLine}~\tty{2}
            \qquad
            l_2 = \mathsf{nil}
            \qquad
            \Eval{(\mathbf{helper}~d~b)}{h_1}{y_2'}{v}
    }
    {\Eval{(\mathbf{helper}~d~b)}{h_1}{\tty{2}}
          {`\textbackslash n`~{+}{+}~v}}
    {}
    \\
    \\
    \infral{
            l_2, y_2' = \mathsf{splitFirstLine}~\tty{2}
            \qquad
            h_2, t_2 = \mathsf{splitFirst}~d~(\mathsf{delPad}~l_2)
            \\
            \Eval{b}{h_1}{h_2}{h}
            \qquad
            v = \mathsf{addPad}~(h~{+}{+}~d~{+}{+}~t_2)
            \qquad
            \Eval{(\mathbf{helper}~d~b)}{h_1}{y_2'}{v'}
    }
    {\Eval{(\mathbf{helper}~d~b)}{h_1}{\tty{2}}
          {v~{+}{+}~`\textbackslash n`~{+}{+}~v'}}
    {}
    \\
    \\
    \infral{
            y_1', l_1 = \mathsf{splitLastNonemptyLine}~\tty{1}
            \qquad
            h_1, t_1 = \mathsf{splitFirst}~d~(\mathsf{delPad}~l_1)
            \\
            \Eval{(\mathbf{helper}~d~b)}{h_1}{\tty{2}}{v}
    }
    {\Eval{(\mathbf{offset}~d~b)}{\tty{1}}{\tty{2}}{\tty{1}~{+}{+}~v}}
    {}
    \end{array}
\end{array}

\\
\vspace{5pt}
\\
b, b_1, b_2 \in \DSLBasic
\qquad
d \in \mathsf{Delim}
\qquad
\tty{1}, \tty{2}, y_1', y_2', v, v', v_1, v_2, h, h_1, h_2, t, t_1, t_2, l_1, l_2 \in \mathsf{String}
\qquad
i_1, i_2 \in \mathsf{Int}
\end{array}
\]
\vspace{-5pt}
\normalsize

%% file: appendix/soundness.tex
\begin{definition}\label{app:def:dsllegal}
For $g \in \DSLCombiner{f}$, $\SetLegal{g}$ denotes the set of legal strings for which $g$ is defined.
\input{proof/legal}
For any $\tty{1},\tty{2} \in \SetLegal{g}$, the evaluation $\Eval{g}{\tty{1}}{\tty{2}}{v}$ succeeds for some $v \in \mathsf{String}$.
\end{definition}

\begin{definition}
A stream is a string that ends with a newline character $`\textbackslash n`$,
$\mathsf{Stream} = \{ x ~{+}{+}~ `\textbackslash n` \mid x \in \mathsf{String} \}$.
\end{definition}

\begin{definition}
A command $f: \mathsf{Stream} \rightarrow \mathsf{Stream}$ is a function that takes a stream as input and produces a stream as output.\footnote{
Although this paper focuses on commands whose outputs terminate with newlines, the \sys algorithm applies also to commands whose outputs do not terminate with newlines.
}
\end{definition}

\begin{definition}
An input pair $\langle \ttx{1}, \ttx{2} \rangle$ consists of two strings $\ttx{1},\ttx{2} \in \mathsf{String}$.
An output tuple $\langle \tty{1},\tty{2},\tty{12} \rangle$ consists of three strings $\tty{1},\tty{2},\tty{12} \in \mathsf{String}$.
\end{definition}

\begin{definition}
An input stream pair $\langle \ttx{1},\ttx{2} \rangle$ consists of two streams $\ttx{1}, \ttx{2} \in \mathsf{Stream}$.
An observation $\langle \tty{1},\tty{2},\tty{12} \rangle$ consists of three streams $\tty{1},\tty{2},\tty{12} \in \mathsf{Stream}$.
\end{definition}

\begin{definition}
Executing command $f$ with an input stream pair $\langle \ttx{1}, \ttx{2} \rangle$ produces the observation $\langle f(\ttx{1}), f(\ttx{2}), f(\ttx{1}~{+}{+}~\ttx{2}) \rangle$.
For a set of input stream pairs $X$, $f(X)$ denotes the set of observations obtained from executing $f$ with $X$,
$f(X) = \{ \langle f(\ttx{1}), f(\ttx{2}), \allowbreak f(\ttx{1}~{+}{+}~\ttx{2}) \rangle \mid \langle \ttx{1},\ttx{2} \rangle \in X\}$.
\end{definition}

\begin{definition}\label{app:def:equivcap}
For $g_1, g_2 \in \DSLCombiner{f}$, $g_1$ and $g_2$ are equivalent by intersection, denoted as $\BoolEquivCap{g_1}{g_2}$, if
for all $\tty{1},\tty{2} \in \SetLegal{g_1} \cap \SetLegal{g_2}$,
$\Eval{g_1}{\tty{1}}{\tty{2}}{v}$ and $\Eval{g_2}{\tty{1}}{\tty{2}}{v}$ for some $v$.
\end{definition}

\begin{example}
For all $d \in \mathsf{Delim}$,
we have
$\BoolEquivCap
{(\mathbf{front}~d~\mathbf{concat})}
{(\mathbf{back}~d~\mathbf{concat})}$
and
$\BoolEquivCap
{(\mathbf{stitch2}~d~\mathbf{first}~\mathbf{first})}
{(\mathbf{stitch}~\mathbf{first})}$.
\end{example}

\begin{definition}\label{app:def:plausible}\label{app:thm:satlegal}
A combiner $g \in \DSLCombiner{f}$ is plausible for output tuples $Y$,
denoted as $\BoolSat{g}{Y}$, if
$\tty{1},\tty{2} \in \SetLegal{g}$
and
$\Eval{g}{\tty{1}}{\tty{2}}{\tty{12}}$
for all $\langle \tty{1},\tty{2},\tty{12} \rangle \in Y$.
\end{definition}

\begin{definition} \label{app:def:correctsat}
A combiner $g \in \DSLCombiner{f}$ is correct for command $f$ if 
$\BoolSat{g}{f(X)}$ holds for all input stream pairs $X$.
\end{definition}
\begin{remark}
Note that this definition does not require $\BoolSat{g}{Y}$ for all sets of output tuples $Y$.
\end{remark}

\begin{definition}
$\Count{d}{y}$ denotes the number of times that $d \in \mathsf{Delim}$ occurs in $y \in \mathsf{String}$.
We write $d \in y$ when $\Count{d}{y} > 0$ and write $d \not\in y$ when $\Count{d}{y} = 0$.
\end{definition}

\begin{definition}
We define two sets of representative combiners for command $f$,
$\GBasic = \{ g_\text{a}, g_\text{c}, g_\text{f}, g_\text{s}, g_\text{ba}, g_\text{fa}, g_\text{bfa}, g_\text{fbfa}, g_\text{fc} \} \subset \DSLBasic$
and
$\GStruct = \{ g_\text{sf}, g_\text{saf}, g_\text{oa} \} \subset \DSLStruct$,
whose elements are defined as follows:
\input{proof/special}
\end{definition}

\input{proof/requirements}

\begin{definition}
For combiner $g \in \GBasic \cup \GStruct$ and any set of output tuples $Y$,
$\BoolEnough{g}{Y}$ denotes a conservative predicate that is true only if $Y$ is sufficient for eliminating incorrect candidates when the correct combiner is $g$.
We define these predicates in \autoref{app:tab:requirements}.
\end{definition}

\begin{definition}
For any set of output tuples $Y$, $\BoolEnoughBasic{Y}$ denotes a conservative predicate that is true only if $Y$ is sufficient for eliminating incorrect candidates when the correct combiner $g \in \GBasic$.
$\BoolEnoughBasic{Y}$ is true if and only if the following conditions hold:
\begin{itemize}
\item There exists $\langle \tty{1},\tty{2},\tty{12} \rangle \in Y$ such that $\tty{1} \ne \tty{2}$.
\item There exists $\langle \tty{1},\tty{2},\tty{12} \rangle \in Y$ and $c \in \tty{1}$ such that $c \not \in \mathsf{Delim} \cup \{`0`\}$.
\item There exists $\langle \tty{1},\tty{2},\tty{12} \rangle \in Y$ and $c \in \tty{2}$ such that $c \not \in \mathsf{Delim} \cup \{`0`\}$.
\end{itemize}
\end{definition}

\begin{definition}
For any set of output tuples $Y$, $\BoolTable{Y}$ denotes a predicate that is true only if $Y$ is interpretable as a table.
$\BoolTable{Y}$ is true if and only if
there exists $p \in [`~`^+~\vert~`\textbackslash t`]$ and $d \in \mathsf{Delim}$ such that
for all $\langle \tty{1},\tty{2},\tty{12} \rangle \in Y$,
each line in $\tty{1},\tty{2},\tty{12}$ is either $\mathsf{nil}$ or of the form
$(p~{+}{+}~h~{+}{+}~d~{+}{+}~t)$ for some $h, t \in \mathsf{String}$.
\end{definition}

\begin{definition}
For any set of output tuples $Y$, $\BoolEnoughStruct{Y}$ denotes a conservative predicate that is true only if $Y$ is sufficient for eliminating incorrect candidates when the correct combiner $g \in \GStruct$.
$\BoolEnoughStruct{Y}$ is true if and only if the following conditions hold:
\begin{itemize}
\item
There exists $\langle \tty{1},\tty{2},\tty{12} \rangle \in Y$ such that
$(\mathsf{splitLastLine}~\tty{1}) = (y_1', l)$,
$(\mathsf{splitFirstLine}~\tty{2}) = (l, y_2')$,
and $(\mathsf{splitFirstLine}\allowbreak~y_2') = (l_2', y_2'')$,
where
$(\mathsf{firstChar}~(\mathsf{delPad}~l)) \not\in \mathsf{Delim} \cup \{`0`\}$,
$(\mathsf{lastChar}\allowbreak~l) \not\in \mathsf{Delim} \cup \{`0`\}$,
and
$l_2' \ne \mathsf{nil}$. 

\item
If $\BoolTable{Y}$ then $\BoolEnoughBasic{Y'}$, where
$Y' = \{ \langle h_1, h_2, y_{12}' \rangle
\mid
\langle \tty{1},\tty{2},\allowbreak\tty{12} \rangle \in Y,
(\mathsf{splitLastLine}~\tty{1}) = (y_1', l_1),
(\mathsf{splitFirstLine}~\tty{2}) = (l_2, y_2'),\allowbreak
(\mathsf{splitFirst}~d~(\mathsf{delPad}~l_1)) = (h_1, t),
\text { and }
(\mathsf{splitFirst}\allowbreak~d\allowbreak~(\mathsf{delPad}~l_2)) = (h_2, t)
\}$.
\end{itemize}
\end{definition}

\begin{definition}
The size of a combiner $g \in \DSLCombiner{f}$ is denoted as $\Size{g}$ and defined as two (each combiner operates on two arguments) plus the number of times that the AST of $g$ applies a production to expand a ``$\DSLBasic$'', ``$\DSLStruct$'', or ``$\DSLRun{f}$'' symbol.
\end{definition}

\begin{example}
We have $\Size{g_\text{a}} = 3$, $\Size{g_\text{fbfa}} = 6$, and $\Size{g_\text{saf}} = 5$.
\end{example}

\begin{definition} \label{app:def:setsat}
For a command $f$, integer $k$, and set of output tuples $Y$,
the set of plausible combiners
$\SetSat{k}{Y} = \{ g \in \DSLCombiner{f} \mid \Size{g} \le k \text{ and } \BoolSat{g}{Y} \}$.
\end{definition}

\begin{proposition}\label{app:thm:correctlegal}
If combiner $g \in \DSLCombiner{f}$ is correct for command $f$,
then
$\tty{1},\tty{2} \in \SetLegal{g}$
holds for all $\langle \tty{1},\tty{2},\tty{12} \rangle \in f(X)$
where $X$ is any set of input stream pairs.
\end{proposition}

\begin{proposition} \label{app:thm:enoughbasic}
For any combiner $g \in \GBasic$ and set of output tuples $Y$, if $\BoolSat{g}{Y}$ and $\BoolEnoughBasic{Y}$ then $\BoolEnough{g}{Y}$.
\end{proposition}
\begin{proof}
By induction on the derivation of $g$.
\end{proof}

\begin{proposition}
For any $d \in \mathsf{Delim}$ and $b_1,b_2 \in \DSLBasic$,
$Y \subseteq \SetLegal{\mathbf{stitch2}~d~b_1~b_2}$ implies $\BoolTable{Y}$.
For any $d \in \mathsf{Delim}$ and $b \in \DSLBasic$,
$Y \subseteq \SetLegal{\mathbf{offset}~d~b}$ implies $\BoolTable{Y}$.
\end{proposition}

\begin{proposition} \label{app:thm:enoughstruct}
For any combiner $g \in \GStruct$ and set of output tuples $Y$, if $\BoolSat{g}{Y}$ and $\BoolEnoughStruct{Y}$ then $\BoolEnough{g}{Y}$.
\end{proposition}

\begin{proposition}
For any integers $k_1,k_2$ such that $0 < k_1 < k_2$, we have
$\SetSat{k_1}{Y} \subseteq \SetSat{k_2}{Y}$
for all sets of output tuples $Y$.
\end{proposition}

\begin{proposition}
If combiner $g \in \DSLCombiner{f}$ is correct for command $f$,
then
$g \in \SetSat{\Size{g}}{f(X)}$ for all sets of input stream pairs $X$.
\end{proposition}

\begin{proposition} \label{app:thm:correctisplausible}
For any integer $k \ge 6$ and set of input stream pairs $X$, we have:
\begin{itemize}
\item If combiner $g \in \GBasic$ is correct for command $f$,
then $g \in \SetSat{k}{f(X)} \cap \DSLBasic$.
\item If combiner $g \in \GBasic$ is correct for command $f$,
then $g \in \SetSat{k}{f(X)} \cap \DSLStruct$.
\end{itemize}
\end{proposition}

\begin{lemma} \label{app:thm:basicdnotin}
For any $g \in \DSLBasic$, $\tty{1},\tty{2} \in \mathsf{String}$, and $d \in \mathsf{Delim}$,
if $\Eval{g}{\tty{1}}{\tty{2}}{v}$ and $d \not\in \tty{1}$ and $d \not\in \tty{2}$ then $d \not\in v$.
\end{lemma}
\begin{proof}
By induction on the derivation of $g$.
\end{proof}

\begin{lemma}\label{app:thm:basicnoty1zy2}
For any $g \in \DSLBasic$, $\tty{1},\tty{2} \in \mathsf{String}$, and $z \in \mathsf{String}$,
if $\Eval{g}{\tty{1}}{\tty{2}}{v}$ and $z \ne \mathsf{nil}$ then $v \ne \tty{1}~{+}{+}~z~{+}{+}~\tty{2}$.
\end{lemma}
\begin{proof}
By induction on the derivation of $g$, where the case of ``$g = \mathbf{fuse}~d~b$'' uses \autoref{app:thm:basicdnotin}.
\end{proof}

\begin{lemma}\label{app:thm:fusedcount}
For any $d \in \mathsf{Delim}$, $b \in \DSLBasic$, $g = \mathbf{fuse}~d~b$, $\tty{1},\tty{2} \in \SetLegal{g}$, and $\tty{12} \in \mathsf{String}$ such that $\Eval{g}{\tty{1}}{\tty{2}}{\tty{12}}$,
we have $\Count{d}{\tty{1}} = \Count{d}{\tty{2}} = \Count{d}{\tty{12}}$.
\end{lemma}
\begin{proof}
By \autoref{app:thm:satlegal} and \autoref{app:def:dsllegal}, $\tty{1},\tty{2} \in \SetLegal{g'} = \{ y_1~{+}{+}~d~{+}{+}\allowbreak~y_2~{+}{+}\allowbreak~d~{+}{+}\allowbreak~\ldots~{+}{+}~d~{+}{+}~y_k \mid y_i \in \SetLegal{b} \text{ and } d \not\in y_i \text{ for all }i=1,\ldots,k, \text{ where}~\allowbreak k \ge 2 \}$.
Let $k = \Count{d}{\tty{1}} + 1$.
By \autoref{app:fig:semantics}, $\Count{d}{\tty{2}} = \Count{d}{\tty{1}} = k - 1$.
Let $\tty{1} = y^1_1~{+}{+}~d~{+}{+}~\ldots~{+}{+}~d~{+}{+}~y^1_k$ and $\tty{2} = y^2_1~{+}{+}~d~{+}{+}~\ldots~{+}{+}\allowbreak~d~{+}{+}\allowbreak~y^2_k$ where $y^1_i,y^2_i \in \SetLegal{b}$, $d \not\in y^1_i$, and $d \not\in y^2_i$ for all $i=1,\ldots,k$.
By \autoref{app:fig:dsl}, $d \in \mathsf{Delim}$ and $b \in \DSLBasic$.
By \autoref{app:fig:semantics}, there exists $v_1,\ldots,v_k \in \mathsf{String}$ such that $\tty{12} = v_1~{+}{+}~d~{+}{+}\allowbreak~\ldots~{+}{+}~d~{+}{+}~v_k$ and $\Eval{b}{y^1_i}{y^2_i}{v_i}$ for all $i=1,\ldots,k$.
By \autoref{app:thm:basicdnotin}, $d \not\in v_i$ for all $i=1,\ldots,k$.
Hence $\Count{d}{\tty{12}} = k-1$.
\end{proof}

\begin{lemma}\label{app:thm:maxcount}
For any $d \in \mathsf{Delim}$, $g \in \DSLBasic$, and $\tty{1},\allowbreak\tty{2},\allowbreak\tty{12} \in \mathsf{String}$ such that $\Eval{g}{\tty{1}}{\tty{2}}{\tty{12}}$,
we have $\Count{d}{\tty{12}} \le \Count{d}{\tty{1}} + \Count{d}{\tty{2}}$.
\end{lemma}
\begin{proof}
By induction on the derivation of $g$.
\end{proof}

\begin{theorem} \label{app:thm:equivcapbasic}
For any combiner $g \in \GBasic$,
set of output tuples $Y$ such that $\BoolSat{g}{Y}$ and $\BoolEnough{g}{Y}$,
and
$g' \in \DSLBasic$,
we have
$\BoolSat{g'}{Y}$
implies
$\BoolEquivCap{g'}{g}$.
\end{theorem}
\begin{proof}
\input{proof/basicsatweaker}
\end{proof}

\begin{theorem} \label{app:thm:enoughbasicequiv}
For any combiners $g \in \GBasic, g' \in \DSLBasic$
and set of output tuples $Y$,
if $\BoolEnoughBasic{Y}$, $\BoolSat{g}{Y}$, and $\BoolSat{g'}{Y}$,
then $\BoolEquivCap{g'}{g}$.
\end{theorem}
\begin{proof}
By \autoref{app:thm:equivcapbasic} and \autoref{app:thm:enoughbasic}.
\end{proof}

\begin{theorem} \label{app:thm:plausibleequivcorrectbasic}
For any command $f$, set of input streams $X$, combiner $g \in \GBasic$, and combiner $g' \in \DSLBasic$,
if the following conditions hold:
\begin{itemize}
\item $\BoolEnoughBasic{f(X)}$,
\item $g$ is correct for $f$, and
\item $\tty{1},\tty{2} \in \SetLegal{g'}$ for all $\langle \tty{1},\tty{2},\tty{12} \rangle \in f(X)$,
\end{itemize}
then
$\BoolSat{g'}{f(X)}$ if and only if $\BoolEquivCap{g'}{g}$.
\end{theorem}
\begin{proof}
To show that ``if $\BoolSat{g'}{f(X)}$ then $\BoolEquivCap{g'}{g}$'':
By \autoref{app:def:correctsat}, $\BoolSat{g}{f(X)}$.
By \autoref{app:thm:enoughbasicequiv}, $\BoolEquivCap{g'}{g}$.

To show that ``if $\BoolEquivCap{g'}{g}$ then $\BoolSat{g'}{f(X)}$'':
For all $\langle \tty{1},\tty{2},\tty{12} \rangle \in f(X)$, by \autoref{app:thm:correctlegal}, $\tty{1},\tty{2} \in \SetLegal{g}$.
Hence $\tty{1},\tty{2} \in \SetLegal{g} \cap \SetLegal{g'}$.
By \autoref{app:def:equivcap}, $\Eval{g}{\tty{1}}{\tty{2}}{v}$ and $\Eval{g'}{\tty{1}}{\tty{2}}{v}$ for some $v$.
By \autoref{app:def:correctsat}, $\BoolSat{g}{f(X)}$.
By \autoref{app:def:plausible}, $\Eval{g}{\tty{1}}{\tty{2}}{\tty{12}}$.
Hence $v = \tty{12}$.
By \autoref{app:def:plausible}, $\BoolSat{g'}{f(X)}$.
\end{proof}

\begin{theorem} \label{app:thm:synthesizebasic}
For any command $f$, set of input streams $X$, combiner $g \in \GBasic$, combiner $g' \in \DSLBasic$, and integer $k$,
if the following conditions hold:
\begin{itemize}
\item $\BoolEnoughBasic{f(X)}$,
\item $g$ is correct for $f$,
\item $\tty{1},\tty{2} \in \SetLegal{g'}$ for all $\langle \tty{1},\tty{2},\tty{12} \rangle \in f(X)$, and
\item $k \ge \Size{g'}$,
\end{itemize}
then
$g' \in \SetSat{k}{f(X)} \cap \DSLBasic$ if and only if $\BoolEquivCap{g'}{g}$.
\end{theorem}
\begin{proof}
By \autoref{app:thm:plausibleequivcorrectbasic} and \autoref{app:def:setsat}.
\end{proof}

\begin{remark}
 \autoref{app:thm:synthesizebasic} states that, if the specified size is large enough, if correct combiner is among $\GBasic$, and if the synthesizer has collected sufficient observations, then the synthesizer must return either the correct combiner or its equivalent.
By \autoref{app:thm:correctisplausible}, we know that as long as the specified size $k \ge 6$, the synthesizer is guaranteed to return a correct combiner if the correct combiner is among $\GBasic$.
\end{remark}

\begin{lemma} \label{app:thm:equivcapfirstsecond}
For any set of output tuples $Y$, let $E(Y)$ be a predicate that is true if and only if the following conditions hold:
\begin{itemize}
\item There exists $\langle \tty{1},\tty{2},\tty{12} \rangle \in Y$ and $c \in \tty{1}$ such that $c \not \in \mathsf{Delim} \cup \{`0`\}$.
\item There exists $\langle \tty{1},\tty{2},\tty{12} \rangle \in Y$ and $c \in \tty{2}$ such that $c \not \in \mathsf{Delim} \cup \{`0`\}$.
\end{itemize}
For any combiner $g \in \{ g_\text{f}, g_\text{s} \}$,
set of output tuples $Y$ such that $\BoolSat{g}{Y}$ and $E(Y)$,
and
$g' \in \DSLBasic$,
if
$\BoolSat{g'}{Y}$
then
either $\BoolEquivCap{g'}{g_\text{f}}$ or $\BoolEquivCap{g'}{g_\text{s}}$.
\end{lemma}
\begin{proof}
\begin{mycases}
\item
There exists $\langle \tty{1},\tty{2},\tty{12} \rangle \in Y$ such that $\tty{1} \ne \tty{2}$.
By \autoref{app:thm:equivcapbasic}, $\BoolEquivCap{g'}{g}$.
\item
For all $\langle \tty{1},\tty{2},\tty{12} \rangle \in Y$, $\tty{1} = \tty{2}$.
The proof is similar to the proof of \autoref{app:proof:basicsatweaker:first} in \autoref{app:thm:equivcapbasic}.
\end{mycases}
\end{proof}

\begin{theorem} \label{app:thm:equivcapstruct}
For any combiner $g \in \GStruct$,
set of output tuples $Y$ such that $\BoolSat{g}{Y}$ and $\BoolEnough{g}{Y}$,
and
$g' \in \DSLStruct$,
we have
$\BoolSat{g'}{Y}$
implies
$\BoolEquivCap{g'}{g}$.
\end{theorem}
\begin{proof}
\input{proof/structsatequiv}
\end{proof}

\begin{theorem}
For any combiners $g \in \GStruct, g' \in \DSLStruct$
and set of output tuples $Y$,
if $\BoolEnoughStruct{Y}$, $\BoolSat{g}{Y}$, and $\BoolSat{g'}{Y}$,
then $\BoolEquivCap{g'}{g}$.
\end{theorem}
\begin{proof}
By \autoref{app:thm:equivcapstruct} and \autoref{app:thm:enoughstruct}.
\end{proof}

\begin{theorem} \label{app:thm:plausibleequivcorrectstruct}
For any command $f$, set of input streams $X$, combiner $g \in \GStruct$, and combiner $g' \in \DSLStruct$,
if the following conditions hold:
\begin{itemize}
\item $\BoolEnoughStruct{f(X)}$,
\item $g$ is correct for $f$, and
\item $\tty{1},\tty{2} \in \SetLegal{g'}$ for all $\langle \tty{1},\tty{2},\tty{12} \rangle \in f(X)$,
\end{itemize}
then
$\BoolSat{g'}{f(X)}$ if and only if $\BoolEquivCap{g'}{g}$.
\end{theorem}
\begin{proof}
The proof is similar to the proof of \autoref{app:thm:plausibleequivcorrectbasic}.
\end{proof}

\begin{theorem} \label{app:thm:synthesizestruct}
For any command $f$, set of input streams $X$, combiner $g \in \GStruct$, combiner $g' \in \DSLStruct$, and integer $k$,
if the following conditions hold:
\begin{itemize}
\item $\BoolEnoughStruct{f(X)}$,
\item $g$ is correct for $f$,
\item $\tty{1},\tty{2} \in \SetLegal{g'}$ for all $\langle \tty{1},\tty{2},\tty{12} \rangle \in f(X)$, and
\item $k \ge \Size{g'}$,
\end{itemize}
then
$g' \in \SetSat{k}{f(X)} \cap \DSLStruct$ if and only if $\BoolEquivCap{g'}{g}$.
\end{theorem}
\begin{proof}
By \autoref{app:thm:plausibleequivcorrectstruct} and \autoref{app:def:setsat}.
\end{proof}

\begin{remark}
 \autoref{app:thm:synthesizestruct} states that, if the specified size is large enough, if correct combiner is among $\GStruct$, and if the synthesizer has collected sufficient observations and eliminated $\DSLBasic$ candidates, then the synthesizer must return either the correct combiner or its equivalent.
By \autoref{app:thm:correctisplausible}, we know that as long as the specified size $k \ge 6$, the synthesizer is guaranteed to return a correct combiner if the correct combiner is among $\GStruct$.
\end{remark}

%% file: proof/legal.tex
\begin{align*}
\SetLegal{\mathbf{add}} &= [`0`-`9`]^+ \\
\SetLegal{\mathbf{concat}} &= \mathsf{String} \\
\SetLegal{\mathbf{first}} &= \mathsf{String} \\
\SetLegal{\mathbf{second}} &= \mathsf{String} \\
\SetLegal{\mathbf{front}~d~b} &= \{ d~{+}{+}~y \mid y \in \SetLegal{b} \} \\
\SetLegal{\mathbf{back}~d~b} &= \{ y~{+}{+}~d \mid y \in \SetLegal{b} \} \\
\SetLegal{\mathbf{fuse}~d~b} &= \{ y_1~{+}{+}~d~{+}{+}~y_2~{+}{+}~d~{+}{+}~\ldots~{+}{+}~d~{+}{+}~y_k \\
                             &\phantom{= \{}
                              \mid y_1 \ne \mathsf{nil}, y_k \ne \mathsf{nil}, \text{ and } y_i \in \SetLegal{b} \text{ and } d \not\in y_i \\
                             &\phantom{= \{ \mid~}
                              \text{ for all }i=1,\ldots,k,
                              \text{ where } k \ge 2\} \\
\SetLegal{\mathbf{stitch}~b} &= \{ y_1~{+}{+}~`\textbackslash n`~{+}{+}~\ldots~{+}{+}~y_k~{+}{+}~`\textbackslash n` \\ 
                             &\phantom{= \{}
                              \mid y_i \in \SetLegal{b} \text{ and } `\textbackslash n` \not\in y_i \text{ for all }i=1,\ldots,k \\
                             &\phantom{= \{ \mid~}
                              \text{ where } k \ge 1\} \\
                             &\phantom{=}
                              \cup \{`\textbackslash n`\}\\ 
\SetLegal{\mathbf{stitch2}~d~b_1~b_2} &= \{ y_1~{+}{+}~`\textbackslash n`~{+}{+}~\ldots~{+}{+}~y_k~{+}{+}~`\textbackslash n` \\ 
                             &\phantom{= \{}
                              \mid y_i = p~{+}{+}~h_i~{+}{+}~d~{+}{+}~t_i
                              \text{ and } `\textbackslash n` \not\in y_i\\
                             &\phantom{= \{ \mid~}
                              \text{for all }i=1,\ldots,k,
                              \text{ where }
                              k \ge 1, p \in [`~`^+~\vert~`\textbackslash t`],\\
                             &\phantom{= \{ \mid~}
                              h_i \in \SetLegal{b_1}, d \not\in h_i, \text{ and } t_i \in \SetLegal{b_2} \} \\
                             &\phantom{=}
                              \cup \{`\textbackslash n`\}\\ 
\SetLegal{\mathbf{offset}~d~b} &= \{ y_1~{+}{+}~`\textbackslash n`~{+}{+}~\ldots~{+}{+}~y_k~{+}{+}~`\textbackslash n` \\ 
                             &\phantom{= \{}
                              \mid y_i \in \{\mathsf{nil}, (p~{+}{+}~h_i~{+}{+}~d~{+}{+}~t_i) \}
                              \text{ and } `\textbackslash n` \not\in y_i\\
                             &\phantom{= \{ \mid~}
                              \text{for all }i=1,\ldots,k,
                              \text{ where }
                              k \ge 1, p \in [`~`^+~\vert~`\textbackslash t`],\\
                             &\phantom{= \{ \mid~}
                              h_i \in \SetLegal{b}, d \not\in h_i, \text{ and } t_i \in \mathsf{String}\} \\
\SetLegal{\mathbf{rerun}_f} &= \{ \text{legal inputs for } f \} \\ 
\SetLegal{\mathbf{merge}~\text{<flags>}} &= \{ \text{legal inputs for } (\mathsf{unixMerge}~\text{<flags>})\} \\
\end{align*}

%% file: proof/special.tex
\begin{align*}
g_\text{a} &= \mathbf{add} \in \DSLBasic \\
g_\text{c} &= \mathbf{concat} \in \DSLBasic \\
g_\text{f} &= \mathbf{first} \in \DSLBasic \\
g_\text{s} &= \mathbf{second} \in \DSLBasic \\
g_\text{ba} &= (\mathbf{back}~d~\mathbf{add}) \in \DSLBasic \\
g_\text{fa} &= (\mathbf{fuse}~d~\mathbf{add}) \in \DSLBasic \\
g_\text{bfa} &= (\mathbf{back}~d_1~(\mathbf{fuse}~d_2~\mathbf{add})) \in \DSLBasic \\
g_\text{fbfa} &= (\mathbf{front}~d_1~(\mathbf{back}~d_2~(\mathbf{fuse}~d_3~\mathbf{add}))) \in \DSLBasic \\
g_\text{fc} &= (\mathbf{front}~d~\mathbf{concat}) \in \DSLBasic \\
g_\text{sf} &= (\mathbf{stitch}~\mathbf{first}) \in \DSLStruct \\
g_\text{saf} &= (\mathbf{stitch2}~d~\mathbf{add}~\mathbf{first}) \in \DSLStruct \\
g_\text{oa} &= (\mathbf{offset}~d~\mathbf{add}) \in \DSLStruct
\end{align*}

%% file: proof/requirements.tex
\begin{table*}[t] \footnotesize \centering
\caption{Representative combiners}
\vspace{-5pt}
\label{app:tab:requirements}
\begin{tabular}{llp{11cm}} \toprule
\textbf{Combiner $g$} & \textbf{Stage} & \textbf{Conditions for $\BoolEnough{g}{Y}$ to be true} \\ \midrule
$g_\text{a} = \mathbf{add}$ & $\DSLBasic$ &
The following conditions hold:
(1) There exists $\langle \tty{1},\tty{2},\tty{12} \rangle \in Y$ such that $\tty{1} \not\in `0`^+$.
(2) There exists $\langle \tty{1},\tty{2},\tty{12} \rangle \in Y$ such that $\tty{2} \not\in `0`^+$.
\\

$g_\text{c} = \mathbf{concat}$ & $\DSLBasic$ &
The following conditions hold:
(1) There exists $\langle \tty{1},\tty{2},\tty{12} \rangle \in Y$ such that $\tty{1} \ne \mathsf{nil}$.
(2) There exists $\langle \tty{1},\tty{2},\tty{12} \rangle \in Y$ such that $\tty{2} \ne \mathsf{nil}$.
\\

$g_\text{f} = \mathbf{first}$ & $\DSLBasic$ &
The following conditions hold:
(1) There exists $\langle \tty{1},\tty{2},\tty{12} \rangle \in Y$ such that $\tty{1} \ne \tty{2}$.
(2) There exists $\langle \tty{1},\tty{2},\tty{12} \rangle \in Y$ and $c \in \tty{2}$ such that $c \not \in \mathsf{Delim} \cup \{`0`\}$.
\\

$g_\text{s} = \mathbf{second}$ & $\DSLBasic$ &
The following conditions hold:
(1) There exists $\langle \tty{1},\tty{2},\tty{12} \rangle \in Y$ such that $\tty{1} \ne \tty{2}$.
(2) There exists $\langle \tty{1},\tty{2},\tty{12} \rangle \in Y$ and $c \in \tty{1}$ such that $c \not \in \mathsf{Delim} \cup \{`0`\}$.
\\

$g_\text{ba} = (\mathbf{back}~d~\mathbf{add})$ & $\DSLBasic$ &
$\BoolEnough{g_\text{a}}{Y'}$ where
$Y' = \{\langle \tty{1},\tty{2},\tty{12} \rangle \mid \langle (\tty{1}~{+}{+}~d),(\tty{2}~{+}{+}~d),(\tty{12}~{+}{+}~d) \rangle \in Y\}$.
\\

$g_\text{fa} = (\mathbf{fuse}~d~\mathbf{add})$ & $\DSLBasic$ &
$\BoolEnough{g_\text{a}}{Y'}$ where
$Y' = \{\langle \tty{1,i},\tty{2,i},\tty{12,i} \rangle \mid \langle (\tty{1,1}~{+}{+}~d~{+}{+}~\ldots~{+}{+}~d~{+}{+}~\tty{1,n}),(\tty{2,1}~{+}{+}~d~{+}{+}\allowbreak~\ldots~{+}{+}\allowbreak~d~{+}{+}\allowbreak~\tty{2,n}),(\tty{12,1}~{+}{+}~d~{+}{+}~\ldots~{+}{+}~d~{+}{+}~\tty{12,n}) \rangle \in Y,
\text{ where } d \not\in y^1_j \text{ and } d \not\in y^2_j \text{ for all } j=1,\ldots,n,
\text{ and } i\in\{1,\ldots,n\} \}$.
\\

$g_\text{bfa} = (\mathbf{back}~d_1~(\mathbf{fuse}~d_2~\mathbf{add}))$ & $\DSLBasic$ &
$\BoolEnough{g_\text{fa}}{Y'}$ where
$Y' = \{\langle \tty{1},\tty{2},\tty{12} \rangle \mid \langle (\tty{1}~{+}{+}~d),(\tty{2}~{+}{+}~d),(\tty{12}~{+}{+}~d) \rangle \in Y\}$.
\\

$g_\text{fbfa} = (\mathbf{front}~d_1~(\mathbf{back}~d_2~(\mathbf{fuse}~d_3~\mathbf{add})))$ & $\DSLBasic$ &
$\BoolEnough{g_\text{bfa}}{Y'}$ where
$Y' = \{\langle \tty{1},\tty{2},\tty{12} \rangle \mid \langle (d~{+}{+}~\tty{1}),(d~{+}{+}~\tty{2}),(d~{+}{+}~\tty{12}) \rangle \in Y\}$.
\\

$g_\text{fc} = (\mathbf{front}~d~\mathbf{concat})$ & $\DSLBasic$ &
$\BoolEnough{g_\text{c}}{Y'}$ where
$Y' = \{\langle \tty{1},\tty{2},\tty{12} \rangle \mid \langle (d~{+}{+}~\tty{1}),(d~{+}{+}~\tty{2}),(d~{+}{+}~\tty{12}) \rangle \in Y\}$.
\\

$g_\text{sf} = (\mathbf{stitch}~\mathbf{first})$ & $\DSLStruct$ &
The following conditions hold:
(1) There exists $\langle \tty{1},\tty{2},\tty{12} \rangle \in Y$ such that
$(\mathsf{splitLastLine}~\tty{1}) = (y_1', l)$
and
$(\mathsf{splitFirstLine}~\tty{2}) = (l, y_2')$
where
$(\mathsf{firstChar}~(\mathsf{delPad}~l)) \not\in \mathsf{Delim} \cup \{`0`\}$
and $(\mathsf{lastChar}~l) \not\in \mathsf{Delim} \cup \{`0`\}$.
(2) If $Y \subseteq \SetLegal{\mathbf{stitch2}~d~b_1~b_2}$ for some $d \in \mathsf{Delim}$ and $b_1,b_2 \in \DSLBasic$,
then there exists $\langle \tty{1},\tty{2},\tty{12} \rangle \in Y$ such that
$h_1 \ne h_2$, where
$(\mathsf{splitLastLine}~\tty{1}) = (y_1', l_1)$,
$(\mathsf{splitFirstLine}~\tty{2}) = (l_2, y_2')$,
$(\mathsf{splitFirst}~d~(\mathsf{delPad}~l_1)) = (h_1, t)$,
and
$(\mathsf{splitFirst}~d~(\mathsf{delPad}~l_2)) = (h_2, t)$.
\\

$g_\text{saf} = (\mathbf{stitch2}~d~\mathbf{add}~\mathbf{first})$ & $\DSLStruct$ &
There exists $\langle \tty{1},\tty{2},\tty{12} \rangle \in Y$ such that
$(\mathsf{splitLastLine}~\tty{1}) = (y_1', l)$
and
$(\mathsf{splitFirstLine}~\tty{2}) = (l, y_2')$
where
$(\mathsf{firstChar}~(\mathsf{delPad}~l)) \not\in \mathsf{Delim} \cup \{`0`\}$
and $(\mathsf{lastChar}~l) \not\in \mathsf{Delim} \cup \{`0`\}$.
\\

$g_\text{oa} = (\mathbf{offset}~d~\mathbf{add})$ & $\DSLStruct$ &
The following conditions hold:
(1)
There exists $\langle \tty{1},\tty{2},\tty{12} \rangle \in Y$ such that
$(\mathsf{splitLastLine}~\tty{1}) = (y_1', l_1)$,
$(\mathsf{splitFirstLine}~\tty{2}) = (l_2, y_2')$,
and
$(\mathsf{splitFirstLine}~y_2') = (l_2', y_2'')$,
where $(\mathsf{firstChar}~(\mathsf{delPad}~l_1)) \not\in \mathsf{Delim} \cup \{`0`\}$,
$l_2 \ne \mathsf{nil}$,
and
$l_2' \ne \mathsf{nil}$.
(2)
$\BoolEnough{g_\text{a}}{Y'}$ where
$Y' = \{ \langle h_1, h_2, y_{12}' \rangle
\mid
\langle \tty{1},\tty{2},\tty{12} \rangle \in Y,
(\mathsf{splitLastLine}~\tty{1}) = (y_1', l_1),
(\mathsf{splitFirstLine}~\tty{2}) = (l_2, y_2'),
(\mathsf{splitFirst}~d~(\mathsf{delPad}~l_1)) = (h_1, t_1),
\text { and }
(\mathsf{splitFirst}~d~(\mathsf{delPad}~l_2)) = (h_2, t_2)
\}$.
\\

\bottomrule
\end{tabular}
\end{table*}

%% file: proof/basicsatweaker.tex
The proof is by induction on the derivation of $g$.

\begin{mycases}
\item $g = g_\text{a}$.
The proof performs case analysis of the values of $g'$.
By \autoref{app:tab:requirements}, there exists $\langle \tty{1},\tty{2},\tty{12} \rangle \in Y$.
By \autoref{app:thm:satlegal} and \autoref{app:def:dsllegal}, $\tty{1},\tty{2} \in \SetLegal{g} = [`0`-`9`]^+$.
By \autoref{app:def:plausible}, $\Eval{g}{\tty{1}}{\tty{2}}{\tty{12}}$.
By \autoref{app:fig:semantics},
$\tty{12} = \mathsf{intToStr}~((\mathsf{strToInt}~\tty{1})+(\mathsf{strToInt}~\tty{2}))$.

\begin{mycases}
\item $g' = \mathbf{add} = g$.
By \autoref{app:def:equivcap}, $\BoolEquivCap{g'}{g}$.

\item $g' = \mathbf{concat}$.
\label{app:proof:basicsatweaker:add:concat}
We show that $\BoolSat{g'}{Y}$ never holds in this case.
If $\tty{1} \not\in `0`^+$, we have
$(\mathsf{strToInt}~\tty{1}) > 0$.
Hence
$\mathsf{strToInt}~(\tty{1}~{+}{+}~\tty{2})
\ge 10 \cdot (\mathsf{strToInt}~\tty{1})+(\mathsf{strToInt}~\tty{2})
> (\mathsf{strToInt}~\tty{1})+(\mathsf{strToInt}~\tty{2})$.
If $\tty{1} \in `0`^+$, the string $(\tty{1}~{+}{+}~\tty{2})$ contains leading `0` characters that are absent in the results of $\mathsf{intToStr}$.
Either case, we have $\mathsf{intToStr}~((\mathsf{strToInt}\allowbreak~\tty{1})+(\mathsf{strToInt}~\tty{2})) \ne \tty{1}~{+}{+}~\tty{2}$.
Assume the opposite that $\BoolSat{g'}{Y}$.
By \autoref{app:def:plausible}, $\Eval{g'}{\tty{1}}{\tty{2}}{\tty{12}}$.
By \autoref{app:fig:semantics},
$\tty{12} = \tty{1}~{+}{+}~\tty{2}$.
We have the desired contradiction.

\item $g' = \mathbf{first}$.
\label{app:proof:basicsatweaker:add:first}
We show that $\BoolSat{g'}{Y}$ never holds in this case.
By \autoref{app:tab:requirements}, there exists $\langle \tty{1},\tty{2},\tty{12} \rangle \in Y$ such that $\tty{2} \not\in `0`^+$.
Hence $(\mathsf{strToInt}~\tty{2}) > 0$.
We have $\mathsf{intToStr}~((\mathsf{strToInt}~\tty{1})+(\mathsf{strToInt}~\tty{2})) \ne \tty{1}$.
Assume the opposite that $\BoolSat{g'}{Y}$.
By \autoref{app:def:plausible}, $\Eval{g'}{\tty{1}}{\tty{2}}{\tty{12}}$.
By \autoref{app:fig:semantics},
$\tty{12} = \tty{1}$.
We have the desired contradiction.

\item $g' = \mathbf{second}$.
The proof is similar to the proof of \autoref{app:proof:basicsatweaker:add:first}.

\item $g' = \mathbf{front}~d~b$.
\label{app:proof:basicsatweaker:add:front}
We show that $\BoolSat{g'}{Y}$ never holds in this case.
Assume the opposite that $\BoolSat{g'}{Y}$.
By \autoref{app:thm:satlegal} and \autoref{app:def:dsllegal}, $\tty{1},\tty{2} \in \SetLegal{g'} = \{ d~{+}{+}~y \mid y \in \SetLegal{b} \}$.
By \autoref{fig:dsl}, $d \in \mathsf{Delim}$ and $b \in \DSLBasic$.
Thus, $d \not\in [`0`-`9`]$ and $\SetLegal{g} \cap \SetLegal{g'} = \emptyset$.
We have the desired contradiction.

\item $g' = \mathbf{back}~d~b$.
The proof is similar to the proof of \autoref{app:proof:basicsatweaker:add:front}.

\item $g' = \mathbf{fuse}~d~b$.
The proof is similar to the proof of \autoref{app:proof:basicsatweaker:add:front}.
\end{mycases}

\item $g = g_\text{c}$.
The proof performs case analysis of the values of $g'$.
By \autoref{app:tab:requirements}, there exists $\langle \tty{1},\tty{2},\tty{12} \rangle \in Y$.
By \autoref{app:def:plausible}, $\Eval{g}{\tty{1}}{\tty{2}}{\tty{12}}$.
By \autoref{app:fig:semantics},
$\tty{12} = \tty{1}~{+}{+}~\tty{2}$.

\begin{mycases}
\item $g' = \mathbf{add}$.
The proof is similar to the proof of \autoref{app:proof:basicsatweaker:add:concat}.

\item $g' = \mathbf{concat} = g$.
By \autoref{app:def:equivcap}, $\BoolEquivCap{g'}{g}$.

\item $g' = \mathbf{first}$.
\label{app:proof:basicsatweaker:concat:first}
We show that $\BoolSat{g'}{Y}$ never holds in this case.
By \autoref{app:tab:requirements}, there exists $\langle \tty{1},\tty{2},\tty{12} \rangle \in Y$ such that $\tty{2} \ne \mathsf{nil}$.
Hence $\tty{1}~{+}{+}~\tty{2} \ne \tty{1}$.
Assume the opposite that $\BoolSat{g'}{Y}$.
By \autoref{app:def:plausible}, $\Eval{g'}{\tty{1}}{\tty{2}}{\tty{12}}$.
By \autoref{app:fig:semantics},
$\tty{12} = \tty{1}$.
We have the desired contradiction.

\item $g' = \mathbf{second}$.
The proof is similar to the proof of \autoref{app:proof:basicsatweaker:concat:first}.

\item $g' = \mathbf{front}~d~b$.
\label{app:proof:basicsatweaker:concat:front}
We show that $\BoolSat{g'}{Y}$ never holds in this case.
Assume the opposite that $\BoolSat{g'}{Y}$.
By \autoref{app:thm:satlegal} and \autoref{app:def:dsllegal}, $\tty{1},\tty{2} \in \SetLegal{g'} = \{ d~{+}{+}~y \mid y \in \SetLegal{b} \}$.
Let $\tty{1} = d~{+}{+}~y_1'$, $\tty{2} = d~{+}{+}~y_2'$ where $y_1',y_2' \in \SetLegal{b}$.
By \autoref{fig:dsl}, $d \in \mathsf{Delim}$ and $b \in \DSLBasic$.
By \autoref{app:def:plausible}, $\Eval{g'}{\tty{1}}{\tty{2}}{\tty{12}}$.
By \autoref{app:fig:semantics}, there exists $v \in \mathsf{String}$ such that $\tty{12} = d~{+}{+}~v$ and $\Eval{b}{y_1'}{y_2'}{v}$.
Since $\tty{12} = \tty{1}~{+}{+}~\tty{2} = d~{+}{+}~y_1'~{+}{+}~d~{+}{+}~y_2'$, we have $v = y_1'~{+}{+}~d~{+}{+}~y_2'$.
Since $d \ne \mathsf{nil}$, by \autoref{app:thm:basicnoty1zy2}, $v \ne y_1'~{+}{+}~d~{+}{+}~y_2'$.
We have the desired contradiction.

\item $g' = \mathbf{back}~d~b$.
The proof is similar to the proof of \autoref{app:proof:basicsatweaker:concat:front}.

\item $g' = \mathbf{fuse}~d~b$.
\label{app:proof:basicsatweaker:concat:fuse}
We show that $\BoolSat{g'}{Y}$ never holds in this case.
Assume the opposite that $\BoolSat{g'}{Y}$.
By \autoref{app:thm:satlegal} and \autoref{app:def:dsllegal}, $\tty{1},\tty{2} \in \SetLegal{g'} = \{ y_1'~{+}{+}~d~{+}{+}\allowbreak~y_2'~{+}{+}\allowbreak~d~{+}{+}\allowbreak~\ldots~{+}{+}~d~{+}{+}~y_k' \mid y_1' \ne \mathsf{nil}, y_k' \ne \mathsf{nil}, \text{ and } y_i' \in \SetLegal{b} \text{ and } d \not\in y_i' \text{ for all }i=1,\ldots,k,~\allowbreak\text{where}~\allowbreak k \ge 2 \}$.
Let $k = \Count{d}{\tty{1}} + 1$.
By \autoref{app:fig:semantics}, $\Count{d}{\tty{1}} \ge 1$ and $k \ge 2$.
By \autoref{app:thm:fusedcount}, $\Count{d}{\tty{2}} = \Count{d}{\tty{12}} = \Count{d}{\tty{1}} = k - 1$.
Since $\tty{12} = \tty{1}~{+}{+}~\tty{2}$, $\Count{d}{\tty{12}} = 2 \cdot (k-1)$.
Since $k \ge 2$, we have $k-1 < 2 \cdot (k-1)$.
We have the desired contradiction.
\end{mycases}

\item $g = g_\text{f}$.
\label{app:proof:basicsatweaker:first}
The proof is by induction on the derivation of $g'$.
By \autoref{app:tab:requirements}, there exists $\langle \tty{1},\tty{2},\tty{12} \rangle \in Y$.
By \autoref{app:def:plausible}, $\Eval{g}{\tty{1}}{\tty{2}}{\tty{12}}$.
By \autoref{app:fig:semantics}, $\tty{12} = \tty{1}$.

\begin{mycases}
\item $g' = \mathbf{add}$.
The proof is similar to the proof of \autoref{app:proof:basicsatweaker:add:first}.

\item $g' = \mathbf{concat}$.
The proof is similar to the proof of \autoref{app:proof:basicsatweaker:concat:first}.

\item $g' = \mathbf{first} = g$.
By \autoref{app:def:equivcap}, $\BoolEquivCap{g'}{g}$.

\item $g' = \mathbf{second}$.
We show that $\BoolSat{g'}{Y}$ never holds in this case.
By \autoref{app:tab:requirements}, there exists $\langle \tty{1},\tty{2},\tty{12} \rangle \in Y$ such that $\tty{1} \ne \tty{2}$.
Assume the opposite that $\BoolSat{g'}{Y}$.
By \autoref{app:def:plausible}, $\Eval{g'}{\tty{1}}{\tty{2}}{\tty{12}}$.
By \autoref{app:fig:semantics}, $\tty{12} = \tty{2}$.
We have the desired contradiction.

\item $g' = \mathbf{front}~d~b$.
\label{app:proof:basicsatweaker:first:front}
By \autoref{app:thm:satlegal} and \autoref{app:def:dsllegal}, $\tty{1},\tty{2} \in \SetLegal{g'} = \{ d~{+}{+}~y \mid y \in \SetLegal{b} \}$.
Let $\tty{1} = d~{+}{+}~y_1'$, $\tty{2} = d~{+}{+}~y_2'$ where $y_1',y_2' \in \SetLegal{b}$.
By \autoref{fig:dsl}, $d \in \mathsf{Delim}$ and $b \in \DSLBasic$.
By \autoref{app:def:plausible}, $\Eval{g'}{\tty{1}}{\tty{2}}{\tty{12}}$.
By \autoref{app:fig:semantics}, there exists $v \in \mathsf{String}$ such that $\tty{12} = d~{+}{+}~v$ and $\Eval{b}{y_1'}{y_2'}{v}$.
Since $\tty{12} = \tty{1} = d~{+}{+}~y_1'$, we have $\Eval{b}{y_1'}{y_2'}{y_1'}$.

Let $Y' = \{ \langle y_1', y_2', y_{12}' \rangle \mid \langle (d~{+}{+}~y_1'), (d~{+}{+}~y_2'),\allowbreak (d~{+}{+}~y_{12}') \rangle \in Y \}$.
We have $\Eval{b}{y_1'}{y_2'}{y_{12}'}$ and $y_{12}' = y_1'$ for all $\langle y_1', y_2', y_{12}' \rangle \in Y'$.
By \autoref{app:def:plausible}, $\BoolSat{b}{Y'}$ and $\BoolSat{g}{Y'}$.

By \autoref{app:tab:requirements}, there exists $\langle \tty{1},\tty{2},\tty{12} \rangle \in Y$ such that $\tty{1} \ne \tty{2}$.
Let $\tty{1} = d~{+}{+}~y_1'$, $\tty{2} = d~{+}{+}~y_2'$, $\tty{12} = d~{+}{+}~y_{12}'$.
We have $y_1' \ne y_2'$ and $\langle y_1', y_2', y_{12}' \rangle \in Y'$.
By \autoref{app:tab:requirements}, there exists $\langle \tty{1},\tty{2},\tty{12} \rangle \in Y$ and $c \in \tty{2}$ such that $c \not \in \mathsf{Delim} \cup \{`0`\}$.
Let $\tty{2} = d~{+}{+}~y_2'$.
Since $d \in \mathsf{Delim}$, we have $c \in y_2'$.

By \autoref{app:tab:requirements}, $\BoolEnough{g}{Y'}$.
By the induction hypothesis, $\BoolEquivCap{b}{g}$.

For all $y_1',y_2' \in \SetLegal{b} \cap \SetLegal{g}$, by \autoref{app:def:equivcap}, $\Eval{b}{y_1'}{y_2'}{v}$ and $\Eval{g}{y_1'}{y_2'}{v}$ for some $v$.
By \autoref{app:fig:semantics}, $v = y_1'$.
By \autoref{app:def:dsllegal}, $\SetLegal{g} = \mathsf{String}$ and $\SetLegal{b} \cap \SetLegal{g} = \SetLegal{b}$.
Hence $\Eval{b}{y_1'}{y_2'}{y_1'}$ for all $y_1',y_2' \in \SetLegal{b}$.

For all $\tty{1},\tty{2} \in \SetLegal{g'}$, by \autoref{app:def:dsllegal}, $\tty{1} = (d~{+}{+}~y_1')$ and $\tty{2} = (d~{+}{+}~y_2')$ for some $y_1', y_2' \in \SetLegal{b}$.
By \autoref{app:fig:semantics}, $\Eval{g'}{(d~{+}{+}~y_1')}{(d~{+}{+}~y_2')}{d~{+}{+}~y_1'}$.
Hence $\Eval{g'}{\tty{1}}{\tty{2}}{\tty{1}}$ for all $\tty{1},\tty{2} \in \SetLegal{g'}$.
Since $\SetLegal{g} = \mathsf{String}$, $\SetLegal{g'} \cap \SetLegal{g} = \SetLegal{g'}$.
By \autoref{app:def:equivcap}, $\BoolEquivCap{g'}{g}$.

\item $g' = \mathbf{back}~d~b$.
The proof is similar to the proof of \autoref{app:proof:basicsatweaker:first:front}.

\item $g' = \mathbf{fuse}~d~b$.
\label{app:proof:basicsatweaker:first:fuse}
By \autoref{app:thm:satlegal} and \autoref{app:def:dsllegal}, $\tty{1},\tty{2} \in \SetLegal{g'} = \{ y_1'~{+}{+}~d~{+}{+}\allowbreak~y_2'~{+}{+}\allowbreak~d~{+}{+}\allowbreak~\ldots~{+}{+}~d~{+}{+}~y_k' \mid y_1' \ne \mathsf{nil}, y_k' \ne \mathsf{nil}, \text{ and } y_i' \in \SetLegal{b} \text{ and } d \not\in y_i' \text{ for all }i=1,\ldots,k, \text{ where } k \ge 2 \}$.
Let $\tty{1} = y^1_1~{+}{+}~d~{+}{+}\allowbreak~\ldots~{+}{+}\allowbreak~d~{+}{+}~y^1_k$ for some $y^1_i \in \SetLegal{b}$, where $d \not\in y^1_i$ for all $i=1,\ldots,k$.
By \autoref{app:thm:fusedcount},
$\tty{2} = y^2_1~{+}{+}~d~{+}{+}\allowbreak~\ldots~{+}{+}\allowbreak~d~{+}{+}~y^2_k$
and
$\tty{12} = y^{12}_1~{+}{+}~d~{+}{+}\allowbreak~\ldots~{+}{+}\allowbreak~d~{+}{+}~y^{12}_k$
for some $y^2_i \in \SetLegal{b}$, $y^{12}_i \in \mathsf{String}$, where $d \not\in y^2_i$ and $d \not\in y^{12}_i$ for all $i=1,\ldots,k$.
By \autoref{app:def:plausible}, $\Eval{g'}{\tty{1}}{\tty{2}}{\tty{12}}$.
By \autoref{app:fig:semantics}, $\Eval{b}{y^1_i}{y^2_i}{y^{12}_i}$ for all $i=1,\ldots,k$.
Since $\tty{12} = \tty{1}$, we have $y^{12}_i = y^1_i$ for all $i=1,\ldots,k$.

Let $Y' = \{ \langle y^1_i, y^2_i, y^{12}_i \rangle \mid \langle
(y^1_1~{+}{+}~d~{+}{+}~\ldots~{+}{+}\allowbreak~d~{+}{+}\allowbreak~y^1_k),
\allowbreak
(y^2_1~{+}{+}~d~{+}{+}~\ldots~{+}{+}~d~{+}{+}~y^2_k),
\allowbreak
(y^{12}_1~{+}{+}\allowbreak~d~{+}{+}\allowbreak~\ldots~{+}{+}~d~{+}{+}~y^{12}_k)
\rangle \in Y,
\text{ where } d \not\in y^1_j ~\allowbreak\text{and}~\allowbreak d \not\in y^2_j \text{ for all } j=1,\ldots,k,
\text{ and } i=\in\{1,\ldots,k\} \}$.
We have $\Eval{b}{y^1_i}{y^2_i}{y^{12}_i}$ and $y^{12}_i = y^1_i$ for all $\langle y^1_i, y^2_i, y^{12}_i \rangle \in Y'$.
By \autoref{app:def:plausible}, $\BoolSat{b}{Y'}$ and $\BoolSat{g}{Y'}$.

The rest of the proof is similar to the proof of \autoref{app:proof:basicsatweaker:first:front}.
\end{mycases}

\item $g = g_\text{s}$.
The proof is similar to the proof of \autoref{app:proof:basicsatweaker:first}.

\item $g = g_\text{ba}$.
\label{app:proof:basicsatweaker:backadd}
The proof is by induction on the derivation of $g'$.
By \autoref{app:tab:requirements}, there exists $\langle \tty{1},\tty{2},\tty{12} \rangle \in Y$.
By \autoref{app:thm:satlegal} and \autoref{app:def:dsllegal}, $\tty{1},\tty{2} \in \SetLegal{g} = [`0`-`9`]^+~d$.

\begin{mycases}
\item $g' = \mathbf{add}$.
The proof is similar to the proof of \autoref{app:proof:basicsatweaker:add:front}.

\item $g' = \mathbf{concat}$.
We show that $\BoolSat{g'}{Y}$ never holds in this case.
By \autoref{app:def:plausible}, $\Eval{g}{\tty{1}}{\tty{2}}{\tty{12}}$.
By \autoref{app:fig:semantics}, $\tty{12} \in \SetLegal{g}$.
Assume the opposite that $\BoolSat{g'}{Y}$.
By \autoref{app:def:plausible}, $\Eval{g'}{\tty{1}}{\tty{2}}{\tty{12}}$.
By \autoref{app:fig:semantics}, $\tty{12} = \tty{1}~{+}{+}~\tty{2}$.
Hence $\tty{12} \in [`0`-`9`]^+~d~[`0`-`9`]^+~d$.
By \autoref{fig:dsl}, $d \in \mathsf{Delim}$.
Thus, $d \not\in [`0`-`9`]$ and $\SetLegal{g} \cap [`0`-`9`]^+~d~[`0`-`9`]^+~d = \emptyset$.
We have the desired contradiction.

\item $g' = \mathbf{first}$.
\label{app:proof:basicsatweaker:backadd:front}
We show that $\BoolSat{g'}{Y}$ never holds in this case.
Let $\tty{1} = y_1'~{+}{+}~d$, $\tty{2} = y_2'~{+}{+}~d$ where $y_1',y_2' \in [`0`-`9`]^+$.
By \autoref{app:def:plausible}, $\Eval{g}{\tty{1}}{\tty{2}}{\tty{12}}$.
By \autoref{app:fig:semantics}, $\Eval{g_\text{a}}{y_1'}{y_2'}{y_1'}$.
Assume the opposite that $\BoolSat{g'}{Y}$.
By \autoref{app:def:plausible}, $\Eval{g'}{\tty{1}}{\tty{2}}{\tty{12}}$.
By \autoref{app:fig:semantics}, $\tty{12} = \tty{1} = y_1'~{+}{+}~d$.

Let $Y' = \{ \langle y_1', y_2', y_{12}' \rangle \mid \langle (y_1'~{+}{+}~d), (y_2'~{+}{+}~d),\allowbreak (y_{12}'~{+}{+}~d) \rangle \in Y \}$.
We have $\Eval{g_\text{a}}{y_1'}{y_2'}{y_{12}'}$ and $y_{12}' = y_1'$ for all $\langle y_1', y_2', y_{12}' \rangle \in Y'$.
By \autoref{app:def:plausible}, $\BoolSat{g_\text{a}}{Y'}$ and $\BoolSat{g'}{Y'}$.

By \autoref{app:tab:requirements}, $\BoolEnough{g_\text{a}}{Y'}$.
We show the desired contradiction similar to the proof of \autoref{app:proof:basicsatweaker:add:first}.

\item $g' = \mathbf{second}$.
The proof is similar to the proof of \autoref{app:proof:basicsatweaker:backadd:front}.

\item $g' = \mathbf{front}~d'~b$.
The proof is similar to the proof of \autoref{app:proof:basicsatweaker:add:front}.

\item $g' = \mathbf{back}~d'~b$ where $d' \ne d$.
The proof is similar to the proof of \autoref{app:proof:basicsatweaker:add:front}.

\item $g' = \mathbf{back}~d~b$.
\label{app:proof:basicsatweaker:backadd:back}
Let $\tty{1} = y_1'~{+}{+}~d$, $\tty{2} = y_2'~{+}{+}~d$ where $y_1',y_2' \in [`0`-`9`]^+$.
By \autoref{app:def:plausible}, $\Eval{g}{\tty{1}}{\tty{2}}{\tty{12}}$ and $\Eval{g'}{\tty{1}}{\tty{2}}{\tty{12}}$.
By \autoref{app:fig:semantics}, there exists $v \in \mathsf{String}$ such that $\tty{12} = v~{+}{+}~d$, $\Eval{g_a}{y_1'}{y_2'}{v}$, and $\Eval{b}{y_1'}{y_2'}{v}$.

Let $Y' = \{ \langle y_1', y_2', y_{12}' \rangle \mid \langle (y_1'~{+}{+}~d), (y_2'~{+}{+}~d),\allowbreak (y_{12}'~{+}{+}~d) \rangle \in Y \}$.
We have $\Eval{g_a}{y_1'}{y_2'}{y_{12}'}$ and $\Eval{b}{y_1'}{y_2'}{y_{12}'}$ for all $\langle y_1', y_2', y_{12}' \rangle \in Y'$.
By \autoref{app:def:plausible}, $\BoolSat{g_\text{a}}{Y'}$ and $\BoolSat{b}{Y'}$.

By \autoref{app:tab:requirements}, $\BoolEnough{g_\text{a}}{Y'}$.
By the induction hypothesis, $\BoolEquivCap{b}{g_\text{a}}$.

For all $y_1',y_2' \in \SetLegal{b} \cap \SetLegal{g_\text{a}}$, by \autoref{app:def:equivcap}, $\Eval{b}{y_1'}{y_2'}{v}$ and $\Eval{g_\text{a}}{y_1'}{y_2'}{v}$ for some $v$.
For all $\tty{1},\tty{2} \in \SetLegal{g'} \cap \SetLegal{g}$, by \autoref{app:fig:semantics}, we have $\Eval{g'}{\tty{1}}{\tty{2}}{\tty{12}}$ and $\Eval{g}{\tty{1}}{\tty{2}}{\tty{12}}$ for some $\tty{12}$.
By \autoref{app:def:equivcap}, $\BoolEquivCap{g'}{g}$.

\item $g' = \mathbf{fuse}~d'~b$ where $d' \ne d$.
The proof is similar to the proof of \autoref{app:proof:basicsatweaker:add:front}.

\item $g' = \mathbf{fuse}~d~b$.
The proof is similar to the proof of \autoref{app:proof:basicsatweaker:backadd:back}.
\end{mycases}

\item $g = g_\text{fa}$.
The proof is by induction on the derivation of $g'$.

\begin{mycases}
\item $g' = \mathbf{add}$.
The proof is similar to the proof of \autoref{app:proof:basicsatweaker:add:front}.

\item $g' = \mathbf{concat}$.
The proof is similar to the proof of \autoref{app:proof:basicsatweaker:concat:fuse}.

\item $g' = \mathbf{first}$.
The proof is similar to the proof of \autoref{app:proof:basicsatweaker:first:fuse}.

\item $g' = \mathbf{second}$.
The proof is similar to the proof of \autoref{app:proof:basicsatweaker:first:fuse}.

\item $g' = \mathbf{front}~d'~b$ where $d' \ne d$.
The proof is similar to the proof of \autoref{app:proof:basicsatweaker:add:front}.

\item $g' = \mathbf{front}~d~b$.
The proof is similar to the proof of \autoref{app:proof:basicsatweaker:backadd:back}.

\item $g' = \mathbf{back}~d'~b$ where $d' \ne d$.
The proof is similar to the proof of \autoref{app:proof:basicsatweaker:add:front}.

\item $g' = \mathbf{back}~d~b$.
The proof is similar to the proof of \autoref{app:proof:basicsatweaker:backadd:back}.

\item $g' = \mathbf{fuse}~d'~b$ where $d' \ne d$.
The proof is similar to the proof of \autoref{app:proof:basicsatweaker:add:front}.

\item $g' = \mathbf{fuse}~d~b$.
By \autoref{app:thm:satlegal} and \autoref{app:def:dsllegal}, $\tty{1},\tty{2} \in \SetLegal{g} \cap \SetLegal{g'} = \{ y_1'~{+}{+}~d~{+}{+}\allowbreak~y_2'~{+}{+}\allowbreak~d~{+}{+}\allowbreak~\ldots~{+}{+}~d~{+}{+}~y_k' \mid y_1' \ne \mathsf{nil}, y_k' \ne \mathsf{nil}, \text{ and } y_i' \in \SetLegal{g_\text{a}} \cap \SetLegal{b} \text{ and } d \not\in y_i' \text{ for all }i=1,\ldots,k, \text{ where } k \ge 2 \}$.
Let $\tty{1} = y^1_1~{+}{+}~d~{+}{+}~\ldots~{+}{+}\allowbreak~d~{+}{+}\allowbreak~y^1_k$ for some $y^1_i \in \SetLegal{b}$, where $d \not\in y^1_i$ for all $i=1,\ldots,k$.
By \autoref{app:thm:fusedcount},
$\tty{2} = y^2_1~{+}{+}~d~{+}{+}~\ldots~{+}{+}~d~{+}{+}~y^2_k$
and
$\tty{12} = y^{12}_1~{+}{+}~d~{+}{+}~\ldots~{+}{+}~d~{+}{+}~y^{12}_k$
for some $y^2_i \in \SetLegal{b}$, $y^{12}_i \in \mathsf{String}$, where $d \not\in y^1_i$ and $d \not\in y^2_i$ for all $i=1,\ldots,k$.
By \autoref{app:def:plausible}, $\Eval{g}{\tty{1}}{\tty{2}}{\tty{12}}$ and $\Eval{g'}{\tty{1}}{\tty{2}}{\tty{12}}$.
By \autoref{app:fig:semantics}, $\Eval{g_\text{a}}{y^1_i}{y^2_i}{y^{12}_i}$ and $\Eval{b}{y^1_i}{y^2_i}{y^{12}_i}$ for all $i=1,\ldots,k$.

Let $Y' = \{ \langle y^1_i, y^2_i, y^{12}_i \rangle \mid \langle
(y^1_1~{+}{+}~d~{+}{+}~\ldots~{+}{+}\allowbreak~d~{+}{+}\allowbreak~y^1_k),
\allowbreak
(y^2_1~{+}{+}~d~{+}{+}~\ldots~{+}{+}~d~{+}{+}~y^2_k),
\allowbreak
(y^{12}_1~{+}{+}\allowbreak~d~{+}{+}\allowbreak~\ldots~{+}{+}~d~{+}{+}~y^{12}_k)
\rangle \in Y,
\text{ where } d \not\in y^1_j \text{ and}~\allowbreak d \not\in y^2_j \text{ for all } j=1,\ldots,k,
\text{ and } i\in\{1,\ldots,k\} \}$.
We have $\Eval{g_\text{a}}{y^1_i}{y^2_i}{y^{12}_i}$ and $\Eval{b}{y^1_i}{y^2_i}{y^{12}_i}$ for all $\langle y^1_i, y^2_i, y^{12}_i \rangle \in Y'$.
By \autoref{app:def:plausible}, $\BoolSat{g_\text{a}}{Y'}$ and $\BoolSat{b}{Y'}$.

By \autoref{app:tab:requirements}, $\BoolEnough{g_\text{a}}{Y'}$.
By the induction hypothesis, $\BoolEquivCap{b}{g_\text{a}}$.

For all $y_1',y_2' \in \SetLegal{b} \cap \SetLegal{g_\text{a}}$, by \autoref{app:def:equivcap}, $\Eval{b}{y_1'}{y_2'}{v}$ and $\Eval{g_\text{a}}{y_1'}{y_2'}{v}$ for some $v$.
For all $\tty{1},\tty{2} \in \SetLegal{g'} \cap \SetLegal{g}$, by \autoref{app:fig:semantics}, we have $\Eval{g'}{\tty{1}}{\tty{2}}{\tty{12}}$ and $\Eval{g}{\tty{1}}{\tty{2}}{\tty{12}}$ for some $\tty{12}$.
By \autoref{app:def:equivcap}, $\BoolEquivCap{g'}{g}$.
\end{mycases}

\item $g = g_\text{bfa}$.
The proof is similar to the proof of \autoref{app:proof:basicsatweaker:backadd}.

\item $g = g_\text{fbfa}$.
The proof is similar to the proof of \autoref{app:proof:basicsatweaker:backadd}.

\item $g = g_\text{fc}$.
By \autoref{app:tab:requirements}, there exists $\langle \tty{1},\tty{2},\tty{12} \rangle \in Y$.
By \autoref{app:thm:satlegal} and \autoref{app:def:dsllegal}, $\tty{1},\tty{2} \in \SetLegal{g} = \{ d~{+}{+}~y \mid y \in \mathsf{String} \}$.
Let $\tty{1} = d~{+}{+}~y_1'$, $\tty{2} = d~{+}{+}~y_2'$ where $y_1',y_2' \in \mathsf{String}$.
By \autoref{fig:dsl}, $d \in \mathsf{Delim}$ and $b \in \DSLBasic$.
By \autoref{app:def:plausible}, $\Eval{g}{\tty{1}}{\tty{2}}{\tty{12}}$.
By \autoref{app:fig:semantics}, there exists $v \in \mathsf{String}$ such that $\tty{12} = d~{+}{+}~v$ and $\Eval{g_\text{c}}{y_1'}{y_2'}{v}$.
Hence $v = y_1'~{+}{+}~y_2'$ and $\tty{12} = d~{+}{+}~y_1'~{+}{+}~y_2'$.

\begin{mycases}
\item $g' = \mathbf{add}$.
The proof is similar to the proof of \autoref{app:proof:basicsatweaker:add:front}.

\item $g' = \mathbf{concat}$.
We show that $\BoolSat{g'}{Y}$ never holds in this case.
Assume the opposite that $\BoolSat{g'}{Y}$.
By \autoref{app:def:plausible}, $\Eval{g'}{\tty{1}}{\tty{2}}{\tty{12}}$.
By \autoref{app:fig:semantics}, $\tty{12} = \tty{1}~{+}{+}~\tty{2} = d~{+}{+}~y_1'~{+}{+}~d~{+}{+}~y_2'$.
Since $d \ne \mathsf{nil}$, we have the desired contradiction.

\item $g' = \mathbf{first}$.
\label{app:proof:basicsatweaker:frontconcat:first}
We show that $\BoolSat{g'}{Y}$ never holds in this case.
By \autoref{app:tab:requirements}, there exists $y_1', y_2', y_{12}' \in \mathsf{String}$ such that $\langle (d~{+}{+}~y_1'),(d~{+}{+}~y_2'),(d~{+}{+}~y_{12}') \rangle \in Y$ and $y_2' \ne \mathsf{nil}$.
By \autoref{app:def:plausible}, $\Eval{g}{(d~{+}{+}~y_1')}{(d~{+}{+}~y_2')}{(d~{+}{+}~y_{12}')}$.
By \autoref{app:fig:semantics}, $y_{12}' = y_1'~{+}{+}~y_2'$.
Assume the opposite that $\BoolSat{g'}{Y}$.
By \autoref{app:def:plausible}, $\Eval{g'}{(d~{+}{+}~y_1')}{(d~{+}{+}\allowbreak~y_2')}{(d~{+}{+}~y_{12}')}$.
By \autoref{app:fig:semantics}, $y_{12}' = y_1'$.
Since $y_2' \ne \mathsf{nil}$, we have the desired contradiction.

\item $g' = \mathbf{second}$.
The proof is similar to the proof of \autoref{app:proof:basicsatweaker:frontconcat:first}.

\item $g' = \mathbf{front}~d'~b$ where $d' \ne d$.
The proof is similar to the proof of \autoref{app:proof:basicsatweaker:add:front}.

\item $g' = \mathbf{front}~d~b$.
The proof is similar to the proof of \autoref{app:proof:basicsatweaker:backadd:back}.

\item $g' = \mathbf{back}~d'~b$ where $d' \ne d$.
We show that $\BoolSat{g'}{Y}$ never holds in this case.
By \autoref{app:thm:satlegal} and \autoref{app:def:dsllegal}, $\tty{1},\tty{2} \in \SetLegal{g} \cap \SetLegal{g'} = \{ d~{+}{+}~y \mid y \in \SetLegal{g_\text{c}} \} \cap \{ y~{+}{+}~d' \mid y \in \SetLegal{b} \}$.
For all $\langle \tty{1},\tty{2},\tty{12}\rangle \in Y$ there exists $y_1', y_2' \in \mathsf{String}$ such that $\tty{1} = d~{+}{+}~y_1'~{+}{+}~d'$, $\tty{2} = d~{+}{+}~y_2'~{+}{+}~d$', $(y_1'~{+}{+}~d') \in \SetLegal{g_\text{c}}$, $(y_2'~{+}{+}~d') \in \SetLegal{g_\text{c}}$, $(d~{+}{+}~y_1') \in \SetLegal{b}$, and $(d~{+}{+}~y_2') \in \SetLegal{b}$.

By \autoref{app:def:plausible}, $\Eval{g}{\tty{1}}{\tty{2}}{\tty{12}}$.
By \autoref{app:fig:semantics}, $\tty{12} = d~{+}{+}~y_1'~{+}{+}\allowbreak~d'~{+}{+}\allowbreak~y_2'~{+}{+}~d'$.
By \autoref{app:def:plausible}, $\Eval{g'}{\tty{1}}{\tty{2}}{\tty{12}}$.
By \autoref{app:fig:semantics}, $\Eval{b}{(d~{+}{+}~y_1')}{(d~{+}{+}~y_2')}{(d~{+}{+}~y_1'~{+}{+}~d'~{+}{+}~y_2')}$.

Since $d' \ne d$, we have
$\Count{d'}{(d~{+}{+}~y_1')} = \Count{d'}{y_1'}$,
$\Count{d'}{(d~{+}{+}~y_2')} = \Count{d'}{y_2'}$,
and
$\Count{d'}{(d~{+}{+}\allowbreak~y_1'~{+}{+}\allowbreak~d'~{+}{+}~y_2')} = \Count{d'}{y_1'} + 1 + \Count{d'}{y_2'} > \Count{d'}{y_1'} + \Count{d'}{y_2'}$,
By \autoref{fig:dsl}, $d' \in \mathsf{Delim}$ and $b \in \DSLBasic$.
By \autoref{app:thm:maxcount}, we have the desired contradiction.

\item $g' = \mathbf{back}~d~b$.
By \autoref{app:thm:satlegal} and \autoref{app:def:dsllegal}, $\tty{1},\tty{2} \in \SetLegal{g} \cap \SetLegal{g'} = \{ d~{+}{+}~y \mid y \in \SetLegal{g_\text{c}} \} \cap \{ y~{+}{+}~d \mid y \in \SetLegal{b} \}$.
For all $\langle \tty{1},\tty{2},\tty{12}\rangle \in Y$ there exists $y_1', y_2' \in \mathsf{String}$ such that $\tty{1} = d~{+}{+}~y_1'~{+}{+}~d$, $\tty{2} = d~{+}{+}~y_2'~{+}{+}~d$, $(y_1'~{+}{+}~d) \in \SetLegal{g_\text{c}}$, $(y_2'~{+}{+}~d) \in \SetLegal{g_\text{c}}$, $(d~{+}{+}~y_1') \in \SetLegal{b}$, and $(d~{+}{+}~y_2') \in \SetLegal{b}$.

By \autoref{app:def:plausible}, $\Eval{g}{\tty{1}}{\tty{2}}{\tty{12}}$.
By \autoref{app:fig:semantics}, $\tty{12} = d~{+}{+}~y_1'~{+}{+}\allowbreak~d~{+}{+}\allowbreak~y_2'~{+}{+}~d$.
By \autoref{app:def:plausible}, $\Eval{g'}{\tty{1}}{\tty{2}}{\tty{12}}$.
By \autoref{app:fig:semantics}, $\Eval{b}{(d~{+}{+}~y_1')}{(d~{+}{+}~y_2')}{(d~{+}{+}~y_1'~{+}{+}~d~{+}{+}~y_2')}$.

Let $Y' = \{ \langle y_1'', y_2'', y_{12}'' \rangle \mid \langle (y_1''~{+}{+}~d), (y_2''~{+}{+}~d),\allowbreak (y_{12}''~{+}{+}~d) \rangle \in Y \}$.
We have $\Eval{b}{y_1''}{y_2''}{y_{12}''}$ and $y_{12}'' = y_1''~{+}{+}~y_2''$ for all $\langle y_1'', y_2'', y_{12}'' \rangle \in Y'$.
By \autoref{app:def:plausible}, $\BoolSat{g_\text{c}}{Y'}$ and $\BoolSat{b}{Y'}$.

Since $Y \ne \emptyset$, we have $Y' \ne \emptyset$.
Also, for all $\langle y_1'', y_2'', y_{12}'' \rangle \in Y'$ there exists $y_1', y_2' \in \mathsf{String}$ such that $y_1'' = d~{+}{+}~y_1'$ and $y_2'' = d~{+}{+}~y_2'$.
Hence $y_1'' \ne \mathsf{nil}$ and $y_2'' \ne \mathsf{nil}$.

By \autoref{app:tab:requirements}, $\BoolEnough{g_\text{c}}{Y'}$.
By the induction hypothesis, $\BoolEquivCap{b}{g_\text{c}}$.

For all $y_1'',y_2'' \in \SetLegal{b} \cap \SetLegal{g_\text{c}}$, by \autoref{app:def:equivcap}, $\Eval{b}{y_1''}{y_2''}{v}$ and $\Eval{g_\text{c}}{y_1''}{y_2''}{v}$ for some $v$.
By \autoref{app:fig:semantics}, we have $\Eval{b}{y_1''}{y_2''}{(y_1''~{+}{+}~y_2'')}$ for all $y_1'', y_2'' \in \SetLegal{b} \cap \SetLegal{g_\text{c}}$.
By \autoref{app:def:dsllegal}, $\SetLegal{g_\text{c}} = \mathsf{String}$ and $\SetLegal{b} \cap \SetLegal{g_\text{c}} = \SetLegal{b}$.
Hence $\Eval{b}{y_1''}{y_2''}{(y_1''~{+}{+}~y_2'')}$ for all $y_1'', y_2'' \in \SetLegal{b}$.

For all $\tty{1},\tty{2} \in \SetLegal{g} \cap \SetLegal{g'}$, by \autoref{app:def:dsllegal}, $\tty{1} = d~{+}{+}\allowbreak~y_1'~{+}{+}\allowbreak~d$ and $\tty{2} = d~{+}{+}~y_2'~{+}{+}~d$ for some $y_1',y_2' \in \mathsf{String}$.
Also, $(d~{+}{+}~y_1') \in \SetLegal{b}$ and $(d~{+}{+}~y_2') \in \SetLegal{b}$.
Hence $\Eval{b}{(d~{+}{+}~y_1')}{(d~{+}{+}~y_2')}{(d~{+}{+}\allowbreak~y_1'~{+}{+}\allowbreak~d~{+}{+}~y_2')}$.
By \autoref{app:fig:semantics}, $\Eval{g'}{\tty{1}}{\tty{2}}{d~{+}{+}\allowbreak~y_1'~{+}{+}\allowbreak~d~{+}{+}~y_2'~{+}{+}~d}$.
Hence $\Eval{g'}{\tty{1}}{\tty{2}}{\tty{12}}$ and $\Eval{g'}{\tty{1}}{\tty{2}}{\tty{12}}$, where $\tty{12} = d~{+}{+}~y_1'~{+}{+}~d~{+}{+}\allowbreak~y_2'~{+}{+}~d$, for all $\tty{1},\tty{2} \in \SetLegal{g} \cap \SetLegal{g'}$.
By \autoref{app:def:equivcap}, $\BoolEquivCap{g'}{g}$.

\item $g' = \mathbf{fuse}~d'~b$ where $d' \ne d$.
We show that $\BoolSat{g'}{Y}$ never holds in this case.
Recall that $\tty{1} = d~{+}{+}~y_1'$ and $\tty{2} = d~{+}{+}~y_2'$ for some $y_1', y_2' \in \mathsf{String}$.
Since $d' \ne d$, we have $\Count{d'}{\tty{1}} = \Count{d'}{y_1'}$ and $\Count{d'}{\tty{2}} = \Count{d'}{y_2'}$.

By \autoref{app:def:plausible}, $\Eval{g}{\tty{1}}{\tty{2}}{\tty{12}}$.
By \autoref{app:fig:semantics}, there exists $v \in \mathsf{String}$ such that $\tty{12} = d~{+}{+}~v$ and $\Eval{g_\text{c}}{y_1'}{y_2'}{v}$.
We have $v = y_1'~{+}{+}~y_2'$ and $\tty{12} = d~{+}{+}~y_1'~{+}{+}~y_2'$.
Since $d' \ne d$, we have $\Count{d'}{\tty{12}} = \Count{d'}{y_1'} + \Count{d'}{y_2'} = \Count{d'}{\tty{1}} + \Count{d'}{\tty{2}}$.

Assume the opposite that $\BoolSat{g'}{Y}$.
By \autoref{app:def:plausible}, we have $\Eval{g'}{\tty{1}}{\tty{2}}{\tty{12}}$.
By \autoref{app:fig:semantics}, we have $\Count{d'}{\tty{1}} = \Count{d'}{\tty{2}} \ge 1$.
By \autoref{fig:dsl}, $d' \in \mathsf{Delim}$ and $b \in \DSLBasic$.
By \autoref{app:thm:fusedcount}, $\Count{d'}{\tty{12}} = \Count{d'}{\tty{1}} = \Count{d'}{\tty{2}} < \Count{d'}{\tty{1}} + \Count{d'}{\tty{2}}$.
We have the desired contradiction.

\item $g' = \mathbf{fuse}~d~b$.
We show that $\BoolSat{g'}{Y}$ never holds in this case.
Assume the opposite that $\BoolSat{g'}{Y}$.
By \autoref{app:thm:satlegal} and \autoref{app:def:dsllegal}, $\tty{1},\tty{2} \in \SetLegal{g'} = \{ y_1'~{+}{+}~d~{+}{+}\allowbreak~y_2'~{+}{+}\allowbreak~d~{+}{+}\allowbreak~\ldots~{+}{+}~d~{+}{+}~y_k' \mid y_1' \ne \mathsf{nil}, y_k' \ne \mathsf{nil}, \text{ and } y_i' \in \SetLegal{b} \text{ and } d \not\in y_i' \text{ for all }i=1,\ldots,k, ~\allowbreak\text{where}~\allowbreak k \ge 2 \}$.
Let $\tty{1} = y^1_1~{+}{+}~d~{+}{+}~\ldots~{+}{+}~d~{+}{+}\allowbreak~y^1_k$ for some $y^1_1,\ldots,y^1_k \in \SetLegal{b}$.
We have $y^1_1 \ne \mathsf{nil}$ and $d \not\in y^1_1$.
Recall that $\tty{1} = d~{+}{+}~y_1'$.
We have the desired contradiction.

\end{mycases}
\end{mycases}

%% file: proof/structsatequiv.tex
The proof performs case analysis of the values of $g, g'$.

\begin{mycases}
\item $g = g_\text{sf}$:
By \autoref{app:tab:requirements}, there exists $\langle \tty{1},\tty{2},\tty{12} \rangle \in Y$ such that $(\mathsf{splitLastLine}~\tty{1}) = (y_1', l)$ and $(\mathsf{splitFirstLine}~\tty{2}) = (l, y_2')$ for some $l$,
where $(\mathsf{firstChar}~(\mathsf{delPad}~l)) \not\in \mathsf{Delim} \cup \{`0`\}$
and $(\mathsf{lastChar}~l) \not\in \mathsf{Delim} \cup \{`0`\}$.

\begin{mycases}
\item $g' = \mathbf{stitch}~b$:
\label{app:proof:structsatequiv:stitchfirst:stitch}
By \autoref{app:def:plausible}, $\Eval{g}{\tty{1}}{\tty{2}}{\tty{12}}$ and $\Eval{g'}{\tty{1}}{\tty{2}}{\tty{12}}$.
By \autoref{app:fig:semantics}, $\tty{12} = y_1'~{+}{+}\allowbreak~`\textbackslash n`~{+}{+}~v~{+}{+}~`\textbackslash n`~{+}{+}~y_2'$ for some $v$, where $\Eval{g_\text{f}}{l}{l}{v}$ and $\Eval{b}{l}{l}{v}$.
By \autoref{app:fig:semantics}, $v = l$ and $\Eval{b}{l}{l}{l}$.

Let $c = \mathsf{lastChar}~l$.
We have $c \in l$ and $c \not\in \mathsf{Delim} \cup \{`0`\}$.
Let $Y' = \{ \langle l, l, l \rangle \}$.
By \autoref{app:thm:equivcapfirstsecond}, either $\BoolEquivCap{b}{g_\text{f}}$ or $\BoolEquivCap{b}{g_\text{s}}$.
Either case, by \autoref{app:def:equivcap}, we have $\Eval{b}{l'}{l'}{l'}$ for all $l' \in \SetLegal{b} \cap \SetLegal{g_\text{s}} = \SetLegal{b}$.
By \autoref{app:def:dsllegal}, $\SetLegal{g'} = \SetLegal{g'} \cap \SetLegal{g}$.
By \autoref{app:fig:semantics}, for all $\tty{1},\tty{2} \in \SetLegal{g'}$ we have $\Eval{g'}{\tty{1}}{\tty{2}}{v'}$ and $\Eval{g}{\tty{1}}{\tty{2}}{v}$ for some $v$.
By \autoref{app:def:equivcap}, $\BoolEquivCap{g'}{g}$.

\item $g' = \mathbf{stitch2}~d~b_1~b_2$:
By \autoref{app:def:plausible}, $\Eval{g}{\tty{1}}{\tty{2}}{\tty{12}}$ and $\Eval{g'}{\tty{1}}{\tty{2}}{\tty{12}}$.
By \autoref{app:fig:semantics}, $\tty{12} = y_1'~{+}{+}~`\textbackslash n`~{+}{+}~v~{+}{+}~`\textbackslash n`~{+}{+}~y_2'$ for some $v$, where $v = \mathsf{addPad}~(h~{+}{+}~d~{+}{+}~t)$.
Let $(h', t') = \mathsf{splitFirst}~d~(\mathsf{delPad}~l)$.
We have $d \not\in h'$.

By \autoref{app:fig:semantics}, we have $\Eval{g_\text{f}}{(h'~{+}{+}\allowbreak~d~{+}{+}\allowbreak~t')}{(h'~{+}{+}\allowbreak~d\allowbreak~{+}{+}~t')}{(h~{+}{+}~d~{+}{+}~t)}$, $\Eval{b_1}{h'}{h'}{h}$, and $\Eval{b_2}{t'}{t'}{t}$.
Hence $(h'~{+}{+}~d~{+}{+}~t' = h~{+}{+}~d~{+}{+}~t$.
By \autoref{fig:dsl}, $d \in \mathsf{Delim}$ and $b_1 \in \DSLBasic$.
By \autoref{app:thm:basicdnotin}, $d \not\in h$.
Hence $h' = h$ and $t' = t$.
We have $\Eval{b_1}{h}{h}{h}$ and $\Eval{b_2}{t}{t}{t}$.

Let $c_1 = (\mathsf{firstChar}~(\mathsf{delPad}~l))$ and $c_2 = \mathsf{lastChar}~l$.
We have $c_1 \in h$, $c_2 \in t$, and $c_1,c_2 \not\in \mathsf{Delim} \cup \{`0`\}$.

By \autoref{app:thm:satlegal}, for all $\langle \tty{1},\tty{2},\tty{12} \rangle \in Y$ we have $\tty{1},\tty{2} \in \SetLegal{g'}$.
By \autoref{app:tab:requirements},
then there exists $\langle \tty{1},\allowbreak\tty{2},\allowbreak\tty{12} \rangle \in Y$ such that
$h_1 \ne h_2$, where
$(\mathsf{splitLastLine}~\tty{1}) = (y_1', l_1)$,
$(\mathsf{splitFirstLine}\allowbreak~\tty{2}) = (l_2, y_2')$,
$(\mathsf{splitFirst}~d~(\mathsf{delPad}~l_1)) = (h_1, t)$,
and
$(\mathsf{splitFirst}~d~(\mathsf{delPad}~l_2)) = (h_2, t)$.
By \autoref{app:def:plausible},  \autoref{app:fig:semantics},  \autoref{fig:dsl}, and \autoref{app:thm:basicdnotin}, $\Eval{b_1}{h_1}{h_2}{h_1}$.

Let $Y_1' = \{ \langle h, h, h \rangle, \langle h_1, h_2, h_1 \rangle \}$.
By \autoref{app:tab:requirements}, we have $\BoolEnough{g_\text{f}}{Y_1'}$.
By \autoref{app:fig:semantics}, $\Eval{g_\text{f}}{h}{h}{h}$ and $\Eval{g_\text{f}}{h_1}{h_2}{h_1}$.
By \autoref{app:def:plausible}, $\BoolSat{g_\text{f}}{Y_1'}$ and $\BoolSat{b_1}{Y_1'}$.
By \autoref{fig:dsl}, $b_1 \in \DSLBasic$.
By \autoref{app:thm:equivcapbasic}, $\BoolEquivCap{b_1}{g_\text{f}}$.

Let $Y_2' = \{ \langle t, t, t \rangle \}$.
By \autoref{app:thm:equivcapfirstsecond}, either $\BoolEquivCap{b_2}{g_\text{f}}$ or $\BoolEquivCap{b_2}{g_\text{s}}$.
The rest of the proof is similar to the proof of \autoref{app:proof:structsatequiv:stitchfirst:stitch}.

\item $g' = \mathbf{offset}~d~b$:
\label{app:proof:structsatequiv:stitchfirst:offset}
We show that $\BoolSat{g'}{Y}$ never holds in this case.
Let $n_1, n_2, n_{12}$ be the numbers of lines in $\tty{1},\tty{2},\tty{12}$, respectively.
By \autoref{app:def:plausible}, $\Eval{g}{\tty{1}}{\tty{2}}{\tty{12}}$.
By \autoref{app:fig:semantics}, $n_{12} = n_1 + n_2 - 1$.
Assume the opposite that $\BoolSat{g'}{Y}$.
By \autoref{app:def:plausible}, $\Eval{g'}{\tty{1}}{\tty{2}}{\tty{12}}$.
Since $l \ne \mathsf{nil}$, by \autoref{app:fig:semantics}, $n_{12} = n_1 + n_2$.
We have the desired contradiction.
\end{mycases}

\item $g = g_\text{saf}$:
By \autoref{app:tab:requirements}, there exists $\langle \tty{1},\tty{2},\tty{12} \rangle \in Y$ such that $(\mathsf{splitLastLine}~\tty{1}) = (y_1', l)$ and $(\mathsf{splitFirstLine}~\tty{2}) = (l, y_2')$ for some $l$,
where $(\mathsf{firstChar}~(\mathsf{delPad}~l)) \not\in \mathsf{Delim} \cup \{`0`\}$
and $(\mathsf{lastChar}~l) \not\in \mathsf{Delim} \cup \{`0`\}$.
By \autoref{app:thm:satlegal} and \autoref{app:def:dsllegal}, $l = p~{+}{+}~h~{+}{+}~d~{+}{+}~t$ for some $p\in [`~`^+~\vert~`\textbackslash t`]$, $h \in \SetLegal{g_\text{a}} = [`0`-`9`]^+$, and $t \in \SetLegal{g_\text{f}} = \mathsf{String}$.

\begin{mycases}
\item $g' = \mathbf{stitch}~b$:
\label{app:proof:structsatequiv:stitch2addfirst:stitch}
We outline the proof below.
By \autoref{app:def:plausible}, $\Eval{g}{\tty{1}}{\tty{2}}{\tty{12}}$.
By \autoref{app:fig:semantics}, $\tty{12} = y_1'~{+}{+}~`\textbackslash n`~{+}{+}~v~{+}{+}~`\textbackslash n`~{+}{+}~y_2'$ for some $v$, where $v = p~{+}{+}~h_{12}~{+}{+}~d~{+}{+}~t_{12}$, $\Eval{g_\text{a}}{h}{h}{h_{12}}$, and $\Eval{g_\text{f}}{t}{t}{t_{12}}$.
We have $h_{12} = (\mathsf{intToStr}~((\mathsf{strToInt}~h) + (\mathsf{strToInt}~h)))$ and  $t_{12} = t$.
By \autoref{app:def:plausible}, $\Eval{g'}{\tty{1}}{\tty{2}}{\tty{12}}$.
By \autoref{app:fig:semantics}, $\Eval{b}{l}{l}{v}$.
Such $b \in \DSLBasic$ does not exist.
The proof is by induction on the derivation of $b$.

\item $g' = \mathbf{stitch2}~d'~b_1~b_2$, where $d' \ne d$:
We show that such $b_1$ does not exist.
The proof is similar to the proof of \autoref{app:proof:structsatequiv:stitch2addfirst:stitch}.

\item $g' = \mathbf{stitch2}~d~b_1~b_2$:
We outline the proof below.
By \autoref{app:thm:equivcapbasic}, $\BoolEquivCap{b_1}{g_\text{a}}$.
By \autoref{app:thm:equivcapfirstsecond}, either $\BoolEquivCap{b_2}{g_\text{f}}$ or $\BoolEquivCap{b_2}{g_\text{s}}$.
Either case, $\BoolEquivCap{g'}{g}$.

\item $g' = \mathbf{offset}~d'~b$:
The proof is similar to the proof of \autoref{app:proof:structsatequiv:stitchfirst:offset}.
\end{mycases}

\item $g = g_\text{oa}$:
By \autoref{app:tab:requirements},
there exists $\langle \tty{1},\tty{2},\tty{12} \rangle \in Y$ such that
$(\mathsf{splitLastLine}~\tty{1}) = (y_1', l_1)$,
$(\mathsf{splitFirstLine}~\tty{2}) = (l_2, y_2')$,
and
$(\mathsf{splitFirstLine}~y_2') = (l_2', y_2'')$,
where $(\mathsf{firstChar}\allowbreak~(\mathsf{delPad}~l_1)) \not\in \mathsf{Delim} \cup \{`0`\}$,
$l_2 \ne \mathsf{nil}$,
and
$l_2' \ne \mathsf{nil}$.

\begin{mycases}
\item $g' = \mathbf{stitch}~b$:
\label{app:proof:structsatequiv:offsetadd:stitch}
We outline the proof below.
We show that $\BoolSat{g'}{Y}$ never holds in this case.
Assume the opposite that $\BoolSat{g'}{Y}$.
By \autoref{app:def:plausible}, $\Eval{g'}{\tty{1}}{\tty{2}}{\tty{12}}$.
If $l_1 = l_2$, the proof is similar to the proof of \autoref{app:proof:structsatequiv:stitchfirst:offset}.
If $l_1 \ne l_2$, by \autoref{app:fig:semantics}, $\tty{12} = \tty{1}~{+}{+}~\tty{2}$.
By \autoref{app:def:plausible}, $\Eval{g}{\tty{1}}{\tty{2}}{\tty{12}}$.
By \autoref{app:fig:semantics}, $\tty{12} = \tty{1}~{+}{+}~v$ for some $v$.
By \autoref{app:tab:requirements}, $v \ne \tty{2}$.

\item $g' = \mathbf{stitch2}~d'~b_1~b_2$:
The proof is similar to the proof of \autoref{app:proof:structsatequiv:offsetadd:stitch}.

\item $g' = \mathbf{offset}~d'~b$, where $d' \ne d$:
We show that such $b$ does not exist.
The proof is similar to the proof of \autoref{app:proof:structsatequiv:stitch2addfirst:stitch}.

\item $g' = \mathbf{offset}~d~b$:
We outline the proof below.
Let $Y' = \{ \langle h_1, h_2, y_{12}' \rangle
\mid
\langle \tty{1},\tty{2},\tty{12} \rangle \in Y,
(\mathsf{splitLastLine}~\tty{1}) = (y_1',\allowbreak l_1),
(\mathsf{splitFirstLine}\allowbreak~\tty{2}) = (l_2, y_2'),
(\mathsf{splitFirst}~d~(\mathsf{delPad}~l_1)) = (h_1, t_1),
\text { and }
(\mathsf{splitFirst}~d~(\mathsf{delPad}~l_2)) = (h_2,\allowbreak t_2)
\}$.
By \autoref{app:tab:requirements}, $\BoolEnough{g_\text{a}}{Y'}$.
By \autoref{app:thm:equivcapbasic}, $\BoolEquivCap{b}{g_\text{a}}$.
\end{mycases}
\end{mycases}

%% file: tables/performance-all.tex
\footnotesize
\begin{longtable}{llll}
\caption{Pipeline commands that are parallelized with the synthesized combiners}
\label{tab:pipeline-counts}
\\
\toprule
\input{tables/parallel-count}
\bottomrule
\end{longtable}

\footnotesize
\begin{longtable}{llrrrr}
\caption{Performance results for all benchmark scripts, comparing new pipelines with original scripts}
\label{tab:performance-all-speedup}
\\
\toprule
\input{tables/parallel-all-speedup}
\bottomrule
\end{longtable}

\footnotesize
\begin{longtable}{llrrrrr}
\caption{Performance results for all benchmark scripts, where new pipelines are unoptimized}
\label{tab:performance-all-naive}
\\
\toprule
\input{tables/parallel-all-naive}
\bottomrule
\end{longtable}

\footnotesize
\begin{longtable}{llrrrrrr}
\caption{Performance results for all benchmark scripts, where new pipelines are optimized by eliminating intermediate combiners}
\label{tab:performance-all-defer}
\\
\toprule
\input{tables/parallel-all-defer}
\bottomrule
\end{longtable}

\normalsize

\begin{table*}[t]
\footnotesize
\caption{Performance results for benchmark scripts ($u_1 \ge \SI{3}{min}$)}
\label{tab:performance-3min}
\begin{tabular}{lp{4cm}p{2.5cm}p{1.4cm}rrrr}
\toprule

\input{tables/parallel-3min}

\bottomrule

\end{tabular}
\end{table*}

%% file: tables/parallel-count.tex
\textbf{Benchmark} & \textbf{Script Name} & \textbf{Parallelized} & \textbf{Eliminated} \\
\midrule
analytics-mts & 1.sh (vehicles per day) & $7/7$ $(7/7)$ & $3$ $(3)$ \\
analytics-mts & 2.sh (vehicle days on road) & $8/8$ $(8/8)$ & $3$ $(3)$ \\
analytics-mts & 3.sh (vehicle hours on road) & $8/8$ $(8/8)$ & $3$ $(3)$ \\
analytics-mts & 4.sh (hours monitored per day) & $7/7$ $(7/7)$ & $3$ $(3)$ \\
oneliners & bi-grams.sh & $3/5$ $(3/5)$ & $0$ $(0)$ \\
oneliners & diff.sh & $4/7$ $(0/1,2/2,2/2,0/1,0/1)$ & $2$ $(0,1,1,0,0)$ \\
oneliners & nfa-regex.sh & $2/2$ $(2/2)$ & $1$ $(1)$ \\
oneliners & set-diff.sh & $5/8$ $(0/1,3/3,2/2,0/1,0/1)$ & $3$ $(0,2,1,0,0)$ \\
oneliners & shortest-scripts.sh & $6/7$ $(6/7)$ & $5$ $(5)$ \\
oneliners & sort-sort.sh & $3/3$ $(3/3)$ & $1$ $(1)$ \\
oneliners & sort.sh & $1/1$ $(1/1)$ & $0$ $(0)$ \\
oneliners & spell.sh & $6/8$ $(6/8)$ & $3$ $(3)$ \\
oneliners & top-n.sh & $4/6$ $(4/6)$ & $1$ $(1)$ \\
oneliners & wf.sh & $4/5$ $(4/5)$ & $1$ $(1)$ \\
poets & 1_1.sh (count_words) & $4/6$ $(4/6)$ & $1$ $(1)$ \\
poets & 2_1.sh (merge_upper) & $5/7$ $(5/7)$ & $2$ $(2)$ \\
poets & 2_2.sh (count_vowel_seq) & $5/7$ $(5/7)$ & $2$ $(2)$ \\
poets & 3_1.sh (sort) & $5/7$ $(5/7)$ & $1$ $(1)$ \\
poets & 3_2.sh (sort_words_by_folding) & $5/7$ $(5/7)$ & $1$ $(1)$ \\
poets & 3_3.sh (sort_words_by_rhyming) & $7/9$ $(7/9)$ & $2$ $(2)$ \\
poets & 4_3.sh (bigrams) & $4/8$ $(2/4,0/1,2/3)$ & $1$ $(1,0,0)$ \\
poets & 4_3b.sh (count_trigrams) & $4/9$ $(2/4,0/1,0/1,2/3)$ & $1$ $(1,0,0,0)$ \\
poets & 6_1.sh (trigram_rec) & $8/14$ $(4/7,4/7)$ & $4$ $(2,2)$ \\
poets & 6_1_1.sh (uppercase_by_token) & $3/5$ $(3/5)$ & $1$ $(1)$ \\
poets & 6_1_2.sh (uppercase_by_type) & $4/6$ $(4/6)$ & $1$ $(1)$ \\
poets & 6_2.sh (4letter_words) & $7/11$ $(3/5,4/6)$ & $2$ $(1,1)$ \\
poets & 6_3.sh (words_no_vowels) & $5/7$ $(5/7)$ & $2$ $(2)$ \\
poets & 6_4.sh (1syllable_words) & $5/8$ $(5/8)$ & $2$ $(2)$ \\
poets & 6_5.sh (2syllable_words) & $5/8$ $(5/8)$ & $2$ $(2)$ \\
poets & 6_7.sh (verses_2om_3om_2instances) & $10/13$ $(3/4,3/4,4/5)$ & $7$ $(2,2,3)$ \\
poets & 7_2.sh (count_consonant_seq) & $5/7$ $(5/7)$ & $2$ $(2)$ \\
poets & 8.2_1.sh (vowel_sequencies_gr_1K) & $5/8$ $(5/8)$ & $1$ $(1)$ \\
poets & 8.2_2.sh (bigrams_appear_twice) & $4/9$ $(2/4,0/1,2/3,0/1)$ & $1$ $(1,0,0,0)$ \\
poets & 8.3_2.sh (find_anagrams) & $7/9$ $(2/4,1/1,1/1,3/3)$ & $1$ $(1,0,0,0)$ \\
poets & 8.3_3.sh (compare_exodus_genesis) & $6/10$ $(3/5,1/2,2/3)$ & $1$ $(1,0,0)$ \\
poets & 8_1.sh (sort_words_by_n_syllables) & $6/10$ $(3/5,2/2,1/3)$ & $2$ $(1,1,0)$ \\
unix50 & 1.sh (1.0: extract last name) & $1/1$ $(1/1)$ & $0$ $(0)$ \\
unix50 & 10.sh (4.4: histogram by piece) & $9/9$ $(9/9)$ & $6$ $(6)$ \\
unix50 & 11.sh (4.5: histogram by piece and pawn) & $9/9$ $(9/9)$ & $6$ $(6)$ \\
unix50 & 12.sh (4.6: piece used most) & $8/9$ $(8/9)$ & $5$ $(5)$ \\
unix50 & 13.sh (5.1: extract hellow world) & $3/3$ $(3/3)$ & $2$ $(2)$ \\
unix50 & 14.sh (6.1: order bodies) & $3/3$ $(3/3)$ & $1$ $(1)$ \\
unix50 & 15.sh (7.1: number of versions) & $3/3$ $(3/3)$ & $2$ $(2)$ \\
unix50 & 16.sh (7.2: most frequent machine) & $6/7$ $(6/7)$ & $1$ $(1)$ \\
unix50 & 17.sh (7.3: decades unix released) & $5/5$ $(5/5)$ & $2$ $(2)$ \\
unix50 & 18.sh (8.1: count unix birth-year) & $3/3$ $(3/3)$ & $2$ $(2)$ \\
unix50 & 19.sh (8.2: location office) & $4/4$ $(4/4)$ & $3$ $(3)$ \\
unix50 & 2.sh (1.1: extract names and sort) & $2/2$ $(2/2)$ & $1$ $(1)$ \\
unix50 & 20.sh (8.3: four most involved) & $4/4$ $(4/4)$ & $3$ $(3)$ \\
unix50 & 21.sh (8.4: longest words w/o hyphens) & $3/3$ $(3/3)$ & $1$ $(1)$ \\
unix50 & 23.sh (9.1: extract word PORT) & $6/6$ $(6/6)$ & $4$ $(4)$ \\
unix50 & 24.sh (9.2: extract word BELL) & $2/2$ $(2/2)$ & $1$ $(1)$ \\
unix50 & 25.sh (9.3: animal decorate) & $2/2$ $(2/2)$ & $1$ $(1)$ \\
unix50 & 26.sh (9.4: four corners) & $4/5$ $(4/5)$ & $2$ $(2)$ \\
unix50 & 28.sh (9.6: follow directions) & $6/10$ $(6/10)$ & $3$ $(3)$ \\
unix50 & 29.sh (9.7: four corners) & $2/4$ $(2/4)$ & $1$ $(1)$ \\
unix50 & 3.sh (1.2: extract names and sort) & $1/2$ $(1/2)$ & $0$ $(0)$ \\
unix50 & 30.sh (9.8: TELE-communications) & $4/8$ $(4/8)$ & $2$ $(2)$ \\
unix50 & 31.sh (9.9) & $4/9$ $(4/9)$ & $2$ $(2)$ \\
unix50 & 32.sh (10.1: count recipients) & $3/4$ $(3/4)$ & $2$ $(2)$ \\
unix50 & 33.sh (10.2: list recipients) & $2/3$ $(2/3)$ & $1$ $(1)$ \\
unix50 & 34.sh (10.3: extract username) & $7/7$ $(7/7)$ & $4$ $(4)$ \\
unix50 & 35.sh (11.1: year received medal) & $2/2$ $(2/2)$ & $1$ $(1)$ \\
unix50 & 36.sh (11.2: most repeated first name) & $7/8$ $(7/8)$ & $2$ $(2)$ \\
unix50 & 4.sh (1.3: sort top first names) & $4/4$ $(4/4)$ & $1$ $(1)$ \\
unix50 & 5.sh (2.1: all Unix utilities) & $2/2$ $(2/2)$ & $1$ $(1)$ \\
unix50 & 6.sh (3.1: first letter of last names) & $4/4$ $(4/4)$ & $2$ $(2)$ \\
unix50 & 7.sh (4.1: number of rounds) & $3/3$ $(3/3)$ & $2$ $(2)$ \\
unix50 & 8.sh (4.2: pieces captured) & $4/4$ $(4/4)$ & $3$ $(3)$ \\
unix50 & 9.sh (4.3: pieces captured with pawn) & $6/6$ $(6/6)$ & $5$ $(5)$ \\
\midrule
\textbf{Total} &  & $325/427$ & $144$ \\

%% file: tables/parallel-all-speedup.tex
\textbf{Benchmark} & \textbf{Script Name} & \textbf{$T_\text{orig}$} & \textbf{$u_{1}$} & \textbf{$u_{16}$} & \textbf{$T_{16}$} \\
\midrule
analytics-mts & 1.sh (vehicles per day) & $333$\,s $(1.1\times)$ & $376$\,s & $40$\,s $(9.4\times)$ & $29$\,s $(13.1\times)$ \\
analytics-mts & 2.sh (vehicle days on road) & $335$\,s $(1.1\times)$ & $379$\,s & $41$\,s $(9.3\times)$ & $28$\,s $(13.5\times)$ \\
analytics-mts & 3.sh (vehicle hours on road) & $408$\,s $(1.0\times)$ & $427$\,s & $51$\,s $(8.4\times)$ & $38$\,s $(11.3\times)$ \\
analytics-mts & 4.sh (hours monitored per day) & $99$\,s $(1.7\times)$ & $167$\,s & $28$\,s $(6.0\times)$ & $13$\,s $(12.8\times)$ \\
oneliners & bi-grams.sh & $668$\,s $(1.5\times)$ & $1007$\,s & $118$\,s $(8.6\times)$ & $115$\,s $(8.7\times)$ \\
oneliners & diff.sh & $325$\,s $(1.5\times)$ & $478$\,s & $98$\,s $(4.9\times)$ & $83$\,s $(5.8\times)$ \\
oneliners & nfa-regex.sh & $389$\,s $(1.0\times)$ & $391$\,s & $26$\,s $(14.9\times)$ & $27$\,s $(14.7\times)$ \\
oneliners & set-diff.sh & $879$\,s $(1.5\times)$ & $1308$\,s & $144$\,s $(9.1\times)$ & $128$\,s $(10.2\times)$ \\
oneliners & shortest-scripts.sh & $82$\,s $(1.3\times)$ & $110$\,s & $9$\,s $(12.4\times)$ & $7$\,s $(16.2\times)$ \\
oneliners & sort-sort.sh & $137$\,s $(1.2\times)$ & $167$\,s & $31$\,s $(5.4\times)$ & $28$\,s $(6.0\times)$ \\
oneliners & sort.sh & $273$\,s $(1.4\times)$ & $389$\,s & $39$\,s $(10.0\times)$ & $38$\,s $(10.3\times)$ \\
oneliners & spell.sh & $427$\,s $(1.7\times)$ & $736$\,s & $78$\,s $(9.5\times)$ & $61$\,s $(12.1\times)$ \\
oneliners & top-n.sh & $372$\,s $(1.7\times)$ & $622$\,s & $63$\,s $(9.9\times)$ & $50$\,s $(12.4\times)$ \\
oneliners & wf.sh & $1155$\,s $(1.8\times)$ & $2089$\,s & $196$\,s $(10.7\times)$ & $145$\,s $(14.4\times)$ \\
poets & 1_1.sh (count_words) & $360$\,s $(1.8\times)$ & $637$\,s & $84$\,s $(7.6\times)$ & $83$\,s $(7.6\times)$ \\
poets & 2_1.sh (merge_upper) & $307$\,s $(1.8\times)$ & $547$\,s & $79$\,s $(6.9\times)$ & $78$\,s $(7.0\times)$ \\
poets & 2_2.sh (count_vowel_seq) & $112$\,s $(1.2\times)$ & $140$\,s & $27$\,s $(5.2\times)$ & $24$\,s $(5.8\times)$ \\
poets & 3_1.sh (sort) & $391$\,s $(1.7\times)$ & $665$\,s & $89$\,s $(7.4\times)$ & $88$\,s $(7.6\times)$ \\
poets & 3_2.sh (sort_words_by_folding) & $402$\,s $(1.7\times)$ & $681$\,s & $94$\,s $(7.3\times)$ & $94$\,s $(7.2\times)$ \\
poets & 3_3.sh (sort_words_by_rhyming) & $415$\,s $(1.7\times)$ & $699$\,s & $100$\,s $(7.0\times)$ & $100$\,s $(7.0\times)$ \\
poets & 4_3.sh (bigrams) & $635$\,s $(1.4\times)$ & $915$\,s & $173$\,s $(5.3\times)$ & $173$\,s $(5.3\times)$ \\
poets & 4_3b.sh (count_trigrams) & $862$\,s $(1.2\times)$ & $1049$\,s & $275$\,s $(3.8\times)$ & $279$\,s $(3.8\times)$ \\
poets & 6_1.sh (trigram_rec) & $2$\,s $(1.9\times)$ & $5$\,s & $8$\,s $(0.6\times)$ & $2$\,s $(2.4\times)$ \\
poets & 6_1_1.sh (uppercase_by_token) & $38$\,s $(1.2\times)$ & $45$\,s & $14$\,s $(3.3\times)$ & $14$\,s $(3.2\times)$ \\
poets & 6_1_2.sh (uppercase_by_type) & $330$\,s $(1.9\times)$ & $635$\,s & $64$\,s $(10.0\times)$ & $24$\,s $(26.9\times)$ \\
poets & 6_2.sh (4letter_words) & $327$\,s $(2.0\times)$ & $647$\,s & $80$\,s $(8.1\times)$ & $34$\,s $(18.8\times)$ \\
poets & 6_3.sh (words_no_vowels) & $220$\,s $(1.1\times)$ & $235$\,s & $32$\,s $(7.4\times)$ & $31$\,s $(7.7\times)$ \\
poets & 6_4.sh (1syllable_words) & $433$\,s $(1.3\times)$ & $542$\,s & $57$\,s $(9.5\times)$ & $31$\,s $(17.4\times)$ \\
poets & 6_5.sh (2syllable_words) & $397$\,s $(1.1\times)$ & $443$\,s & $48$\,s $(9.2\times)$ & $40$\,s $(11.0\times)$ \\
poets & 6_7.sh (verses_2om_3om_2instances) & $4$\,s $(2.0\times)$ & $7$\,s & $11$\,s $(0.6\times)$ & $5$\,s $(1.5\times)$ \\
poets & 7_2.sh (count_consonant_seq) & $475$\,s $(1.4\times)$ & $678$\,s & $80$\,s $(8.5\times)$ & $48$\,s $(14.2\times)$ \\
poets & 8.2_1.sh (vowel_sequencies_gr_1K) & $417$\,s $(1.4\times)$ & $573$\,s & $73$\,s $(7.9\times)$ & $42$\,s $(13.7\times)$ \\
poets & 8.2_2.sh (bigrams_appear_twice) & $645$\,s $(1.4\times)$ & $921$\,s & $177$\,s $(5.2\times)$ & $91$\,s $(10.2\times)$ \\
poets & 8.3_2.sh (find_anagrams) & $237$\,s $(3.1\times)$ & $724$\,s & $102$\,s $(7.1\times)$ & $50$\,s $(14.5\times)$ \\
poets & 8.3_3.sh (compare_exodus_genesis) & $334$\,s $(2.0\times)$ & $656$\,s & $74$\,s $(8.8\times)$ & $34$\,s $(19.3\times)$ \\
poets & 8_1.sh (sort_words_by_n_syllables) & $346$\,s $(1.9\times)$ & $653$\,s & $69$\,s $(9.5\times)$ & $26$\,s $(24.6\times)$ \\
unix50 & 1.sh (1.0: extract last name) & $12$\,s $(1.0\times)$ & $12$\,s & $3$\,s $(3.6\times)$ & $3$\,s $(3.5\times)$ \\
unix50 & 10.sh (4.4: histogram by piece) & $27$\,s $(1.8\times)$ & $48$\,s & $13$\,s $(3.7\times)$ & $6$\,s $(7.8\times)$ \\
unix50 & 11.sh (4.5: histogram by piece and pawn) & $25$\,s $(1.7\times)$ & $42$\,s & $12$\,s $(3.4\times)$ & $6$\,s $(6.8\times)$ \\
unix50 & 12.sh (4.6: piece used most) & $115$\,s $(1.3\times)$ & $149$\,s & $27$\,s $(5.5\times)$ & $18$\,s $(8.2\times)$ \\
unix50 & 13.sh (5.1: extract hellow world) & $4$\,s $(2.7\times)$ & $12$\,s & $7$\,s $(1.7\times)$ & $2$\,s $(5.1\times)$ \\
unix50 & 14.sh (6.1: order bodies) & $143$\,s $(1.3\times)$ & $185$\,s & $31$\,s $(6.0\times)$ & $25$\,s $(7.5\times)$ \\
unix50 & 15.sh (7.1: number of versions) & $5$\,s $(1.4\times)$ & $8$\,s & $6$\,s $(1.4\times)$ & $3$\,s $(2.5\times)$ \\
unix50 & 16.sh (7.2: most frequent machine) & $80$\,s $(1.2\times)$ & $93$\,s & $15$\,s $(6.4\times)$ & $13$\,s $(7.4\times)$ \\
unix50 & 17.sh (7.3: decades unix released) & $39$\,s $(1.1\times)$ & $43$\,s & $10$\,s $(4.4\times)$ & $8$\,s $(5.1\times)$ \\
unix50 & 18.sh (8.1: count unix birth-year) & $2$\,s $(1.7\times)$ & $3$\,s & $6$\,s $(0.5\times)$ & $3$\,s $(1.1\times)$ \\
unix50 & 19.sh (8.2: location office) & $2$\,s $(1.3\times)$ & $2$\,s & $2$\,s $(1.0\times)$ & $2$\,s $(1.0\times)$ \\
unix50 & 2.sh (1.1: extract names and sort) & $133$\,s $(1.3\times)$ & $171$\,s & $20$\,s $(8.4\times)$ & $17$\,s $(9.9\times)$ \\
unix50 & 20.sh (8.3: four most involved) & $0$\,s (NaN) & $5$\,s & $6$\,s $(0.8\times)$ & $3$\,s $(1.7\times)$ \\
unix50 & 21.sh (8.4: longest words w/o hyphens) & $428$\,s $(1.7\times)$ & $733$\,s & $64$\,s $(11.4\times)$ & $49$\,s $(14.9\times)$ \\
unix50 & 23.sh (9.1: extract word PORT) & $111$\,s $(1.8\times)$ & $202$\,s & $23$\,s $(8.8\times)$ & $10$\,s $(19.8\times)$ \\
unix50 & 24.sh (9.2: extract word BELL) & $4$\,s $(1.1\times)$ & $5$\,s & $2$\,s $(2.1\times)$ & $2$\,s $(2.4\times)$ \\
unix50 & 25.sh (9.3: animal decorate) & $5$\,s $(1.1\times)$ & $6$\,s & $3$\,s $(2.1\times)$ & $2$\,s $(2.3\times)$ \\
unix50 & 26.sh (9.4: four corners) & $11$\,s $(2.8\times)$ & $32$\,s & $18$\,s $(1.7\times)$ & $15$\,s $(2.1\times)$ \\
unix50 & 28.sh (9.6: follow directions) & $87$\,s $(2.2\times)$ & $188$\,s & $54$\,s $(3.5\times)$ & $49$\,s $(3.8\times)$ \\
unix50 & 29.sh (9.7: four corners) & $6$\,s $(3.0\times)$ & $19$\,s & $18$\,s $(1.0\times)$ & $15$\,s $(1.3\times)$ \\
unix50 & 3.sh (1.2: extract names and sort) & $0$\,s (NaN) & $0$\,s & $0$\,s $(0.7\times)$ & $0$\,s $(0.7\times)$ \\
unix50 & 30.sh (9.8: TELE-communications) & $100$\,s $(1.6\times)$ & $154$\,s & $66$\,s $(2.3\times)$ & $62$\,s $(2.5\times)$ \\
unix50 & 31.sh (9.9) & $88$\,s $(1.7\times)$ & $149$\,s & $73$\,s $(2.0\times)$ & $68$\,s $(2.2\times)$ \\
unix50 & 32.sh (10.1: count recipients) & $3$\,s $(1.8\times)$ & $6$\,s & $7$\,s $(0.9\times)$ & $6$\,s $(0.9\times)$ \\
unix50 & 33.sh (10.2: list recipients) & $3$\,s $(1.8\times)$ & $6$\,s & $6$\,s $(1.0\times)$ & $5$\,s $(1.1\times)$ \\
unix50 & 34.sh (10.3: extract username) & $0$\,s (NaN) & $2$\,s & $3$\,s $(0.7\times)$ & $3$\,s $(0.9\times)$ \\
unix50 & 35.sh (11.1: year received medal) & $1$\,s $(0.9\times)$ & $1$\,s & $2$\,s $(0.5\times)$ & $2$\,s $(0.6\times)$ \\
unix50 & 36.sh (11.2: most repeated first name) & $15$\,s $(1.3\times)$ & $19$\,s & $7$\,s $(2.8\times)$ & $6$\,s $(3.5\times)$ \\
unix50 & 4.sh (1.3: sort top first names) & $134$\,s $(1.1\times)$ & $154$\,s & $21$\,s $(7.4\times)$ & $19$\,s $(8.3\times)$ \\
unix50 & 5.sh (2.1: all Unix utilities) & $7$\,s $(1.1\times)$ & $8$\,s & $3$\,s $(2.5\times)$ & $2$\,s $(3.1\times)$ \\
unix50 & 6.sh (3.1: first letter of last names) & $10$\,s $(1.4\times)$ & $14$\,s & $5$\,s $(2.7\times)$ & $3$\,s $(5.2\times)$ \\
unix50 & 7.sh (4.1: number of rounds) & $15$\,s $(1.2\times)$ & $18$\,s & $9$\,s $(2.0\times)$ & $4$\,s $(4.7\times)$ \\
unix50 & 8.sh (4.2: pieces captured) & $6$\,s $(2.1\times)$ & $12$\,s & $8$\,s $(1.6\times)$ & $3$\,s $(4.1\times)$ \\
unix50 & 9.sh (4.3: pieces captured with pawn) & $14$\,s $(2.0\times)$ & $28$\,s & $9$\,s $(2.9\times)$ & $4$\,s $(7.3\times)$ \\
\midrule
\textbf{Max} &  & $1155$\,s ($3.1\times$) & $2089$\,s & $275$\,s ($14.9\times$) & $279$\,s ($26.9\times$) \\
\textbf{Min} &  & $0$\,s ($0.9\times$) & $0$\,s & $0$\,s ($0.5\times$) & $0$\,s ($0.6\times$) \\
\textbf{Mean} &  & $217$\,s ($1.6\times$) & $332$\,s & $48$\,s ($5.5\times$) & $37$\,s ($8.0\times$) \\
\textbf{Median} &  & $114$\,s ($1.5\times$) & $167$\,s & $28$\,s ($5.3\times$) & $24$\,s ($7.1\times$) \\

%% file: tables/parallel-all-naive.tex
\textbf{Benchmark} & \textbf{Script Name} & \textbf{$u_{1}$} & \textbf{$u_{2}$} & \textbf{$u_{4}$} & \textbf{$u_{8}$} & \textbf{$u_{16}$} \\
\midrule
analytics-mts & 1.sh (vehicles per day) & $376$\,s & $200$\,s $(1.9\times)$ & $107$\,s $(3.5\times)$ & $62$\,s $(6.1\times)$ & $40$\,s $(9.4\times)$ \\
analytics-mts & 2.sh (vehicle days on road) & $379$\,s & $199$\,s $(1.9\times)$ & $107$\,s $(3.6\times)$ & $62$\,s $(6.1\times)$ & $41$\,s $(9.3\times)$ \\
analytics-mts & 3.sh (vehicle hours on road) & $427$\,s & $232$\,s $(1.8\times)$ & $126$\,s $(3.4\times)$ & $74$\,s $(5.8\times)$ & $51$\,s $(8.4\times)$ \\
analytics-mts & 4.sh (hours monitored per day) & $167$\,s & $94$\,s $(1.8\times)$ & $54$\,s $(3.1\times)$ & $35$\,s $(4.7\times)$ & $28$\,s $(6.0\times)$ \\
oneliners & bi-grams.sh & $1007$\,s & $539$\,s $(1.9\times)$ & $286$\,s $(3.5\times)$ & $166$\,s $(6.1\times)$ & $118$\,s $(8.6\times)$ \\
oneliners & diff.sh & $478$\,s & $276$\,s $(1.7\times)$ & $167$\,s $(2.9\times)$ & $120$\,s $(4.0\times)$ & $98$\,s $(4.9\times)$ \\
oneliners & nfa-regex.sh & $391$\,s & $197$\,s $(2.0\times)$ & $99$\,s $(3.9\times)$ & $51$\,s $(7.7\times)$ & $26$\,s $(14.9\times)$ \\
oneliners & set-diff.sh & $1308$\,s & $717$\,s $(1.8\times)$ & $376$\,s $(3.5\times)$ & $220$\,s $(6.0\times)$ & $144$\,s $(9.1\times)$ \\
oneliners & shortest-scripts.sh & $110$\,s & $57$\,s $(1.9\times)$ & $29$\,s $(3.8\times)$ & $16$\,s $(7.0\times)$ & $9$\,s $(12.4\times)$ \\
oneliners & sort-sort.sh & $167$\,s & $101$\,s $(1.7\times)$ & $60$\,s $(2.8\times)$ & $40$\,s $(4.2\times)$ & $31$\,s $(5.4\times)$ \\
oneliners & sort.sh & $389$\,s & $207$\,s $(1.9\times)$ & $106$\,s $(3.7\times)$ & $59$\,s $(6.6\times)$ & $39$\,s $(10.0\times)$ \\
oneliners & spell.sh & $736$\,s & $386$\,s $(1.9\times)$ & $208$\,s $(3.5\times)$ & $115$\,s $(6.4\times)$ & $78$\,s $(9.5\times)$ \\
oneliners & top-n.sh & $622$\,s & $328$\,s $(1.9\times)$ & $169$\,s $(3.7\times)$ & $99$\,s $(6.3\times)$ & $63$\,s $(9.9\times)$ \\
oneliners & wf.sh & $2089$\,s & $1065$\,s $(2.0\times)$ & $545$\,s $(3.8\times)$ & $298$\,s $(7.0\times)$ & $196$\,s $(10.7\times)$ \\
poets & 1_1.sh (count_words) & $637$\,s & $443$\,s $(1.4\times)$ & $224$\,s $(2.8\times)$ & $123$\,s $(5.2\times)$ & $84$\,s $(7.6\times)$ \\
poets & 2_1.sh (merge_upper) & $547$\,s & $380$\,s $(1.4\times)$ & $195$\,s $(2.8\times)$ & $114$\,s $(4.8\times)$ & $79$\,s $(6.9\times)$ \\
poets & 2_2.sh (count_vowel_seq) & $140$\,s & $86$\,s $(1.6\times)$ & $52$\,s $(2.7\times)$ & $35$\,s $(4.0\times)$ & $27$\,s $(5.2\times)$ \\
poets & 3_1.sh (sort) & $665$\,s & $455$\,s $(1.5\times)$ & $232$\,s $(2.9\times)$ & $129$\,s $(5.1\times)$ & $89$\,s $(7.4\times)$ \\
poets & 3_2.sh (sort_words_by_folding) & $681$\,s & $467$\,s $(1.5\times)$ & $240$\,s $(2.8\times)$ & $136$\,s $(5.0\times)$ & $94$\,s $(7.3\times)$ \\
poets & 3_3.sh (sort_words_by_rhyming) & $699$\,s & $478$\,s $(1.5\times)$ & $246$\,s $(2.8\times)$ & $140$\,s $(5.0\times)$ & $100$\,s $(7.0\times)$ \\
poets & 4_3.sh (bigrams) & $915$\,s & $635$\,s $(1.4\times)$ & $346$\,s $(2.6\times)$ & $215$\,s $(4.2\times)$ & $173$\,s $(5.3\times)$ \\
poets & 4_3b.sh (count_trigrams) & $1049$\,s & $734$\,s $(1.4\times)$ & $430$\,s $(2.4\times)$ & $311$\,s $(3.4\times)$ & $275$\,s $(3.8\times)$ \\
poets & 6_1.sh (trigram_rec) & $5$\,s & $7$\,s $(0.6\times)$ & $7$\,s $(0.7\times)$ & $6$\,s $(0.8\times)$ & $8$\,s $(0.6\times)$ \\
poets & 6_1_1.sh (uppercase_by_token) & $45$\,s & $33$\,s $(1.3\times)$ & $22$\,s $(2.1\times)$ & $16$\,s $(2.7\times)$ & $14$\,s $(3.3\times)$ \\
poets & 6_1_2.sh (uppercase_by_type) & $635$\,s & $387$\,s $(1.6\times)$ & $188$\,s $(3.4\times)$ & $99$\,s $(6.4\times)$ & $64$\,s $(10.0\times)$ \\
poets & 6_2.sh (4letter_words) & $647$\,s & $399$\,s $(1.6\times)$ & $199$\,s $(3.2\times)$ & $108$\,s $(6.0\times)$ & $80$\,s $(8.1\times)$ \\
poets & 6_3.sh (words_no_vowels) & $235$\,s & $156$\,s $(1.5\times)$ & $83$\,s $(2.8\times)$ & $48$\,s $(4.9\times)$ & $32$\,s $(7.4\times)$ \\
poets & 6_4.sh (1syllable_words) & $542$\,s & $318$\,s $(1.7\times)$ & $164$\,s $(3.3\times)$ & $91$\,s $(6.0\times)$ & $57$\,s $(9.5\times)$ \\
poets & 6_5.sh (2syllable_words) & $443$\,s & $282$\,s $(1.6\times)$ & $143$\,s $(3.1\times)$ & $78$\,s $(5.7\times)$ & $48$\,s $(9.2\times)$ \\
poets & 6_7.sh (verses_2om_3om_2instances) & $7$\,s & $11$\,s $(0.7\times)$ & $10$\,s $(0.7\times)$ & $10$\,s $(0.7\times)$ & $11$\,s $(0.6\times)$ \\
poets & 7_2.sh (count_consonant_seq) & $678$\,s & $370$\,s $(1.8\times)$ & $198$\,s $(3.4\times)$ & $119$\,s $(5.7\times)$ & $80$\,s $(8.5\times)$ \\
poets & 8.2_1.sh (vowel_sequencies_gr_1K) & $573$\,s & $348$\,s $(1.6\times)$ & $186$\,s $(3.1\times)$ & $110$\,s $(5.2\times)$ & $73$\,s $(7.9\times)$ \\
poets & 8.2_2.sh (bigrams_appear_twice) & $921$\,s & $642$\,s $(1.4\times)$ & $351$\,s $(2.6\times)$ & $222$\,s $(4.2\times)$ & $177$\,s $(5.2\times)$ \\
poets & 8.3_2.sh (find_anagrams) & $724$\,s & $440$\,s $(1.6\times)$ & $227$\,s $(3.2\times)$ & $133$\,s $(5.4\times)$ & $102$\,s $(7.1\times)$ \\
poets & 8.3_3.sh (compare_exodus_genesis) & $656$\,s & $401$\,s $(1.6\times)$ & $197$\,s $(3.3\times)$ & $108$\,s $(6.1\times)$ & $74$\,s $(8.8\times)$ \\
poets & 8_1.sh (sort_words_by_n_syllables) & $653$\,s & $403$\,s $(1.6\times)$ & $196$\,s $(3.3\times)$ & $104$\,s $(6.3\times)$ & $69$\,s $(9.5\times)$ \\
unix50 & 1.sh (1.0: extract last name) & $12$\,s & $8$\,s $(1.5\times)$ & $5$\,s $(2.4\times)$ & $4$\,s $(3.2\times)$ & $3$\,s $(3.6\times)$ \\
unix50 & 10.sh (4.4: histogram by piece) & $48$\,s & $32$\,s $(1.5\times)$ & $19$\,s $(2.5\times)$ & $14$\,s $(3.4\times)$ & $13$\,s $(3.7\times)$ \\
unix50 & 11.sh (4.5: histogram by piece and pawn) & $42$\,s & $28$\,s $(1.5\times)$ & $18$\,s $(2.4\times)$ & $13$\,s $(3.1\times)$ & $12$\,s $(3.4\times)$ \\
unix50 & 12.sh (4.6: piece used most) & $149$\,s & $91$\,s $(1.6\times)$ & $53$\,s $(2.8\times)$ & $35$\,s $(4.3\times)$ & $27$\,s $(5.5\times)$ \\
unix50 & 13.sh (5.1: extract hellow world) & $12$\,s & $10$\,s $(1.2\times)$ & $7$\,s $(1.7\times)$ & $6$\,s $(2.0\times)$ & $7$\,s $(1.7\times)$ \\
unix50 & 14.sh (6.1: order bodies) & $185$\,s & $106$\,s $(1.7\times)$ & $61$\,s $(3.0\times)$ & $40$\,s $(4.6\times)$ & $31$\,s $(6.0\times)$ \\
unix50 & 15.sh (7.1: number of versions) & $8$\,s & $7$\,s $(1.2\times)$ & $5$\,s $(1.6\times)$ & $5$\,s $(1.6\times)$ & $6$\,s $(1.4\times)$ \\
unix50 & 16.sh (7.2: most frequent machine) & $93$\,s & $53$\,s $(1.7\times)$ & $30$\,s $(3.1\times)$ & $19$\,s $(4.8\times)$ & $15$\,s $(6.4\times)$ \\
unix50 & 17.sh (7.3: decades unix released) & $43$\,s & $26$\,s $(1.6\times)$ & $17$\,s $(2.6\times)$ & $12$\,s $(3.7\times)$ & $10$\,s $(4.4\times)$ \\
unix50 & 18.sh (8.1: count unix birth-year) & $3$\,s & $5$\,s $(0.6\times)$ & $4$\,s $(0.7\times)$ & $5$\,s $(0.6\times)$ & $6$\,s $(0.5\times)$ \\
unix50 & 19.sh (8.2: location office) & $2$\,s & $3$\,s $(0.9\times)$ & $2$\,s $(1.0\times)$ & $2$\,s $(1.0\times)$ & $2$\,s $(1.0\times)$ \\
unix50 & 2.sh (1.1: extract names and sort) & $171$\,s & $98$\,s $(1.7\times)$ & $51$\,s $(3.3\times)$ & $31$\,s $(5.6\times)$ & $20$\,s $(8.4\times)$ \\
unix50 & 20.sh (8.3: four most involved) & $5$\,s & $6$\,s $(0.8\times)$ & $5$\,s $(1.0\times)$ & $5$\,s $(1.0\times)$ & $6$\,s $(0.8\times)$ \\
unix50 & 21.sh (8.4: longest words w/o hyphens) & $733$\,s & $384$\,s $(1.9\times)$ & $192$\,s $(3.8\times)$ & $104$\,s $(7.0\times)$ & $64$\,s $(11.4\times)$ \\
unix50 & 23.sh (9.1: extract word PORT) & $202$\,s & $109$\,s $(1.9\times)$ & $61$\,s $(3.3\times)$ & $35$\,s $(5.7\times)$ & $23$\,s $(8.8\times)$ \\
unix50 & 24.sh (9.2: extract word BELL) & $5$\,s & $4$\,s $(1.3\times)$ & $3$\,s $(1.8\times)$ & $2$\,s $(2.0\times)$ & $2$\,s $(2.1\times)$ \\
unix50 & 25.sh (9.3: animal decorate) & $6$\,s & $4$\,s $(1.3\times)$ & $3$\,s $(1.7\times)$ & $3$\,s $(2.0\times)$ & $3$\,s $(2.1\times)$ \\
unix50 & 26.sh (9.4: four corners) & $32$\,s & $26$\,s $(1.2\times)$ & $21$\,s $(1.5\times)$ & $18$\,s $(1.7\times)$ & $18$\,s $(1.7\times)$ \\
unix50 & 28.sh (9.6: follow directions) & $188$\,s & $119$\,s $(1.6\times)$ & $82$\,s $(2.3\times)$ & $64$\,s $(2.9\times)$ & $54$\,s $(3.5\times)$ \\
unix50 & 29.sh (9.7: four corners) & $19$\,s & $20$\,s $(0.9\times)$ & $18$\,s $(1.1\times)$ & $18$\,s $(1.0\times)$ & $18$\,s $(1.0\times)$ \\
unix50 & 3.sh (1.2: extract names and sort) & $0$\,s & $0$\,s $(1.0\times)$ & $0$\,s $(0.7\times)$ & $0$\,s $(0.7\times)$ & $0$\,s $(0.7\times)$ \\
unix50 & 30.sh (9.8: TELE-communications) & $154$\,s & $119$\,s $(1.3\times)$ & $85$\,s $(1.8\times)$ & $72$\,s $(2.1\times)$ & $66$\,s $(2.3\times)$ \\
unix50 & 31.sh (9.9) & $149$\,s & $111$\,s $(1.3\times)$ & $89$\,s $(1.7\times)$ & $78$\,s $(1.9\times)$ & $73$\,s $(2.0\times)$ \\
unix50 & 32.sh (10.1: count recipients) & $6$\,s & $6$\,s $(0.9\times)$ & $6$\,s $(1.0\times)$ & $6$\,s $(1.0\times)$ & $7$\,s $(0.9\times)$ \\
unix50 & 33.sh (10.2: list recipients) & $6$\,s & $6$\,s $(1.0\times)$ & $5$\,s $(1.1\times)$ & $6$\,s $(1.1\times)$ & $6$\,s $(1.0\times)$ \\
unix50 & 34.sh (10.3: extract username) & $2$\,s & $3$\,s $(0.9\times)$ & $2$\,s $(1.0\times)$ & $3$\,s $(0.9\times)$ & $3$\,s $(0.7\times)$ \\
unix50 & 35.sh (11.1: year received medal) & $1$\,s & $2$\,s $(0.5\times)$ & $2$\,s $(0.6\times)$ & $2$\,s $(0.5\times)$ & $2$\,s $(0.5\times)$ \\
unix50 & 36.sh (11.2: most repeated first name) & $19$\,s & $12$\,s $(1.6\times)$ & $8$\,s $(2.4\times)$ & $6$\,s $(3.1\times)$ & $7$\,s $(2.8\times)$ \\
unix50 & 4.sh (1.3: sort top first names) & $154$\,s & $89$\,s $(1.7\times)$ & $49$\,s $(3.2\times)$ & $30$\,s $(5.1\times)$ & $21$\,s $(7.4\times)$ \\
unix50 & 5.sh (2.1: all Unix utilities) & $8$\,s & $5$\,s $(1.4\times)$ & $4$\,s $(2.1\times)$ & $3$\,s $(2.7\times)$ & $3$\,s $(2.5\times)$ \\
unix50 & 6.sh (3.1: first letter of last names) & $14$\,s & $10$\,s $(1.4\times)$ & $7$\,s $(2.1\times)$ & $5$\,s $(2.6\times)$ & $5$\,s $(2.7\times)$ \\
unix50 & 7.sh (4.1: number of rounds) & $18$\,s & $14$\,s $(1.3\times)$ & $10$\,s $(1.8\times)$ & $9$\,s $(2.1\times)$ & $9$\,s $(2.0\times)$ \\
unix50 & 8.sh (4.2: pieces captured) & $12$\,s & $11$\,s $(1.2\times)$ & $8$\,s $(1.6\times)$ & $7$\,s $(1.7\times)$ & $8$\,s $(1.6\times)$ \\
unix50 & 9.sh (4.3: pieces captured with pawn) & $28$\,s & $19$\,s $(1.5\times)$ & $12$\,s $(2.2\times)$ & $10$\,s $(2.9\times)$ & $9$\,s $(2.9\times)$ \\
\midrule
\textbf{Max} &  & $2089$\,s & $1065$\,s ($2.0\times$) & $545$\,s ($3.9\times$) & $311$\,s ($7.7\times$) & $275$\,s ($14.9\times$) \\
\textbf{Min} &  & $0$\,s & $0$\,s ($0.5\times$) & $0$\,s ($0.6\times$) & $0$\,s ($0.5\times$) & $0$\,s ($0.5\times$) \\
\textbf{Mean} &  & $332$\,s & $200$\,s ($1.5\times$) & $107$\,s ($2.5\times$) & $65$\,s ($4.0\times$) & $48$\,s ($5.5\times$) \\
\textbf{Median} &  & $167$\,s & $104$\,s ($1.5\times$) & $60$\,s ($2.8\times$) & $38$\,s ($4.2\times$) & $28$\,s ($5.3\times$) \\

%% file: tables/parallel-all-defer.tex
\textbf{Benchmark} & \textbf{Script Name} & \textbf{$u_{1}$} & \textbf{$T_{1}$} & \textbf{$T_{2}$} & \textbf{$T_{4}$} & \textbf{$T_{8}$} & \textbf{$T_{16}$} \\
\midrule
analytics-mts & 1.sh (vehicles per day) & $376$\,s & $330$\,s $(1.1\times)$ & $170$\,s $(2.2\times)$ & $89$\,s $(4.2\times)$ & $48$\,s $(7.9\times)$ & $29$\,s $(13.1\times)$ \\
analytics-mts & 2.sh (vehicle days on road) & $379$\,s & $331$\,s $(1.1\times)$ & $169$\,s $(2.2\times)$ & $88$\,s $(4.3\times)$ & $47$\,s $(8.0\times)$ & $28$\,s $(13.5\times)$ \\
analytics-mts & 3.sh (vehicle hours on road) & $427$\,s & $411$\,s $(1.0\times)$ & $213$\,s $(2.0\times)$ & $112$\,s $(3.8\times)$ & $61$\,s $(6.9\times)$ & $38$\,s $(11.3\times)$ \\
analytics-mts & 4.sh (hours monitored per day) & $167$\,s & $97$\,s $(1.7\times)$ & $53$\,s $(3.2\times)$ & $29$\,s $(5.8\times)$ & $18$\,s $(9.4\times)$ & $13$\,s $(12.8\times)$ \\
oneliners & bi-grams.sh & $1007$\,s & $1015$\,s $(1.0\times)$ & $535$\,s $(1.9\times)$ & $283$\,s $(3.6\times)$ & $168$\,s $(6.0\times)$ & $115$\,s $(8.7\times)$ \\
oneliners & diff.sh & $478$\,s & $332$\,s $(1.4\times)$ & $226$\,s $(2.1\times)$ & $137$\,s $(3.5\times)$ & $100$\,s $(4.8\times)$ & $83$\,s $(5.8\times)$ \\
oneliners & nfa-regex.sh & $391$\,s & $388$\,s $(1.0\times)$ & $196$\,s $(2.0\times)$ & $99$\,s $(3.9\times)$ & $51$\,s $(7.7\times)$ & $27$\,s $(14.7\times)$ \\
oneliners & set-diff.sh & $1308$\,s & $816$\,s $(1.6\times)$ & $495$\,s $(2.6\times)$ & $279$\,s $(4.7\times)$ & $175$\,s $(7.5\times)$ & $128$\,s $(10.2\times)$ \\
oneliners & shortest-scripts.sh & $110$\,s & $82$\,s $(1.3\times)$ & $42$\,s $(2.6\times)$ & $22$\,s $(5.0\times)$ & $12$\,s $(9.5\times)$ & $7$\,s $(16.2\times)$ \\
oneliners & sort-sort.sh & $167$\,s & $137$\,s $(1.2\times)$ & $85$\,s $(2.0\times)$ & $53$\,s $(3.2\times)$ & $35$\,s $(4.7\times)$ & $28$\,s $(6.0\times)$ \\
oneliners & sort.sh & $389$\,s & $391$\,s $(1.0\times)$ & $207$\,s $(1.9\times)$ & $106$\,s $(3.7\times)$ & $59$\,s $(6.6\times)$ & $38$\,s $(10.3\times)$ \\
oneliners & spell.sh & $736$\,s & $484$\,s $(1.5\times)$ & $282$\,s $(2.6\times)$ & $154$\,s $(4.8\times)$ & $90$\,s $(8.1\times)$ & $61$\,s $(12.1\times)$ \\
oneliners & top-n.sh & $622$\,s & $388$\,s $(1.6\times)$ & $228$\,s $(2.7\times)$ & $127$\,s $(4.9\times)$ & $75$\,s $(8.3\times)$ & $50$\,s $(12.4\times)$ \\
oneliners & wf.sh & $2089$\,s & $1196$\,s $(1.7\times)$ & $667$\,s $(3.1\times)$ & $368$\,s $(5.7\times)$ & $223$\,s $(9.4\times)$ & $145$\,s $(14.4\times)$ \\
poets & 1_1.sh (count_words) & $637$\,s & $637$\,s $(1.0\times)$ & $440$\,s $(1.4\times)$ & $223$\,s $(2.9\times)$ & $123$\,s $(5.2\times)$ & $83$\,s $(7.6\times)$ \\
poets & 2_1.sh (merge_upper) & $547$\,s & $543$\,s $(1.0\times)$ & $376$\,s $(1.5\times)$ & $193$\,s $(2.8\times)$ & $109$\,s $(5.0\times)$ & $78$\,s $(7.0\times)$ \\
poets & 2_2.sh (count_vowel_seq) & $140$\,s & $142$\,s $(1.0\times)$ & $83$\,s $(1.7\times)$ & $50$\,s $(2.8\times)$ & $40$\,s $(3.5\times)$ & $24$\,s $(5.8\times)$ \\
poets & 3_1.sh (sort) & $665$\,s & $670$\,s $(1.0\times)$ & $460$\,s $(1.4\times)$ & $232$\,s $(2.9\times)$ & $132$\,s $(5.0\times)$ & $88$\,s $(7.6\times)$ \\
poets & 3_2.sh (sort_words_by_folding) & $681$\,s & $685$\,s $(1.0\times)$ & $472$\,s $(1.4\times)$ & $237$\,s $(2.9\times)$ & $134$\,s $(5.1\times)$ & $94$\,s $(7.2\times)$ \\
poets & 3_3.sh (sort_words_by_rhyming) & $699$\,s & $704$\,s $(1.0\times)$ & $478$\,s $(1.5\times)$ & $250$\,s $(2.8\times)$ & $140$\,s $(5.0\times)$ & $100$\,s $(7.0\times)$ \\
poets & 4_3.sh (bigrams) & $915$\,s & $909$\,s $(1.0\times)$ & $640$\,s $(1.4\times)$ & $343$\,s $(2.7\times)$ & $216$\,s $(4.2\times)$ & $173$\,s $(5.3\times)$ \\
poets & 4_3b.sh (count_trigrams) & $1049$\,s & $1056$\,s $(1.0\times)$ & $733$\,s $(1.4\times)$ & $430$\,s $(2.4\times)$ & $311$\,s $(3.4\times)$ & $279$\,s $(3.8\times)$ \\
poets & 6_1.sh (trigram_rec) & $5$\,s & $2$\,s $(1.9\times)$ & $2$\,s $(2.8\times)$ & $2$\,s $(2.4\times)$ & $1$\,s $(3.6\times)$ & $2$\,s $(2.4\times)$ \\
poets & 6_1_1.sh (uppercase_by_token) & $45$\,s & $40$\,s $(1.1\times)$ & $30$\,s $(1.5\times)$ & $21$\,s $(2.1\times)$ & $16$\,s $(2.8\times)$ & $14$\,s $(3.2\times)$ \\
poets & 6_1_2.sh (uppercase_by_type) & $635$\,s & $158$\,s $(4.0\times)$ & $101$\,s $(6.3\times)$ & $53$\,s $(12.1\times)$ & $31$\,s $(20.5\times)$ & $24$\,s $(26.9\times)$ \\
poets & 6_2.sh (4letter_words) & $647$\,s & $171$\,s $(3.8\times)$ & $114$\,s $(5.7\times)$ & $63$\,s $(10.2\times)$ & $41$\,s $(15.6\times)$ & $34$\,s $(18.8\times)$ \\
poets & 6_3.sh (words_no_vowels) & $235$\,s & $223$\,s $(1.1\times)$ & $148$\,s $(1.6\times)$ & $82$\,s $(2.9\times)$ & $46$\,s $(5.1\times)$ & $31$\,s $(7.7\times)$ \\
poets & 6_4.sh (1syllable_words) & $542$\,s & $203$\,s $(2.7\times)$ & $139$\,s $(3.9\times)$ & $76$\,s $(7.1\times)$ & $45$\,s $(11.9\times)$ & $31$\,s $(17.4\times)$ \\
poets & 6_5.sh (2syllable_words) & $443$\,s & $316$\,s $(1.4\times)$ & $205$\,s $(2.2\times)$ & $109$\,s $(4.1\times)$ & $61$\,s $(7.3\times)$ & $40$\,s $(11.0\times)$ \\
poets & 6_7.sh (verses_2om_3om_2instances) & $7$\,s & $4$\,s $(1.8\times)$ & $3$\,s $(2.8\times)$ & $3$\,s $(2.7\times)$ & $3$\,s $(2.4\times)$ & $5$\,s $(1.5\times)$ \\
poets & 7_2.sh (count_consonant_seq) & $678$\,s & $250$\,s $(2.7\times)$ & $152$\,s $(4.5\times)$ & $88$\,s $(7.7\times)$ & $59$\,s $(11.5\times)$ & $48$\,s $(14.2\times)$ \\
poets & 8.2_1.sh (vowel_sequencies_gr_1K) & $573$\,s & $155$\,s $(3.7\times)$ & $101$\,s $(5.7\times)$ & $66$\,s $(8.7\times)$ & $50$\,s $(11.4\times)$ & $42$\,s $(13.7\times)$ \\
poets & 8.2_2.sh (bigrams_appear_twice) & $921$\,s & $247$\,s $(3.7\times)$ & $182$\,s $(5.0\times)$ & $120$\,s $(7.7\times)$ & $96$\,s $(9.6\times)$ & $91$\,s $(10.2\times)$ \\
poets & 8.3_2.sh (find_anagrams) & $724$\,s & $208$\,s $(3.5\times)$ & $131$\,s $(5.5\times)$ & $76$\,s $(9.5\times)$ & $55$\,s $(13.3\times)$ & $50$\,s $(14.5\times)$ \\
poets & 8.3_3.sh (compare_exodus_genesis) & $656$\,s & $162$\,s $(4.1\times)$ & $106$\,s $(6.2\times)$ & $59$\,s $(11.1\times)$ & $39$\,s $(16.7\times)$ & $34$\,s $(19.3\times)$ \\
poets & 8_1.sh (sort_words_by_n_syllables) & $653$\,s & $169$\,s $(3.9\times)$ & $108$\,s $(6.0\times)$ & $58$\,s $(11.3\times)$ & $35$\,s $(18.4\times)$ & $26$\,s $(24.6\times)$ \\
unix50 & 1.sh (1.0: extract last name) & $12$\,s & $12$\,s $(1.0\times)$ & $8$\,s $(1.5\times)$ & $5$\,s $(2.4\times)$ & $4$\,s $(3.2\times)$ & $3$\,s $(3.5\times)$ \\
unix50 & 10.sh (4.4: histogram by piece) & $48$\,s & $27$\,s $(1.7\times)$ & $17$\,s $(2.9\times)$ & $10$\,s $(4.9\times)$ & $7$\,s $(6.7\times)$ & $6$\,s $(7.8\times)$ \\
unix50 & 11.sh (4.5: histogram by piece and pawn) & $42$\,s & $24$\,s $(1.7\times)$ & $16$\,s $(2.6\times)$ & $10$\,s $(4.3\times)$ & $7$\,s $(5.9\times)$ & $6$\,s $(6.8\times)$ \\
unix50 & 12.sh (4.6: piece used most) & $149$\,s & $112$\,s $(1.3\times)$ & $69$\,s $(2.2\times)$ & $38$\,s $(4.0\times)$ & $24$\,s $(6.2\times)$ & $18$\,s $(8.2\times)$ \\
unix50 & 13.sh (5.1: extract hellow world) & $12$\,s & $5$\,s $(2.6\times)$ & $4$\,s $(3.1\times)$ & $3$\,s $(4.1\times)$ & $3$\,s $(4.7\times)$ & $2$\,s $(5.1\times)$ \\
unix50 & 14.sh (6.1: order bodies) & $185$\,s & $154$\,s $(1.2\times)$ & $92$\,s $(2.0\times)$ & $51$\,s $(3.6\times)$ & $33$\,s $(5.6\times)$ & $25$\,s $(7.5\times)$ \\
unix50 & 15.sh (7.1: number of versions) & $8$\,s & $6$\,s $(1.4\times)$ & $4$\,s $(2.0\times)$ & $3$\,s $(2.9\times)$ & $3$\,s $(3.0\times)$ & $3$\,s $(2.5\times)$ \\
unix50 & 16.sh (7.2: most frequent machine) & $93$\,s & $85$\,s $(1.1\times)$ & $48$\,s $(2.0\times)$ & $27$\,s $(3.4\times)$ & $17$\,s $(5.4\times)$ & $13$\,s $(7.4\times)$ \\
unix50 & 17.sh (7.3: decades unix released) & $43$\,s & $41$\,s $(1.1\times)$ & $24$\,s $(1.8\times)$ & $15$\,s $(3.0\times)$ & $10$\,s $(4.2\times)$ & $8$\,s $(5.1\times)$ \\
unix50 & 18.sh (8.1: count unix birth-year) & $3$\,s & $2$\,s $(1.5\times)$ & $2$\,s $(1.4\times)$ & $2$\,s $(1.6\times)$ & $2$\,s $(1.4\times)$ & $3$\,s $(1.1\times)$ \\
unix50 & 19.sh (8.2: location office) & $2$\,s & $2$\,s $(1.1\times)$ & $2$\,s $(1.0\times)$ & $2$\,s $(1.2\times)$ & $2$\,s $(1.4\times)$ & $2$\,s $(1.0\times)$ \\
unix50 & 2.sh (1.1: extract names and sort) & $171$\,s & $131$\,s $(1.3\times)$ & $74$\,s $(2.3\times)$ & $41$\,s $(4.1\times)$ & $25$\,s $(6.8\times)$ & $17$\,s $(9.9\times)$ \\
unix50 & 20.sh (8.3: four most involved) & $5$\,s & $0$\,s (NaN) & $1$\,s $(3.9\times)$ & $1$\,s $(3.6\times)$ & $2$\,s $(2.5\times)$ & $3$\,s $(1.7\times)$ \\
unix50 & 21.sh (8.4: longest words w/o hyphens) & $733$\,s & $440$\,s $(1.7\times)$ & $257$\,s $(2.8\times)$ & $141$\,s $(5.2\times)$ & $78$\,s $(9.4\times)$ & $49$\,s $(14.9\times)$ \\
unix50 & 23.sh (9.1: extract word PORT) & $202$\,s & $116$\,s $(1.7\times)$ & $59$\,s $(3.4\times)$ & $31$\,s $(6.5\times)$ & $17$\,s $(11.7\times)$ & $10$\,s $(19.8\times)$ \\
unix50 & 24.sh (9.2: extract word BELL) & $5$\,s & $5$\,s $(1.1\times)$ & $4$\,s $(1.4\times)$ & $2$\,s $(2.1\times)$ & $2$\,s $(2.3\times)$ & $2$\,s $(2.4\times)$ \\
unix50 & 25.sh (9.3: animal decorate) & $6$\,s & $5$\,s $(1.1\times)$ & $4$\,s $(1.5\times)$ & $3$\,s $(2.0\times)$ & $2$\,s $(2.6\times)$ & $2$\,s $(2.3\times)$ \\
unix50 & 26.sh (9.4: four corners) & $32$\,s & $28$\,s $(1.1\times)$ & $22$\,s $(1.5\times)$ & $17$\,s $(1.9\times)$ & $16$\,s $(2.0\times)$ & $15$\,s $(2.1\times)$ \\
unix50 & 28.sh (9.6: follow directions) & $188$\,s & $185$\,s $(1.0\times)$ & $114$\,s $(1.7\times)$ & $78$\,s $(2.4\times)$ & $59$\,s $(3.2\times)$ & $49$\,s $(3.8\times)$ \\
unix50 & 29.sh (9.7: four corners) & $19$\,s & $16$\,s $(1.2\times)$ & $16$\,s $(1.2\times)$ & $15$\,s $(1.2\times)$ & $15$\,s $(1.3\times)$ & $15$\,s $(1.3\times)$ \\
unix50 & 3.sh (1.2: extract names and sort) & $0$\,s & $0$\,s $(0.9\times)$ & $0$\,s $(1.0\times)$ & $0$\,s $(0.7\times)$ & $0$\,s $(0.7\times)$ & $0$\,s $(0.7\times)$ \\
unix50 & 30.sh (9.8: TELE-communications) & $154$\,s & $152$\,s $(1.0\times)$ & $105$\,s $(1.5\times)$ & $80$\,s $(1.9\times)$ & $71$\,s $(2.2\times)$ & $62$\,s $(2.5\times)$ \\
unix50 & 31.sh (9.9) & $149$\,s & $145$\,s $(1.0\times)$ & $106$\,s $(1.4\times)$ & $85$\,s $(1.8\times)$ & $74$\,s $(2.0\times)$ & $68$\,s $(2.2\times)$ \\
unix50 & 32.sh (10.1: count recipients) & $6$\,s & $5$\,s $(1.1\times)$ & $6$\,s $(1.0\times)$ & $5$\,s $(1.1\times)$ & $6$\,s $(1.0\times)$ & $6$\,s $(0.9\times)$ \\
unix50 & 33.sh (10.2: list recipients) & $6$\,s & $6$\,s $(1.0\times)$ & $6$\,s $(1.0\times)$ & $5$\,s $(1.1\times)$ & $5$\,s $(1.1\times)$ & $5$\,s $(1.1\times)$ \\
unix50 & 34.sh (10.3: extract username) & $2$\,s & $0$\,s $(19.7\times)$ & $1$\,s $(1.9\times)$ & $1$\,s $(1.8\times)$ & $2$\,s $(1.3\times)$ & $3$\,s $(0.9\times)$ \\
unix50 & 35.sh (11.1: year received medal) & $1$\,s & $1$\,s $(1.1\times)$ & $1$\,s $(0.6\times)$ & $1$\,s $(0.6\times)$ & $2$\,s $(0.6\times)$ & $2$\,s $(0.6\times)$ \\
unix50 & 36.sh (11.2: most repeated first name) & $19$\,s & $15$\,s $(1.3\times)$ & $10$\,s $(1.9\times)$ & $7$\,s $(2.9\times)$ & $5$\,s $(3.6\times)$ & $6$\,s $(3.5\times)$ \\
unix50 & 4.sh (1.3: sort top first names) & $154$\,s & $131$\,s $(1.2\times)$ & $79$\,s $(2.0\times)$ & $43$\,s $(3.6\times)$ & $26$\,s $(5.9\times)$ & $19$\,s $(8.3\times)$ \\
unix50 & 5.sh (2.1: all Unix utilities) & $8$\,s & $7$\,s $(1.1\times)$ & $5$\,s $(1.6\times)$ & $3$\,s $(2.5\times)$ & $2$\,s $(3.2\times)$ & $2$\,s $(3.1\times)$ \\
unix50 & 6.sh (3.1: first letter of last names) & $14$\,s & $10$\,s $(1.4\times)$ & $7$\,s $(2.2\times)$ & $4$\,s $(3.4\times)$ & $3$\,s $(4.1\times)$ & $3$\,s $(5.2\times)$ \\
unix50 & 7.sh (4.1: number of rounds) & $18$\,s & $14$\,s $(1.3\times)$ & $9$\,s $(2.0\times)$ & $5$\,s $(3.5\times)$ & $4$\,s $(4.7\times)$ & $4$\,s $(4.7\times)$ \\
unix50 & 8.sh (4.2: pieces captured) & $12$\,s & $6$\,s $(2.1\times)$ & $4$\,s $(3.0\times)$ & $3$\,s $(4.2\times)$ & $3$\,s $(5.0\times)$ & $3$\,s $(4.1\times)$ \\
unix50 & 9.sh (4.3: pieces captured with pawn) & $28$\,s & $14$\,s $(2.0\times)$ & $8$\,s $(3.3\times)$ & $5$\,s $(5.6\times)$ & $4$\,s $(7.5\times)$ & $4$\,s $(7.3\times)$ \\
\midrule
\textbf{Max} &  & $2089$\,s & $1196$\,s ($19.7\times$) & $733$\,s ($6.3\times$) & $430$\,s ($12.1\times$) & $311$\,s ($20.5\times$) & $279$\,s ($26.9\times$) \\
\textbf{Min} &  & $0$\,s & $0$\,s ($0.9\times$) & $0$\,s ($0.6\times$) & $0$\,s ($0.6\times$) & $0$\,s ($0.6\times$) & $0$\,s ($0.6\times$) \\
\textbf{Mean} &  & $332$\,s & $228$\,s ($1.9\times$) & $142$\,s ($2.4\times$) & $79$\,s ($4.0\times$) & $50$\,s ($6.1\times$) & $37$\,s ($8.0\times$) \\
\textbf{Median} &  & $167$\,s & $140$\,s ($1.2\times$) & $84$\,s ($2.0\times$) & $50$\,s ($3.5\times$) & $32$\,s ($5.1\times$) & $24$\,s ($7.1\times$) \\

%% file: tables/parallel-3min.tex
\textbf{Benchmark} & \textbf{Script Name} & \textbf{Parallelized} & \textbf{Eliminated} & \textbf{$T_\text{orig}$} & \textbf{$u_{1}$} & \textbf{$u_{16}$} & \textbf{$T_{16}$} \\
\midrule
analytics-mts & 1.sh (vehicles per day) & $7/7$ $(7/7)$ & $3$ $(3)$ & $333$\,s $(1.1\times)$ & $376$\,s & $40$\,s $(9.4\times)$ & $29$\,s $(13.1\times)$ \\
analytics-mts & 2.sh (vehicle days on road) & $8/8$ $(8/8)$ & $3$ $(3)$ & $335$\,s $(1.1\times)$ & $379$\,s & $41$\,s $(9.3\times)$ & $28$\,s $(13.5\times)$ \\
analytics-mts & 3.sh (vehicle hours on road) & $8/8$ $(8/8)$ & $3$ $(3)$ & $408$\,s $(1.0\times)$ & $427$\,s & $51$\,s $(8.4\times)$ & $38$\,s $(11.3\times)$ \\
oneliners & bi-grams.sh & $3/5$ $(3/5)$ & $0$ $(0)$ & $668$\,s $(1.5\times)$ & $1007$\,s & $118$\,s $(8.6\times)$ & $115$\,s $(8.7\times)$ \\
oneliners & diff.sh & $4/7$ $(0/1,2/2,2/2,0/1,0/1)$ & $2$ $(0,1,1,0,0)$ & $325$\,s $(1.5\times)$ & $478$\,s & $98$\,s $(4.9\times)$ & $83$\,s $(5.8\times)$ \\
oneliners & nfa-regex.sh & $2/2$ $(2/2)$ & $1$ $(1)$ & $389$\,s $(1.0\times)$ & $391$\,s & $26$\,s $(14.9\times)$ & $27$\,s $(14.7\times)$ \\
oneliners & set-diff.sh & $5/8$ $(0/1,3/3,2/2,0/1,0/1)$ & $3$ $(0,2,1,0,0)$ & $879$\,s $(1.5\times)$ & $1308$\,s & $144$\,s $(9.1\times)$ & $128$\,s $(10.2\times)$ \\
oneliners & sort.sh & $1/1$ $(1/1)$ & $0$ $(0)$ & $273$\,s $(1.4\times)$ & $389$\,s & $39$\,s $(10.0\times)$ & $38$\,s $(10.3\times)$ \\
oneliners & spell.sh & $6/8$ $(6/8)$ & $3$ $(3)$ & $427$\,s $(1.7\times)$ & $736$\,s & $78$\,s $(9.5\times)$ & $61$\,s $(12.1\times)$ \\
oneliners & top-n.sh & $4/6$ $(4/6)$ & $1$ $(1)$ & $372$\,s $(1.7\times)$ & $622$\,s & $63$\,s $(9.9\times)$ & $50$\,s $(12.4\times)$ \\
oneliners & wf.sh & $4/5$ $(4/5)$ & $1$ $(1)$ & $1155$\,s $(1.8\times)$ & $2089$\,s & $196$\,s $(10.7\times)$ & $145$\,s $(14.4\times)$ \\
poets & 1_1.sh (count_words) & $4/6$ $(4/6)$ & $1$ $(1)$ & $360$\,s $(1.8\times)$ & $637$\,s & $84$\,s $(7.6\times)$ & $83$\,s $(7.6\times)$ \\
poets & 2_1.sh (merge_upper) & $5/7$ $(5/7)$ & $2$ $(2)$ & $307$\,s $(1.8\times)$ & $547$\,s & $79$\,s $(6.9\times)$ & $78$\,s $(7.0\times)$ \\
poets & 3_1.sh (sort) & $5/7$ $(5/7)$ & $1$ $(1)$ & $391$\,s $(1.7\times)$ & $665$\,s & $89$\,s $(7.4\times)$ & $88$\,s $(7.6\times)$ \\
poets & 3_2.sh (sort_words_by_folding) & $5/7$ $(5/7)$ & $1$ $(1)$ & $402$\,s $(1.7\times)$ & $681$\,s & $94$\,s $(7.3\times)$ & $94$\,s $(7.2\times)$ \\
poets & 3_3.sh (sort_words_by_rhyming) & $7/9$ $(7/9)$ & $2$ $(2)$ & $415$\,s $(1.7\times)$ & $699$\,s & $100$\,s $(7.0\times)$ & $100$\,s $(7.0\times)$ \\
poets & 4_3.sh (bigrams) & $4/8$ $(2/4,0/1,2/3)$ & $1$ $(1,0,0)$ & $635$\,s $(1.4\times)$ & $915$\,s & $173$\,s $(5.3\times)$ & $173$\,s $(5.3\times)$ \\
poets & 4_3b.sh (count_trigrams) & $4/9$ $(2/4,0/1,0/1,2/3)$ & $1$ $(1,0,0,0)$ & $862$\,s $(1.2\times)$ & $1049$\,s & $275$\,s $(3.8\times)$ & $279$\,s $(3.8\times)$ \\
poets & 6_1_2.sh (uppercase_by_type) & $4/6$ $(4/6)$ & $1$ $(1)$ & $330$\,s $(1.9\times)$ & $635$\,s & $64$\,s $(10.0\times)$ & $24$\,s $(26.9\times)$ \\
poets & 6_2.sh (4letter_words) & $7/11$ $(3/5,4/6)$ & $2$ $(1,1)$ & $327$\,s $(2.0\times)$ & $647$\,s & $80$\,s $(8.1\times)$ & $34$\,s $(18.8\times)$ \\
poets & 6_3.sh (words_no_vowels) & $5/7$ $(5/7)$ & $2$ $(2)$ & $220$\,s $(1.1\times)$ & $235$\,s & $32$\,s $(7.4\times)$ & $31$\,s $(7.7\times)$ \\
poets & 6_4.sh (1syllable_words) & $5/8$ $(5/8)$ & $2$ $(2)$ & $433$\,s $(1.3\times)$ & $542$\,s & $57$\,s $(9.5\times)$ & $31$\,s $(17.4\times)$ \\
poets & 6_5.sh (2syllable_words) & $5/8$ $(5/8)$ & $2$ $(2)$ & $397$\,s $(1.1\times)$ & $443$\,s & $48$\,s $(9.2\times)$ & $40$\,s $(11.0\times)$ \\
poets & 7_2.sh (count_consonant_seq) & $5/7$ $(5/7)$ & $2$ $(2)$ & $475$\,s $(1.4\times)$ & $678$\,s & $80$\,s $(8.5\times)$ & $48$\,s $(14.2\times)$ \\
poets & 8.2_1.sh (vowel_sequencies_gr_1K) & $5/8$ $(5/8)$ & $1$ $(1)$ & $417$\,s $(1.4\times)$ & $573$\,s & $73$\,s $(7.9\times)$ & $42$\,s $(13.7\times)$ \\
poets & 8.2_2.sh (bigrams_appear_twice) & $4/9$ $(2/4,0/1,2/3,0/1)$ & $1$ $(1,0,0,0)$ & $645$\,s $(1.4\times)$ & $921$\,s & $177$\,s $(5.2\times)$ & $91$\,s $(10.2\times)$ \\
poets & 8.3_2.sh (find_anagrams) & $7/9$ $(2/4,1/1,1/1,3/3)$ & $1$ $(1,0,0,0)$ & $237$\,s $(3.1\times)$ & $724$\,s & $102$\,s $(7.1\times)$ & $50$\,s $(14.5\times)$ \\
poets & 8.3_3.sh (compare_exodus_genesis) & $6/10$ $(3/5,1/2,2/3)$ & $1$ $(1,0,0)$ & $334$\,s $(2.0\times)$ & $656$\,s & $74$\,s $(8.8\times)$ & $34$\,s $(19.3\times)$ \\
poets & 8_1.sh (sort_words_by_n_syllables) & $6/10$ $(3/5,2/2,1/3)$ & $2$ $(1,1,0)$ & $346$\,s $(1.9\times)$ & $653$\,s & $69$\,s $(9.5\times)$ & $26$\,s $(24.6\times)$ \\
unix50 & 14.sh (6.1: order bodies) & $3/3$ $(3/3)$ & $1$ $(1)$ & $143$\,s $(1.3\times)$ & $185$\,s & $31$\,s $(6.0\times)$ & $25$\,s $(7.5\times)$ \\
unix50 & 21.sh (8.4: longest words w/o hyphens) & $3/3$ $(3/3)$ & $1$ $(1)$ & $428$\,s $(1.7\times)$ & $733$\,s & $64$\,s $(11.4\times)$ & $49$\,s $(14.9\times)$ \\
unix50 & 23.sh (9.1: extract word PORT) & $6/6$ $(6/6)$ & $4$ $(4)$ & $111$\,s $(1.8\times)$ & $202$\,s & $23$\,s $(8.8\times)$ & $10$\,s $(19.8\times)$ \\
unix50 & 28.sh (9.6: follow directions) & $6/10$ $(6/10)$ & $3$ $(3)$ & $87$\,s $(2.2\times)$ & $188$\,s & $54$\,s $(3.5\times)$ & $49$\,s $(3.8\times)$ \\
\midrule
\textbf{Total} &  & $163/233$ & $55$ &  &  &  &  \\
\textbf{Max} &  &  &  & $1155$\,s ($3.1\times$) & $2089$\,s & $275$\,s ($14.9\times$) & $279$\,s ($26.9\times$) \\
\textbf{Min} &  &  &  & $87$\,s ($1.0\times$) & $185$\,s & $23$\,s ($3.5\times$) & $10$\,s ($3.8\times$) \\
\textbf{Mean} &  &  &  & $420$\,s ($1.6\times$) & $649$\,s & $85$\,s ($8.2\times$) & $67$\,s ($12.0\times$) \\
\textbf{Median} &  &  &  & $389$\,s ($1.5\times$) & $637$\,s & $74$\,s ($8.5\times$) & $49$\,s ($11.3\times$) \\

%% file: tables/synthesis-combiners.tex
\begin{table}[t]
\footnotesize
\caption{Combiners synthesized for all benchmark scripts}
\label{tab:plausible-combiners}
\begin{tabular}{ll}
\toprule
\textbf{Count} & \textbf{Synthesized Plausible Combiner}
\\ \midrule
$81$ & $(\mathbf{concat}~a~b)$ \\
$22$ & $(\mathbf{rerun}~a~b)$ \\
$16$ & $(\mathbf{merge}(\texttt{*})~a~b)$ or $(\mathbf{merge}(\texttt{*})~b~a)$ \\
$12$ & $((\mathbf{back}~`\textbackslash n`~\mathbf{add})~a~b)$ or $((\mathbf{back}~`\textbackslash n`~\mathbf{add})~b~a)$ \\
$8$ & $(\mathbf{rerun}~b~a)$ \\
$2$ & $((\mathbf{back}~`\textbackslash n`~\mathbf{first})~a~b)$ or $((\mathbf{back}~`\textbackslash n`~\mathbf{second})~b~a)$ \\
$2$ & $(\mathbf{first}~a~b)$ or $(\mathbf{second}~b~a)$ \\
$2$ & $((\mathbf{fuse}~`\textbackslash n`~\mathbf{first})~a~b)$ or $((\mathbf{fuse}~`\textbackslash n`~\mathbf{second})~b~a)$ \\
$2$ & $((\mathbf{back}~`\textbackslash n`~\mathbf{second})~a~b)$ or $((\mathbf{back}~`\textbackslash n`~\mathbf{first})~b~a)$ \\
$2$ & $(\mathbf{second}~a~b)$ or $(\mathbf{first}~b~a)$\\
$2$ & $((\mathbf{fuse}~`\textbackslash n`~\mathbf{second})~a~b)$ or $((\mathbf{fuse}~`\textbackslash n`~\mathbf{first})~b~a)$ \\
$2$ & $((\mathbf{stitch2}~`~`~\mathbf{add}~\mathbf{first})~a~b)$ or $((\mathbf{stitch2}~`~`~\mathbf{add}~\mathbf{second})~a~b)$ \\
$2$ & $((\mathbf{stitch}~\mathbf{first})~a~b)$ or $((\mathbf{stitch}~\mathbf{second})~a~b)$ \\
\bottomrule

\end{tabular}
\end{table}

%% file: tables/synthesis-errors.tex
\begin{table}[t]
\footnotesize
\caption{Unsupported commands in all benchmark scripts}
\label{tab:unsupported}
\begin{tabular}{p{3cm}ll}
\toprule
\textbf{Command} & \textbf{Reason Unsupported} & \textbf{Counterexample Input Streams}
\\ \midrule
\texttt{awk "\textbackslash \$1 == 2 {print \textbackslash \$2, \textbackslash \$3}"}
& \sys did not generate inputs for the command to produce nonempty outputs.
& \\

\texttt{sed 1d}
& \textbf{No combiners $g$ exist} such that $f(\ttx{1}~{+}{+}~\ttx{2}) = g(f(\ttx{1}), f(\ttx{2}))$ for all streams $\ttx{1}, \ttx{2}$.
& Each of $\ttx{1},\ttx{2}$ has at least one line.\\
\texttt{sed 2d}
& \textbf{No combiners $g$ exist} such that $f(\ttx{1}~{+}{+}~\ttx{2}) = g(f(\ttx{1}), f(\ttx{2}))$ for all streams $\ttx{1}, \ttx{2}$.
& Each of $\ttx{1},\ttx{2}$ has at least two lines.\\
\texttt{sed 3d}
& \textbf{No combiners $g$ exist} such that $f(\ttx{1}~{+}{+}~\ttx{2}) = g(f(\ttx{1}), f(\ttx{2}))$ for all streams $\ttx{1}, \ttx{2}$.
& Each of $\ttx{1},\ttx{2}$ has at least three lines.\\
\texttt{sed 4d}
& \textbf{No combiners $g$ exist} such that $f(\ttx{1}~{+}{+}~\ttx{2}) = g(f(\ttx{1}), f(\ttx{2}))$ for all streams $\ttx{1}, \ttx{2}$.
& Each of $\ttx{1},\ttx{2}$ has at least four lines.\\
\texttt{sed 5d}
& \textbf{No combiners $g$ exist} such that $f(\ttx{1}~{+}{+}~\ttx{2}) = g(f(\ttx{1}), f(\ttx{2}))$ for all streams $\ttx{1}, \ttx{2}$.
& Each of $\ttx{1},\ttx{2}$ has at least five lines.\\
\texttt{tail +2}
& \textbf{No combiners $g$ exist} such that $f(\ttx{1}~{+}{+}~\ttx{2}) = g(f(\ttx{1}), f(\ttx{2}))$ for all streams $\ttx{1}, \ttx{2}$.
& Each of $\ttx{1},\ttx{2}$ has at least one line.\\
\texttt{tail +3}
& \textbf{No combiners $g$ exist} such that $f(\ttx{1}~{+}{+}~\ttx{2}) = g(f(\ttx{1}), f(\ttx{2}))$ for all streams $\ttx{1}, \ttx{2}$.
& Each of $\ttx{1},\ttx{2}$ has at least two lines.\\
\bottomrule

\end{tabular}
\end{table}

%% file: tables/synthesis.tex
\footnotesize
\begin{longtable}{p{.9cm}p{.9cm}rlllll}
\caption{Synthesis results for unique command/flag combinations}
\label{tab:synthesis}
\\

\toprule

\input{tables/synthesis-size7}

\bottomrule

\end{longtable}
\clearpage
\twocolumn

%% file: tables/synthesis-size7.tex
\textbf{Bench} & \textbf{Script} & \textbf{Idx} & \textbf{Command} & \textbf{Search Space} & \textbf{Time} & \textbf{Synthesized Plausible} & \textbf{\#P}
\\ \midrule
oneliners & spell & $8$ & \makecell[l]{\texttt{IN=\$\{IN:-../benchmarks/pipelines/one}\\ \texttt{liners/input/1G.txt\}}\\ \texttt{dict=\$\{dict:-../in/dict.sorted\}}\\ \texttt{LC_COLLATE=C comm -23 - \$dict}} & $\begin{aligned} & 26404 \\ &(=12440+13960+4) \end{aligned}$ & $331$ s & $\begin{aligned} e_{1} &= (\mathbf{concat}~a~b),\\ e_{2} &= (\mathbf{rerun}~a~b). \end{aligned}$ & $2$
\\ \midrule
poets & vowel_sequencies_gr_1K & $7$ & \makecell[l]{\texttt{awk "\textbackslash \$1 >= 1000"}} & $\begin{aligned} & 26404 \\ &(=12440+13960+4) \end{aligned}$ & $176$ s & $\begin{aligned} e_{1} &= (\mathbf{concat}~a~b),\\ e_{2} &= (\mathbf{rerun}~a~b). \end{aligned}$ & $2$
\\ \midrule
poets & find_anagrams & $8$ & \makecell[l]{\texttt{awk "\textbackslash \$1 >= 2 \{print \textbackslash \$2\}"}} & $\begin{aligned} & 2700 \\ &(=968+1728+4) \end{aligned}$ & $40$ s & $\begin{aligned} e_{1} &= (\mathbf{concat}~a~b). \end{aligned}$ & $1$
\\ \midrule
unix50 & 8.4: longest words w/o hyphens & $3$ & \makecell[l]{\texttt{awk "length >= 16"}} & $\begin{aligned} & 26404 \\ &(=12440+13960+4) \end{aligned}$ & $172$ s & $\begin{aligned} e_{1} &= (\mathbf{concat}~a~b),\\ e_{2} &= (\mathbf{rerun}~a~b). \end{aligned}$ & $2$
\\ \midrule
unix50 & 8.2: location office & $4$ & \makecell[l]{\texttt{awk "\{\textbackslash \$1=\textbackslash \$1\};1"}} & $\begin{aligned} & 26404 \\ &(=12440+13960+4) \end{aligned}$ & $60$ s & $\begin{aligned} e_{1} &= (\mathbf{concat}~a~b),\\ e_{2} &= (\mathbf{rerun}~a~b). \end{aligned}$ & $2$
\\ \midrule
unix50 & 6.1: order bodies & $1$ & \makecell[l]{\texttt{awk "\{print \textbackslash \$2, \textbackslash \$0\}"}} & $\begin{aligned} & 110444 \\ &(=59048+51392+4) \end{aligned}$ & $124$ s & $\begin{aligned} e_{1} &= (\mathbf{concat}~a~b). \end{aligned}$ & $1$
\\ \midrule
unix50 & 8.2: location office & $2$ & \makecell[l]{\texttt{awk 'length <= 45'}} & $\begin{aligned} & 26404 \\ &(=12440+13960+4) \end{aligned}$ & $60$ s & $\begin{aligned} e_{1} &= (\mathbf{concat}~a~b),\\ e_{2} &= (\mathbf{rerun}~a~b). \end{aligned}$ & $2$
\\ \midrule
poets & sort_words_by_n_syllables & $6$ & \makecell[l]{\texttt{awk '\{print NF\}'}} & $\begin{aligned} & 2700 \\ &(=968+1728+4) \end{aligned}$ & $39$ s & $\begin{aligned} e_{1} &= (\mathbf{concat}~a~b). \end{aligned}$ & $1$
\\ \midrule
analytics-mts & vehicles per day & $7$ & \makecell[l]{\texttt{awk -v OFS="\textbackslash t" "\{print \textbackslash \$2,\textbackslash \$1\}"}} & $\begin{aligned} & 110444 \\ &(=59048+51392+4) \end{aligned}$ & $125$ s & $\begin{aligned} e_{1} &= (\mathbf{concat}~a~b). \end{aligned}$ & $1$
\\ \midrule
analytics-mts & vehicles per day & $0$ & \makecell[l]{\texttt{cat}} & $\begin{aligned} & 26404 \\ &(=12440+13960+4) \end{aligned}$ & $59$ s & $\begin{aligned} e_{1} &= (\mathbf{concat}~a~b),\\ e_{2} &= (\mathbf{rerun}~a~b). \end{aligned}$ & $2$
\\ \midrule
oneliners & spell & $2$ & \makecell[l]{\texttt{col -bx}} & $\begin{aligned} & 26404 \\ &(=12440+13960+4) \end{aligned}$ & $59$ s & $\begin{aligned} e_{1} &= (\mathbf{concat}~a~b),\\ e_{2} &= (\mathbf{rerun}~a~b). \end{aligned}$ & $2$
\\ \midrule
unix50 & 4.4: histogram by piece & $6$ & \makecell[l]{\texttt{cut -c 1-1}} & $\begin{aligned} & 26404 \\ &(=12440+13960+4) \end{aligned}$ & $60$ s & $\begin{aligned} e_{1} &= (\mathbf{concat}~a~b),\\ e_{2} &= (\mathbf{rerun}~a~b). \end{aligned}$ & $2$
\\ \midrule
unix50 & 5.1: extract hellow world & $3$ & \makecell[l]{\texttt{cut -c 1-12}} & $\begin{aligned} & 26404 \\ &(=12440+13960+4) \end{aligned}$ & $60$ s & $\begin{aligned} e_{1} &= (\mathbf{concat}~a~b),\\ e_{2} &= (\mathbf{rerun}~a~b). \end{aligned}$ & $2$
\\ \midrule
unix50 & 9.3: animal decorate & $1$ & \makecell[l]{\texttt{cut -c 1-2}} & $\begin{aligned} & 26404 \\ &(=12440+13960+4) \end{aligned}$ & $59$ s & $\begin{aligned} e_{1} &= (\mathbf{concat}~a~b),\\ e_{2} &= (\mathbf{rerun}~a~b). \end{aligned}$ & $2$
\\ \midrule
unix50 & 9.1: extract word PORT & $6$ & \makecell[l]{\texttt{cut -c 1-4}} & $\begin{aligned} & 26404 \\ &(=12440+13960+4) \end{aligned}$ & $60$ s & $\begin{aligned} e_{1} &= (\mathbf{concat}~a~b),\\ e_{2} &= (\mathbf{rerun}~a~b). \end{aligned}$ & $2$
\\ \midrule
unix50 & 7.3: decades unix released & $3$ & \makecell[l]{\texttt{cut -c 3-3}} & $\begin{aligned} & 26404 \\ &(=12440+13960+4) \end{aligned}$ & $59$ s & $\begin{aligned} e_{1} &= (\mathbf{concat}~a~b). \end{aligned}$ & $1$
\\ \midrule
unix50 & 5.1: extract hellow world & $2$ & \makecell[l]{\texttt{cut -d "\textbackslash "" -f 2}} & $\begin{aligned} & 26404 \\ &(=12440+13960+4) \end{aligned}$ & $104$ s & $\begin{aligned} e_{1} &= (\mathbf{concat}~a~b),\\ e_{2} &= (\mathbf{rerun}~a~b). \end{aligned}$ & $2$
\\ \midrule
oneliners & set-diff & $2$ & \makecell[l]{\texttt{cut -d ' ' -f 1}} & $\begin{aligned} & 2700 \\ &(=968+1728+4) \end{aligned}$ & $71$ s & $\begin{aligned} e_{1} &= (\mathbf{concat}~a~b),\\ e_{2} &= (\mathbf{rerun}~a~b). \end{aligned}$ & $2$
\\ \midrule
unix50 & 1.0: extract last name & $1$ & \makecell[l]{\texttt{cut -d ' ' -f 2}} & $\begin{aligned} & 2700 \\ &(=968+1728+4) \end{aligned}$ & $72$ s & $\begin{aligned} e_{1} &= (\mathbf{concat}~a~b),\\ e_{2} &= (\mathbf{rerun}~a~b). \end{aligned}$ & $2$
\\ \midrule
unix50 & 2.1: all Unix utilities & $1$ & \makecell[l]{\texttt{cut -d ' ' -f 4}} & $\begin{aligned} & 2700 \\ &(=968+1728+4) \end{aligned}$ & $71$ s & $\begin{aligned} e_{1} &= (\mathbf{concat}~a~b),\\ e_{2} &= (\mathbf{rerun}~a~b). \end{aligned}$ & $2$
\\ \midrule
unix50 & 8.3: four most involved & $2$ & \makecell[l]{\texttt{cut -d '(' -f 2}} & $\begin{aligned} & 26404 \\ &(=12440+13960+4) \end{aligned}$ & $102$ s & $\begin{aligned} e_{1} &= (\mathbf{concat}~a~b),\\ e_{2} &= (\mathbf{rerun}~a~b). \end{aligned}$ & $2$
\\ \midrule
unix50 & 8.3: four most involved & $3$ & \makecell[l]{\texttt{cut -d ')' -f 1}} & $\begin{aligned} & 26404 \\ &(=12440+13960+4) \end{aligned}$ & $102$ s & $\begin{aligned} e_{1} &= (\mathbf{concat}~a~b),\\ e_{2} &= (\mathbf{rerun}~a~b). \end{aligned}$ & $2$
\\ \midrule
analytics-mts & vehicles per day & $4$ & \makecell[l]{\texttt{cut -d ',' -f 1}} & $\begin{aligned} & 26404 \\ &(=12440+13960+4) \end{aligned}$ & $102$ s & $\begin{aligned} e_{1} &= (\mathbf{concat}~a~b),\\ e_{2} &= (\mathbf{rerun}~a~b). \end{aligned}$ & $2$
\\ \midrule
analytics-mts & hours monitored per day & $2$ & \makecell[l]{\texttt{cut -d ',' -f 1,2}} & $\begin{aligned} & 110444 \\ &(=59048+51392+4) \end{aligned}$ & $126$ s & $\begin{aligned} e_{1} &= (\mathbf{concat}~a~b),\\ e_{2} &= (\mathbf{rerun}~a~b). \end{aligned}$ & $2$
\\ \midrule
analytics-mts & vehicle hours on road & $2$ & \makecell[l]{\texttt{cut -d ',' -f 1,2,4}} & $\begin{aligned} & 110444 \\ &(=59048+51392+4) \end{aligned}$ & $126$ s & $\begin{aligned} e_{1} &= (\mathbf{concat}~a~b). \end{aligned}$ & $1$
\\ \midrule
analytics-mts & vehicles per day & $2$ & \makecell[l]{\texttt{cut -d ',' -f 1,3}} & $\begin{aligned} & 110444 \\ &(=59048+51392+4) \end{aligned}$ & $125$ s & $\begin{aligned} e_{1} &= (\mathbf{concat}~a~b). \end{aligned}$ & $1$
\\ \midrule
analytics-mts & vehicle days on road & $4$ & \makecell[l]{\texttt{cut -d ',' -f 2}} & $\begin{aligned} & 26404 \\ &(=12440+13960+4) \end{aligned}$ & $103$ s & $\begin{aligned} e_{1} &= (\mathbf{concat}~a~b),\\ e_{2} &= (\mathbf{rerun}~a~b). \end{aligned}$ & $2$
\\ \midrule
analytics-mts & vehicle hours on road & $4$ & \makecell[l]{\texttt{cut -d ',' -f 3}} & $\begin{aligned} & 26404 \\ &(=12440+13960+4) \end{aligned}$ & $101$ s & $\begin{aligned} e_{1} &= (\mathbf{concat}~a~b),\\ e_{2} &= (\mathbf{rerun}~a~b). \end{aligned}$ & $2$
\\ \midrule
analytics-mts & vehicle days on road & $2$ & \makecell[l]{\texttt{cut -d ',' -f 3,1}} & $\begin{aligned} & 110444 \\ &(=59048+51392+4) \end{aligned}$ & $124$ s & $\begin{aligned} e_{1} &= (\mathbf{concat}~a~b). \end{aligned}$ & $1$
\\ \midrule
unix50 & 4.4: histogram by piece & $4$ & \makecell[l]{\texttt{cut -d '.' -f 2}} & $\begin{aligned} & 26404 \\ &(=12440+13960+4) \end{aligned}$ & $103$ s & $\begin{aligned} e_{1} &= (\mathbf{concat}~a~b),\\ e_{2} &= (\mathbf{rerun}~a~b). \end{aligned}$ & $2$
\\ \midrule
oneliners & shortest-scripts & $3$ & \makecell[l]{\texttt{cut -d: -f1}} & $\begin{aligned} & 26404 \\ &(=12440+13960+4) \end{aligned}$ & $61$ s & $\begin{aligned} e_{1} &= (\mathbf{concat}~a~b),\\ e_{2} &= (\mathbf{rerun}~a~b). \end{aligned}$ & $2$
\\ \midrule
unix50 & 7.1: number of versions & $1$ & \makecell[l]{\texttt{cut -f 1}} & $\begin{aligned} & 26404 \\ &(=12440+13960+4) \end{aligned}$ & $102$ s & $\begin{aligned} e_{1} &= (\mathbf{concat}~a~b),\\ e_{2} &= (\mathbf{rerun}~a~b). \end{aligned}$ & $2$
\\ \midrule
unix50 & 7.2: most frequent machine & $1$ & \makecell[l]{\texttt{cut -f 2}} & $\begin{aligned} & 26404 \\ &(=12440+13960+4) \end{aligned}$ & $103$ s & $\begin{aligned} e_{1} &= (\mathbf{concat}~a~b),\\ e_{2} &= (\mathbf{rerun}~a~b). \end{aligned}$ & $2$
\\ \midrule
unix50 & 7.3: decades unix released & $1$ & \makecell[l]{\texttt{cut -f 4}} & $\begin{aligned} & 26404 \\ &(=12440+13960+4) \end{aligned}$ & $102$ s & $\begin{aligned} e_{1} &= (\mathbf{concat}~a~b),\\ e_{2} &= (\mathbf{rerun}~a~b). \end{aligned}$ & $2$
\\ \midrule
unix50 & 10.3: extract username & $4$ & \makecell[l]{\texttt{fmt -w1}} & $\begin{aligned} & 26404 \\ &(=12440+13960+4) \end{aligned}$ & $60$ s & $\begin{aligned} e_{1} &= (\mathbf{concat}~a~b),\\ e_{2} &= (\mathbf{rerun}~a~b). \end{aligned}$ & $2$
\\ \midrule
unix50 & 9.4: four corners & $2$ & \makecell[l]{\texttt{grep "\textbackslash ""}} & $\begin{aligned} & 26404 \\ &(=12440+13960+4) \end{aligned}$ & $61$ s & $\begin{aligned} e_{1} &= (\mathbf{concat}~a~b),\\ e_{2} &= (\mathbf{rerun}~a~b). \end{aligned}$ & $2$
\\ \midrule
oneliners & shortest-scripts & $2$ & \makecell[l]{\texttt{grep "shell script"}} & $\begin{aligned} & 26404 \\ &(=12440+13960+4) \end{aligned}$ & $62$ s & $\begin{aligned} e_{1} &= (\mathbf{concat}~a~b),\\ e_{2} &= (\mathbf{rerun}~a~b). \end{aligned}$ & $2$
\\ \midrule
unix50 & 8.3: four most involved & $1$ & \makecell[l]{\texttt{grep '('}} & $\begin{aligned} & 26404 \\ &(=12440+13960+4) \end{aligned}$ & $60$ s & $\begin{aligned} e_{1} &= (\mathbf{concat}~a~b),\\ e_{2} &= (\mathbf{rerun}~a~b). \end{aligned}$ & $2$
\\ \midrule
unix50 & 7.1: number of versions & $2$ & \makecell[l]{\texttt{grep 'AT\&T'}} & $\begin{aligned} & 26404 \\ &(=12440+13960+4) \end{aligned}$ & $61$ s & $\begin{aligned} e_{1} &= (\mathbf{concat}~a~b),\\ e_{2} &= (\mathbf{rerun}~a~b). \end{aligned}$ & $2$
\\ \midrule
poets & trigram_rec & $10$ & \makecell[l]{\texttt{grep 'And he said'}} & $\begin{aligned} & 26404 \\ &(=12440+13960+4) \end{aligned}$ & $61$ s & $\begin{aligned} e_{1} &= (\mathbf{concat}~a~b),\\ e_{2} &= (\mathbf{rerun}~a~b). \end{aligned}$ & $2$
\\ \midrule
unix50 & 8.2: location office & $1$ & \makecell[l]{\texttt{grep 'Bell'}} & $\begin{aligned} & 26404 \\ &(=12440+13960+4) \end{aligned}$ & $61$ s & $\begin{aligned} e_{1} &= (\mathbf{concat}~a~b),\\ e_{2} &= (\mathbf{rerun}~a~b). \end{aligned}$ & $2$
\\ \midrule
unix50 & 11.1: year received medal & $1$ & \makecell[l]{\texttt{grep 'UNIX'}} & $\begin{aligned} & 26404 \\ &(=12440+13960+4) \end{aligned}$ & $61$ s & $\begin{aligned} e_{1} &= (\mathbf{concat}~a~b),\\ e_{2} &= (\mathbf{rerun}~a~b). \end{aligned}$ & $2$
\\ \midrule
unix50 & 9.1: extract word PORT & $2$ & \makecell[l]{\texttt{grep '[A-Z]'}} & $\begin{aligned} & 26404 \\ &(=12440+13960+4) \end{aligned}$ & $59$ s & $\begin{aligned} e_{1} &= (\mathbf{concat}~a~b),\\ e_{2} &= (\mathbf{rerun}~a~b). \end{aligned}$ & $2$
\\ \midrule
unix50 & 4.4: histogram by piece & $5$ & \makecell[l]{\texttt{grep '[KQRBN]'}} & $\begin{aligned} & 26404 \\ &(=12440+13960+4) \end{aligned}$ & $60$ s & $\begin{aligned} e_{1} &= (\mathbf{concat}~a~b),\\ e_{2} &= (\mathbf{rerun}~a~b). \end{aligned}$ & $2$
\\ \midrule
oneliners & nfa-regex & $2$ & \makecell[l]{\texttt{grep '\textbackslash (.\textbackslash ).*\textbackslash 1\textbackslash (.\textbackslash ).*\textbackslash 2\textbackslash (.\textbackslash ).*\textbackslash 3\textbackslash (.}\\ \texttt{\textbackslash ).*\textbackslash 4'}} & $\begin{aligned} & 110444 \\ &(=59048+51392+4) \end{aligned}$ & $129$ s & $\begin{aligned} e_{1} &= (\mathbf{concat}~a~b),\\ e_{2} &= (\mathbf{rerun}~a~b). \end{aligned}$ & $2$
\\ \midrule
unix50 & 4.4: histogram by piece & $3$ & \makecell[l]{\texttt{grep '\textbackslash .'}} & $\begin{aligned} & 26404 \\ &(=12440+13960+4) \end{aligned}$ & $165$ s & $\begin{aligned} e_{1} &= (\mathbf{concat}~a~b),\\ e_{2} &= (\mathbf{rerun}~a~b). \end{aligned}$ & $2$
\\ \midrule
poets & verses_2om_3om_2instances & $11$ & \makecell[l]{\texttt{grep 'light.\textbackslash *light'}} & $\begin{aligned} & 26404 \\ &(=12440+13960+4) \end{aligned}$ & $172$ s & $\begin{aligned} e_{1} &= (\mathbf{concat}~a~b),\\ e_{2} &= (\mathbf{rerun}~a~b). \end{aligned}$ & $2$
\\ \midrule
unix50 & 5.1: extract hellow world & $1$ & \makecell[l]{\texttt{grep 'print'}} & $\begin{aligned} & 26404 \\ &(=12440+13960+4) \end{aligned}$ & $61$ s & $\begin{aligned} e_{1} &= (\mathbf{concat}~a~b),\\ e_{2} &= (\mathbf{rerun}~a~b). \end{aligned}$ & $2$
\\ \midrule
poets & trigram_rec & $3$ & \makecell[l]{\texttt{grep 'the land of'}} & $\begin{aligned} & 26404 \\ &(=12440+13960+4) \end{aligned}$ & $61$ s & $\begin{aligned} e_{1} &= (\mathbf{concat}~a~b),\\ e_{2} &= (\mathbf{rerun}~a~b). \end{aligned}$ & $2$
\\ \midrule
unix50 & 4.4: histogram by piece & $2$ & \makecell[l]{\texttt{grep 'x'}} & $\begin{aligned} & 26404 \\ &(=12440+13960+4) \end{aligned}$ & $179$ s & $\begin{aligned} e_{1} &= (\mathbf{concat}~a~b),\\ e_{2} &= (\mathbf{rerun}~a~b). \end{aligned}$ & $2$
\\ \midrule
poets & 4letter_words & $10$ & \makecell[l]{\texttt{grep -c '\textasciicircum ....\$'}} & $\begin{aligned} & 2700 \\ &(=968+1728+4) \end{aligned}$ & $38$ s & $\begin{aligned} e_{1} &= ((\mathbf{back}~`\textbackslash n`~\mathbf{add})~a~b),\\ e_{2} &= ((\mathbf{back}~`\textbackslash n`~\mathbf{add})~b~a). \end{aligned}$ & $2$
\\ \midrule
poets & uppercase_by_token & $4$ & \makecell[l]{\texttt{grep -c '\textasciicircum [A-Z]'}} & $\begin{aligned} & 2700 \\ &(=968+1728+4) \end{aligned}$ & $39$ s & $\begin{aligned} e_{1} &= ((\mathbf{back}~`\textbackslash n`~\mathbf{add})~a~b),\\ e_{2} &= ((\mathbf{back}~`\textbackslash n`~\mathbf{add})~b~a). \end{aligned}$ & $2$
\\ \midrule
poets & verses_2om_3om_2instances & $3$ & \makecell[l]{\texttt{grep -c 'light.\textbackslash *light'}} & $\begin{aligned} & 2700 \\ &(=968+1728+4) \end{aligned}$ & $39$ s & $\begin{aligned} e_{1} &= ((\mathbf{back}~`\textbackslash n`~\mathbf{add})~a~b),\\ e_{2} &= ((\mathbf{back}~`\textbackslash n`~\mathbf{add})~b~a). \end{aligned}$ & $2$
\\ \midrule
poets & verses_2om_3om_2instances & $7$ & \makecell[l]{\texttt{grep -c 'light.\textbackslash *light.\textbackslash *light'}} & $\begin{aligned} & 2700 \\ &(=968+1728+4) \end{aligned}$ & $39$ s & $\begin{aligned} e_{1} &= ((\mathbf{back}~`\textbackslash n`~\mathbf{add})~a~b),\\ e_{2} &= ((\mathbf{back}~`\textbackslash n`~\mathbf{add})~b~a). \end{aligned}$ & $2$
\\ \midrule
poets & 1syllable_words & $4$ & \makecell[l]{\texttt{grep -i '\textasciicircum [\textasciicircum aeiou]*[aeiou][\textasciicircum aeiou]*\$}\\ \texttt{'}} & $\begin{aligned} & 110444 \\ &(=59048+51392+4) \end{aligned}$ & $177$ s & $\begin{aligned} e_{1} &= (\mathbf{concat}~a~b),\\ e_{2} &= (\mathbf{merge}~a~b),\\ e_{3} &= (\mathbf{merge}~b~a),\\ e_{4} &= (\mathbf{rerun}~a~b). \end{aligned}$ & $4$
\\ \midrule
poets & 2syllable_words & $4$ & \makecell[l]{\texttt{grep -i '\textasciicircum [\textasciicircum aeiou]*[aeiou][\textasciicircum aeiou]*[}\\ \texttt{aeiou][\textasciicircum aeiou]\$'}} & $\begin{aligned} & 110444 \\ &(=59048+51392+4) \end{aligned}$ & $315$ s & $\begin{aligned} e_{1} &= (\mathbf{concat}~a~b),\\ e_{2} &= (\mathbf{rerun}~a~b). \end{aligned}$ & $2$
\\ \midrule
unix50 & 4.3: pieces captured with pawn & $5$ & \makecell[l]{\texttt{grep -v '[KQRBN]'}} & $\begin{aligned} & 26404 \\ &(=12440+13960+4) \end{aligned}$ & $60$ s & $\begin{aligned} e_{1} &= (\mathbf{concat}~a~b),\\ e_{2} &= (\mathbf{rerun}~a~b). \end{aligned}$ & $2$
\\ \midrule
oneliners & shortest-scripts & $5$ & \makecell[l]{\texttt{grep -v '\textasciicircum 0\$'}} & $\begin{aligned} & 26404 \\ &(=12440+13960+4) \end{aligned}$ & $60$ s & $\begin{aligned} e_{1} &= (\mathbf{concat}~a~b),\\ e_{2} &= (\mathbf{rerun}~a~b). \end{aligned}$ & $2$
\\ \midrule
poets & verses_2om_3om_2instances & $12$ & \makecell[l]{\texttt{grep -vc 'light.\textbackslash *light.\textbackslash *light'}} & $\begin{aligned} & 2700 \\ &(=968+1728+4) \end{aligned}$ & $39$ s & $\begin{aligned} e_{1} &= ((\mathbf{back}~`\textbackslash n`~\mathbf{add})~a~b),\\ e_{2} &= ((\mathbf{back}~`\textbackslash n`~\mathbf{add})~b~a). \end{aligned}$ & $2$
\\ \midrule
poets & words_no_vowels & $4$ & \makecell[l]{\texttt{grep -vi '[aeiou]'}} & $\begin{aligned} & 26404 \\ &(=12440+13960+4) \end{aligned}$ & $59$ s & $\begin{aligned} e_{1} &= (\mathbf{concat}~a~b),\\ e_{2} &= (\mathbf{rerun}~a~b). \end{aligned}$ & $2$
\\ \midrule
unix50 & 8.1: count unix birth-year & $2$ & \makecell[l]{\texttt{grep 1969}} & $\begin{aligned} & 26404 \\ &(=12440+13960+4) \end{aligned}$ & $60$ s & $\begin{aligned} e_{1} &= (\mathbf{concat}~a~b),\\ e_{2} &= (\mathbf{rerun}~a~b). \end{aligned}$ & $2$
\\ \midrule
poets & compare_exodus_genesis & $9$ & \makecell[l]{\texttt{head}} & $\begin{aligned} & 26404 \\ &(=12440+13960+4) \end{aligned}$ & $163$ s & $\begin{aligned} e_{1} &= (\mathbf{rerun}~a~b). \end{aligned}$ & $1$
\\ \midrule
oneliners & shortest-scripts & $7$ & \makecell[l]{\texttt{head -15}} & $\begin{aligned} & 26404 \\ &(=12440+13960+4) \end{aligned}$ & $185$ s & $\begin{aligned} e_{1} &= (\mathbf{rerun}~a~b). \end{aligned}$ & $1$
\\ \midrule
unix50 & 7.2: most frequent machine & $5$ & \makecell[l]{\texttt{head -n 1}} & $\begin{aligned} & 26404 \\ &(=12440+13960+4) \end{aligned}$ & $102$ s & $\begin{aligned} e_{1} &= (\mathbf{first}~a~b),\\ e_{2} &= (\mathbf{second}~b~a),\\ e_{3} &= ((\mathbf{back}~`\textbackslash n`~\mathbf{first})~a~b),\\ e_{4} &= ((\mathbf{fuse}~`\textbackslash n`~\mathbf{first})~a~b),\\ e_{5} &= ((\mathbf{back}~`\textbackslash n`~\mathbf{second})~b~a),\\ e_{6} &= ((\mathbf{fuse}~`\textbackslash n`~\mathbf{second})~b~a),\\ e_{7} &= (\mathbf{rerun}~a~b). \end{aligned}$ & $7$
\\ \midrule
unix50 & 1.2: extract names and sort & $1$ & \makecell[l]{\texttt{head -n 2}} & $\begin{aligned} & 26404 \\ &(=12440+13960+4) \end{aligned}$ & $101$ s & $\begin{aligned} e_{1} &= (\mathbf{rerun}~a~b). \end{aligned}$ & $1$
\\ \midrule
unix50 & 4.6: piece used most & $8$ & \makecell[l]{\texttt{head -n 3}} & $\begin{aligned} & 26404 \\ &(=12440+13960+4) \end{aligned}$ & $101$ s & $\begin{aligned} e_{1} &= (\mathbf{rerun}~a~b). \end{aligned}$ & $1$
\\ \midrule
oneliners & spell & $1$ & \makecell[l]{\texttt{iconv -f utf-8 -t ascii//translit}} & $\begin{aligned} & 26404 \\ &(=12440+13960+4) \end{aligned}$ & $59$ s & $\begin{aligned} e_{1} &= (\mathbf{concat}~a~b),\\ e_{2} &= (\mathbf{rerun}~a~b). \end{aligned}$ & $2$
\\ \midrule
poets & sort_words_by_rhyming & $6$ & \makecell[l]{\texttt{rev}} & $\begin{aligned} & 26404 \\ &(=12440+13960+4) \end{aligned}$ & $59$ s & $\begin{aligned} e_{1} &= (\mathbf{concat}~a~b). \end{aligned}$ & $1$
\\ \midrule
poets & count_words & $1$ & \makecell[l]{\texttt{IN=\$\{IN:-../benchmarks/pipelines/poe}\\ \texttt{ts/input/pg/\}}\\ \texttt{sed "s;\textasciicircum ;\$IN;"}} & $\begin{aligned} & 26404 \\ &(=12440+13960+4) \end{aligned}$ & $59$ s & $\begin{aligned} e_{1} &= (\mathbf{concat}~a~b). \end{aligned}$ & $1$
\\ \midrule
analytics-mts & vehicles per day & $1$ & \makecell[l]{\texttt{sed 's/T..:..:..//'}} & $\begin{aligned} & 26404 \\ &(=12440+13960+4) \end{aligned}$ & $59$ s & $\begin{aligned} e_{1} &= (\mathbf{concat}~a~b). \end{aligned}$ & $1$
\\ \midrule
analytics-mts & vehicle hours on road & $1$ & \makecell[l]{\texttt{sed 's/T\textbackslash (..\textbackslash ):..:../,\textbackslash 1/'}} & $\begin{aligned} & 110444 \\ &(=59048+51392+4) \end{aligned}$ & $126$ s & $\begin{aligned} e_{1} &= (\mathbf{concat}~a~b),\\ e_{2} &= (\mathbf{rerun}~a~b). \end{aligned}$ & $2$
\\ \midrule
oneliners & top-n & $6$ & \makecell[l]{\texttt{sed 100q}} & $\begin{aligned} & 26404 \\ &(=12440+13960+4) \end{aligned}$ & $258$ s & $\begin{aligned} e_{1} &= (\mathbf{rerun}~a~b). \end{aligned}$ & $1$
\\ \midrule
poets & trigram_rec & $13$ & \makecell[l]{\texttt{sed 5q}} & $\begin{aligned} & 26404 \\ &(=12440+13960+4) \end{aligned}$ & $183$ s & $\begin{aligned} e_{1} &= (\mathbf{rerun}~a~b). \end{aligned}$ & $1$
\\ \midrule
unix50 & 7.3: decades unix released & $5$ & \makecell[l]{\texttt{sed s/\textbackslash \$/'0s'/}} & $\begin{aligned} & 26404 \\ &(=12440+13960+4) \end{aligned}$ & $60$ s & $\begin{aligned} e_{1} &= (\mathbf{concat}~a~b). \end{aligned}$ & $1$
\\ \midrule
analytics-mts & vehicles per day & $5$ & \makecell[l]{\texttt{sort}} & $\begin{aligned} & 26404 \\ &(=12440+13960+4) \end{aligned}$ & $65$ s & $\begin{aligned} e_{1} &= (\mathbf{merge}~a~b),\\ e_{2} &= (\mathbf{merge}~b~a),\\ e_{3} &= (\mathbf{rerun}~a~b),\\ e_{4} &= (\mathbf{rerun}~b~a). \end{aligned}$ & $4$
\\ \midrule
poets & sort_words_by_folding & $6$ & \makecell[l]{\texttt{sort -f}} & $\begin{aligned} & 26404 \\ &(=12440+13960+4) \end{aligned}$ & $65$ s & $\begin{aligned} e_{1} &= (\mathbf{merge}(\texttt{'-f'})~a~b),\\ e_{2} &= (\mathbf{merge}(\texttt{'-f'})~b~a),\\ e_{3} &= (\mathbf{rerun}~a~b),\\ e_{4} &= (\mathbf{rerun}~b~a). \end{aligned}$ & $4$
\\ \midrule
analytics-mts & vehicle days on road & $7$ & \makecell[l]{\texttt{sort -k1n}} & $\begin{aligned} & 26404 \\ &(=12440+13960+4) \end{aligned}$ & $65$ s & $\begin{aligned} e_{1} &= (\mathbf{merge}(\texttt{'-k1n'})~a~b),\\ e_{2} &= (\mathbf{merge}(\texttt{'-k1n'})~b~a),\\ e_{3} &= (\mathbf{rerun}~a~b),\\ e_{4} &= (\mathbf{rerun}~b~a). \end{aligned}$ & $4$
\\ \midrule
oneliners & shortest-scripts & $6$ & \makecell[l]{\texttt{sort -n}} & $\begin{aligned} & 26404 \\ &(=12440+13960+4) \end{aligned}$ & $64$ s & $\begin{aligned} e_{1} &= (\mathbf{merge}(\texttt{'-n'})~a~b),\\ e_{2} &= (\mathbf{merge}(\texttt{'-n'})~b~a),\\ e_{3} &= (\mathbf{rerun}~a~b),\\ e_{4} &= (\mathbf{rerun}~b~a). \end{aligned}$ & $4$
\\ \midrule
poets & sort & $6$ & \makecell[l]{\texttt{sort -nr}} & $\begin{aligned} & 26404 \\ &(=12440+13960+4) \end{aligned}$ & $65$ s & $\begin{aligned} e_{1} &= (\mathbf{merge}(\texttt{'-nr'})~a~b),\\ e_{2} &= (\mathbf{merge}(\texttt{'-nr'})~b~a),\\ e_{3} &= (\mathbf{rerun}~a~b),\\ e_{4} &= (\mathbf{rerun}~b~a). \end{aligned}$ & $4$
\\ \midrule
oneliners & sort-sort & $3$ & \makecell[l]{\texttt{sort -r}} & $\begin{aligned} & 26404 \\ &(=12440+13960+4) \end{aligned}$ & $66$ s & $\begin{aligned} e_{1} &= (\mathbf{merge}(\texttt{'-r'})~a~b),\\ e_{2} &= (\mathbf{merge}(\texttt{'-r'})~b~a),\\ e_{3} &= (\mathbf{rerun}~a~b),\\ e_{4} &= (\mathbf{rerun}~b~a). \end{aligned}$ & $4$
\\ \midrule
oneliners & top-n & $5$ & \makecell[l]{\texttt{sort -rn}} & $\begin{aligned} & 26404 \\ &(=12440+13960+4) \end{aligned}$ & $65$ s & $\begin{aligned} e_{1} &= (\mathbf{merge}(\texttt{'-rn'})~a~b),\\ e_{2} &= (\mathbf{merge}(\texttt{'-rn'})~b~a),\\ e_{3} &= (\mathbf{rerun}~a~b),\\ e_{4} &= (\mathbf{rerun}~b~a). \end{aligned}$ & $4$
\\ \midrule
analytics-mts & vehicles per day & $3$ & \makecell[l]{\texttt{sort -u}} & $\begin{aligned} & 26404 \\ &(=12440+13960+4) \end{aligned}$ & $64$ s & $\begin{aligned} e_{1} &= (\mathbf{merge}(\texttt{'-u'})~a~b),\\ e_{2} &= (\mathbf{merge}(\texttt{'-u'})~b~a),\\ e_{3} &= (\mathbf{rerun}~a~b),\\ e_{4} &= (\mathbf{rerun}~b~a). \end{aligned}$ & $4$
\\ \midrule
unix50 & 4.6: piece used most & $9$ & \makecell[l]{\texttt{tail -n 1}} & $\begin{aligned} & 26404 \\ &(=12440+13960+4) \end{aligned}$ & $104$ s & $\begin{aligned} e_{1} &= (\mathbf{first}~b~a),\\ e_{2} &= (\mathbf{second}~a~b),\\ e_{3} &= ((\mathbf{back}~`\textbackslash n`~\mathbf{first})~b~a),\\ e_{4} &= ((\mathbf{fuse}~`\textbackslash n`~\mathbf{first})~b~a),\\ e_{5} &= ((\mathbf{back}~`\textbackslash n`~\mathbf{second})~a~b),\\ e_{6} &= ((\mathbf{fuse}~`\textbackslash n`~\mathbf{second})~a~b),\\ e_{7} &= (\mathbf{rerun}~a~b). \end{aligned}$ & $7$
\\ \midrule
unix50 & 4.4: histogram by piece & $1$ & \makecell[l]{\texttt{tr ' ' '\textbackslash n'}} & $\begin{aligned} & 2700 \\ &(=968+1728+4) \end{aligned}$ & $41$ s & $\begin{aligned} e_{1} &= (\mathbf{concat}~a~b),\\ e_{2} &= (\mathbf{rerun}~a~b). \end{aligned}$ & $2$
\\ \midrule
unix50 & 10.3: extract username & $7$ & \makecell[l]{\texttt{tr '[A-Z]' '[a-z]'}} & $\begin{aligned} & 26404 \\ &(=12440+13960+4) \end{aligned}$ & $60$ s & $\begin{aligned} e_{1} &= (\mathbf{concat}~a~b),\\ e_{2} &= (\mathbf{rerun}~a~b). \end{aligned}$ & $2$
\\ \midrule
unix50 & 4.5: histogram by piece and pawn & $6$ & \makecell[l]{\texttt{tr '[a-z]' 'P'}} & $\begin{aligned} & 26404 \\ &(=12440+13960+4) \end{aligned}$ & $60$ s & $\begin{aligned} e_{1} &= (\mathbf{concat}~a~b),\\ e_{2} &= (\mathbf{rerun}~a~b). \end{aligned}$ & $2$
\\ \midrule
poets & merge_upper & $3$ & \makecell[l]{\texttt{tr '[a-z]' '[A-Z]'}} & $\begin{aligned} & 26404 \\ &(=12440+13960+4) \end{aligned}$ & $60$ s & $\begin{aligned} e_{1} &= (\mathbf{concat}~a~b),\\ e_{2} &= (\mathbf{rerun}~a~b). \end{aligned}$ & $2$
\\ \midrule
unix50 & 9.1: extract word PORT & $3$ & \makecell[l]{\texttt{tr '[a-z]' '\textbackslash n'}} & $\begin{aligned} & 26404 \\ &(=12440+13960+4) \end{aligned}$ & $59$ s & $\begin{aligned} e_{1} &= (\mathbf{concat}~a~b),\\ e_{2} &= (\mathbf{rerun}~a~b). \end{aligned}$ & $2$
\\ \midrule
poets & count_vowel_seq & $3$ & \makecell[l]{\texttt{tr 'a-z' '[A-Z]'}} & $\begin{aligned} & 26404 \\ &(=12440+13960+4) \end{aligned}$ & $60$ s & $\begin{aligned} e_{1} &= (\mathbf{concat}~a~b),\\ e_{2} &= (\mathbf{rerun}~a~b). \end{aligned}$ & $2$
\\ \midrule
unix50 & 8.4: longest words w/o hyphens & $1$ & \makecell[l]{\texttt{tr -c "[a-z][A-Z]" '\textbackslash n'}} & $\begin{aligned} & 2700 \\ &(=968+1728+4) \end{aligned}$ & $40$ s & $\begin{aligned} e_{1} &= (\mathbf{concat}~a~b),\\ e_{2} &= (\mathbf{rerun}~a~b). \end{aligned}$ & $2$
\\ \midrule
unix50 & 9.6: follow directions & $9$ & \makecell[l]{\texttt{tr -c '[A-Z]' '\textbackslash n'}} & $\begin{aligned} & 2700 \\ &(=968+1728+4) \end{aligned}$ & $40$ s & $\begin{aligned} e_{1} &= (\mathbf{concat}~a~b),\\ e_{2} &= (\mathbf{rerun}~a~b). \end{aligned}$ & $2$
\\ \midrule
unix50 & 9.8: TELE-communications & $1$ & \makecell[l]{\texttt{tr -c '[a-z][A-Z]' '\textbackslash n'}} & $\begin{aligned} & 2700 \\ &(=968+1728+4) \end{aligned}$ & $40$ s & $\begin{aligned} e_{1} &= (\mathbf{concat}~a~b),\\ e_{2} &= (\mathbf{rerun}~a~b). \end{aligned}$ & $2$
\\ \midrule
oneliners & bi-grams & $1$ & \makecell[l]{\texttt{tr -cs A-Za-z '\textbackslash n'}} & $\begin{aligned} & 2700 \\ &(=968+1728+4) \end{aligned}$ & $40$ s & $\begin{aligned} e_{1} &= (\mathbf{rerun}~a~b). \end{aligned}$ & $1$
\\ \midrule
unix50 & 2.1: all Unix utilities & $2$ & \makecell[l]{\texttt{tr -d ','}} & $\begin{aligned} & 26404 \\ &(=12440+13960+4) \end{aligned}$ & $60$ s & $\begin{aligned} e_{1} &= (\mathbf{concat}~a~b),\\ e_{2} &= (\mathbf{rerun}~a~b). \end{aligned}$ & $2$
\\ \midrule
oneliners & spell & $5$ & \makecell[l]{\texttt{tr -d '[:punct:]'}} & $\begin{aligned} & 26404 \\ &(=12440+13960+4) \end{aligned}$ & $60$ s & $\begin{aligned} e_{1} &= (\mathbf{concat}~a~b),\\ e_{2} &= (\mathbf{rerun}~a~b). \end{aligned}$ & $2$
\\ \midrule
unix50 & 9.1: extract word PORT & $5$ & \makecell[l]{\texttt{tr -d '\textbackslash n'}} & $\begin{aligned} & 2700 \\ &(=968+1728+4) \end{aligned}$ & $40$ s & $\begin{aligned} e_{1} &= (\mathbf{concat}~a~b),\\ e_{2} &= (\mathbf{rerun}~a~b). \end{aligned}$ & $2$
\\ \midrule
unix50 & 7.2: most frequent machine & $6$ & \makecell[l]{\texttt{tr -s ' ' '\textbackslash n'}} & $\begin{aligned} & 2700 \\ &(=968+1728+4) \end{aligned}$ & $41$ s & $\begin{aligned} e_{1} &= (\mathbf{rerun}~a~b). \end{aligned}$ & $1$
\\ \midrule
poets & count_vowel_seq & $4$ & \makecell[l]{\texttt{tr -sc 'AEIOU' '[\textbackslash 012*]'}} & $\begin{aligned} & 2700 \\ &(=968+1728+4) \end{aligned}$ & $40$ s & $\begin{aligned} e_{1} &= (\mathbf{rerun}~a~b). \end{aligned}$ & $1$
\\ \midrule
poets & vowel_sequencies_gr_1K & $4$ & \makecell[l]{\texttt{tr -sc 'AEIOUaeiou' '[\textbackslash 012*]'}} & $\begin{aligned} & 2700 \\ &(=968+1728+4) \end{aligned}$ & $40$ s & $\begin{aligned} e_{1} &= (\mathbf{rerun}~a~b). \end{aligned}$ & $1$
\\ \midrule
poets & count_consonant_seq & $4$ & \makecell[l]{\texttt{tr -sc 'BCDFGHJKLMNPQRSTVWXYZ' '[\textbackslash 01}\\ \texttt{2*]'}} & $\begin{aligned} & 2700 \\ &(=968+1728+4) \end{aligned}$ & $41$ s & $\begin{aligned} e_{1} &= (\mathbf{rerun}~a~b). \end{aligned}$ & $1$
\\ \midrule
poets & merge_upper & $4$ & \makecell[l]{\texttt{tr -sc '[A-Z]' '[\textbackslash 012*]'}} & $\begin{aligned} & 2700 \\ &(=968+1728+4) \end{aligned}$ & $40$ s & $\begin{aligned} e_{1} &= (\mathbf{rerun}~a~b). \end{aligned}$ & $1$
\\ \midrule
poets & 2syllable_words & $3$ & \makecell[l]{\texttt{tr -sc '[A-Z][a-z]' ' [\textbackslash 012*]'}} & $\begin{aligned} & 2700 \\ &(=968+1728+4) \end{aligned}$ & $40$ s & $\begin{aligned} e_{1} &= (\mathbf{rerun}~a~b). \end{aligned}$ & $1$
\\ \midrule
poets & count_words & $3$ & \makecell[l]{\texttt{tr -sc '[A-Z][a-z]' '[\textbackslash 012*]'}} & $\begin{aligned} & 2700 \\ &(=968+1728+4) \end{aligned}$ & $40$ s & $\begin{aligned} e_{1} &= (\mathbf{rerun}~a~b). \end{aligned}$ & $1$
\\ \midrule
poets & sort_words_by_n_syllables & $5$ & \makecell[l]{\texttt{tr -sc '[AEIOUaeiou\textbackslash 012]' ' '}} & $\begin{aligned} & 26404 \\ &(=12440+13960+4) \end{aligned}$ & $60$ s & $\begin{aligned} e_{1} &= (\mathbf{concat}~a~b),\\ e_{2} &= (\mathbf{rerun}~a~b). \end{aligned}$ & $2$
\\ \midrule
oneliners & bi-grams & $2$ & \makecell[l]{\texttt{tr A-Z a-z}} & $\begin{aligned} & 26404 \\ &(=12440+13960+4) \end{aligned}$ & $60$ s & $\begin{aligned} e_{1} &= (\mathbf{concat}~a~b),\\ e_{2} &= (\mathbf{rerun}~a~b). \end{aligned}$ & $2$
\\ \midrule
oneliners & diff & $2$ & \makecell[l]{\texttt{tr [:lower:] [:upper:]}} & $\begin{aligned} & 26404 \\ &(=12440+13960+4) \end{aligned}$ & $60$ s & $\begin{aligned} e_{1} &= (\mathbf{concat}~a~b),\\ e_{2} &= (\mathbf{rerun}~a~b). \end{aligned}$ & $2$
\\ \midrule
oneliners & diff & $5$ & \makecell[l]{\texttt{tr [:upper:] [:lower:]}} & $\begin{aligned} & 26404 \\ &(=12440+13960+4) \end{aligned}$ & $59$ s & $\begin{aligned} e_{1} &= (\mathbf{concat}~a~b),\\ e_{2} &= (\mathbf{rerun}~a~b). \end{aligned}$ & $2$
\\ \midrule
oneliners & bi-grams & $5$ & \makecell[l]{\texttt{uniq}} & $\begin{aligned} & 26404 \\ &(=12440+13960+4) \end{aligned}$ & $60$ s & $\begin{aligned} e_{1} &= ((\mathbf{stitch}~\mathbf{first})~a~b),\\ e_{2} &= ((\mathbf{stitch}~\mathbf{second})~a~b),\\ e_{3} &= (\mathbf{rerun}~a~b). \end{aligned}$ & $3$
\\ \midrule
analytics-mts & vehicles per day & $6$ & \makecell[l]{\texttt{uniq -c}} & $\begin{aligned} & 26404 \\ &(=12440+13960+4) \end{aligned}$ & $59$ s & $\begin{aligned} e_{1} &= ((\mathbf{stitch2}~`~`~\mathbf{add}~\mathbf{first})~a~b),\\ e_{2} &= ((\mathbf{stitch2}~`~`~\mathbf{add}~\mathbf{second})~a~b). \end{aligned}$ & $2$
\\ \midrule
unix50 & 7.1: number of versions & $3$ & \makecell[l]{\texttt{wc -l}} & $\begin{aligned} & 2700 \\ &(=968+1728+4) \end{aligned}$ & $39$ s & $\begin{aligned} e_{1} &= ((\mathbf{back}~`\textbackslash n`~\mathbf{add})~a~b),\\ e_{2} &= ((\mathbf{back}~`\textbackslash n`~\mathbf{add})~b~a). \end{aligned}$ & $2$
\\ \midrule
oneliners & shortest-scripts & $4$ & \makecell[l]{\texttt{xargs -L 1 wc -l}} & $\begin{aligned} & 26404 \\ &(=12440+13960+4) \end{aligned}$ & $103$ s & $\begin{aligned} e_{1} &= (\mathbf{concat}~a~b),\\ e_{2} &= ((\mathbf{offset}~`~`~\mathbf{first})~a~b),\\ e_{3} &= ((\mathbf{offset}~`~`~\mathbf{second})~a~b). \end{aligned}$ & $3$
\\ \midrule
poets & count_words & $2$ & \makecell[l]{\texttt{xargs cat}} & $\begin{aligned} & 26404 \\ &(=12440+13960+4) \end{aligned}$ & $60$ s & $\begin{aligned} e_{1} &= (\mathbf{concat}~a~b),\\ e_{2} &= ((\mathbf{offset}~`~`~\mathbf{first})~a~b),\\ e_{3} &= ((\mathbf{offset}~`~`~\mathbf{second})~a~b). \end{aligned}$ & $3$
\\ \midrule
oneliners & shortest-scripts & $1$ & \makecell[l]{\texttt{xargs file}} & $\begin{aligned} & 26404 \\ &(=12440+13960+4) \end{aligned}$ & $61$ s & $\begin{aligned} e_{1} &= (\mathbf{concat}~a~b),\\ e_{2} &= ((\mathbf{offset}~`~`~\mathbf{second})~a~b). \end{aligned}$ & $2$
\\